\documentclass[a4paper,11pt]{article}
\pdfoutput=1 % if your are submitting a pdflatex (i.e. if you have
\usepackage{jcappub} % for details on the use of the package, please
\usepackage{lineno}
\usepackage{orcidlink}

\usepackage{array}

\usepackage{sectsty}
\allsectionsfont{\boldmath}

\usepackage{xspace}

\usepackage{tikz}
\usetikzlibrary{shapes.geometric, arrows}

\usepackage{float}

\usepackage{booktabs}
\usepackage{multirow}

\usepackage{graphicx}	% Including figure files
\usepackage{amsmath}	% Advanced maths commands
\usepackage{mathtools}
\usepackage{physics}
\usepackage[dvipsnames]{xcolor}
\usepackage{subcaption}
\usepackage[noabbrev,nameinlink]{cleveref}
\creflabelformat{equation}{#2#1#3} % removes the ( )
\usepackage{hyperref}
\crefname{appendix}{Appendix}{appendices}
\crefname{section}{Section}{sections}
\crefname{figure}{Figure}{figures}
\crefname{table}{Table}{tables}

\newcommand{\pollux}{{\tt Pollux}\xspace}

\usepackage{ulem}

\title{\boldmath $4\times3$ Point Correlation Functions in Galaxy Surveys: Impact of Baryonic Feedback
}

\author[a]{Avijit Bera,}
\author[b]{Joachim Harnois-D\'eraps,}
\author[c,d]{Juan Mena-Fern\'andez,}
\author[e]{Mike Jarvis,}
\author[c]{Cyrille Doux,}
\author[f]{Katrin Heitmann,}
\author[a]{Mustapha Ishak,}
\author[]{and The LSST Dark Energy Science Collaboration}

\affiliation[a]{Department of Physics, The University of Texas at Dallas, Richardson, TX 75080, USA}
\affiliation[b]{School of Mathematics, Statistics and Physics, Newcastle University, Herschel building, NE1 7RU, Newcastle-upon-Tyne, UK}
\affiliation[c]{Univ. Grenoble Alpes, CNRS, LPSC-IN2P3, 38000 Grenoble, France}
\affiliation[d]{Centre de Physique des Particules de Marseille, Aix Marseille Univ, CNRS/IN2P3, CPPM, Marseille, France}
\affiliation[e]{Department of Physics and Astronomy, The University of Pennsylvania, Philadelphia, PA 19104, USA}
\affiliation[f]{High Energy Physics Division, Argonne National Laboratory, 9700 S Cass Ave, Lemont, IL 60439, USA}

\emailAdd{avijit.bera@utdallas.edu}
\emailAdd{joachim.harnois-deraps@ncl.ac.uk}

\abstract{

We investigate the impact of baryonic feedback on 
two-point and three-point correlation functions (2PCFs and 3PCFs hereafter, respectively) involving galaxy density fields (g) and weak lensing shear fields (G), from simulated photometric catalogs of galaxies. Specifically, 
we baryonify high-resolution simulation using a baryonic correction model (BCM) and explore the consequences down to sub-arcminute (arcmin) scales, 
varying two model parameters with the largest impact on our probes: $M_{\rm c}$, which governs the amount of gas expelled beyond the halo boundary, and $\theta_{\rm ej}$, which encodes the maximal ejection radius relative to halo boundary. % We present nine baryonified maps: one for the fiducial case and four each corresponding to variations in the two parameters. 
We create lensing maps and galaxy catalogs assuming survey properties of the upcoming Year-10 data for the Vera C. Rubin Observatory's Legacy Survey of Space and Time (LSST), %Convergence maps are being created from these baryonified maps using Born approximation. In the next step, the galaxy catalogs are generated using Poisson sampling, where the galaxies are divided into five tomographic bins in the SRD Y1 setup for the LSST Dark Energy Science Collaboration (DESC). 
and investigate the impact of baryonic feedback on the observed correlations, 
including the galaxy--galaxy--shear (ggG) and the galaxy--shear--shear (gGG) 3PCFs, which are measured, for the first time from simulations, with \textsc{TreeCorr}. Focusing on equilateral 3PCFs, we find that small scales are more heavily affected by baryonic effects than the corresponding 2PCFs, by up to 90 percent depending on the probe, redshift and BCM model. 
The galaxy--galaxy--galaxy (ggg) 3PCF is significantly affected at scales smaller than about 4 arcmin; a similar effect occurs at 10 arcmin for the ggG 3PCF, at 40 arcmin for the gGG 3PCF, and at about a degree for the shear--shear--shear (GGG) 3PCF. This is consistent with the scales and redshift probed by these different statistics and the amount of projection. These four three-point statistics, which are collectively referred to as the $4\times3$PCFs, can be used at large scales to robustly constrain cosmological parameters. At smaller scales, their enhanced sensitivity to baryonic effects provides valuable leverage for constraining the BCM parameters and supplying informative priors.}

\begin{document}
\maketitle
\flushbottom

% \newpage
\section{Introduction}\label{sec:intro}
\par\noindent

%\textcolor{red}{Joachim}
Cosmic shear, defined as the weak gravitational lensing of background galaxy shapes by foreground large-scale structure LSS), is one of the best probes of dark matter \citep[see][for a review]{2015RPPh...78h6901K}. Recent measurements from the Dark Energy Survey\footnote{DES: \url{https://www.darkenergysurvey.org/}}, the Kilo Degree Survey\footnote{KiDS:  \url{https://kids.strw.leidenuniv.nl/}}, and the Hyper-Suprime Cam Survey\footnote{HSC:  \url{https://hsc.mtk.nao.ac.jp/ssp/survey/}} provide constraints on key cosmological parameters that reach a precision of a few percent \citep{desy6_3x2pt, desy6_cosmic_shear, DESY3_Secco, KiDSLegacy_Wright, HSCY3_Cl, HSCY3_2PCF}. The clustering of foreground galaxies, which trace the same LSS,
% large-scale structures, 
is highly complementary, especially when used in combination with galaxy-galaxy lensing to break the degeneracy between cosmology and galaxy bias \citep{ DESY3_2x2pta,DESY3_2x2ptb,KiDS1000_2x2, HSCY3_2x2}. The full combination of galaxy clustering, galaxy--galaxy lensing and cosmic shear data has been successfully analysed with two-point (2pt) statistics, leading to the so-called $3\times2$pt method \citep{KiDS1000_3x2, DESY3_3x2, HSCY3_3x2}, which has now become one of the golden probes of photometric galaxy surveys. 
% \footnote{LSST DESC- https://lsstdesc.org/}
Although, it can capture very well the cosmological information stored in the linear scales of the LSS, the main drawback of the 2pt statistics resides in their inability to access the higher-order information, mainly, encoded in small scales, where the non-linear regime of gravitational collapse produces a rich environment characterized by filaments, clusters and voids. Many ongoing efforts aim to develop alternative estimators that can probe this evasive information and complement the $3\times2$pt method. For cosmic shear alone, examples of `beyond-2pt' methods include lensing peak count \citep{DESY3_Zuercher, KiDS1000_JHD, HSCY1_peaks_sims, HSCY1_peaks_th}, marked correlation functions \citep{HSCmarked_correl2025}, Minkowski functionals \citep{2015PhRvD..91j3511P, HSCY1_Armijo}, persistent homology \citep{DESY1_Heydenreich, DESY3_Prat}, three-point correlation functions (3PCFs) \citep{KiDS1000_Map3, DESY3_3pt}, integrated three-point correlation functions (I3PCFs) \citep{Halder2021, Gong2023, gebauer2025}, lensing PDFs \citep{Friedrich_2022, barthelemy2024makingleapimodelling}, lensing one-point statistics \citep{DESY3_PDF, HSCY1_PDF}, etc. Projected galaxy clustering is mostly analysed with 2pt statistics \citep[see e.g.,][for a recent measurement and analysis from the first data release of the Dark Energy Spectroscopic Instrument]{DESIY1_Heydenreich, porredon2025desidr13times2pt}. Alternative probes that include both types of data are less common but include density-split statistics \citep{Gruen2017, KiDS1000_Burger}, 
% integrated bispectra \citep{Halder2023} 
and three-point (3pt) statistics \citep{2004ApJ...607..140J,2014ApJ...780..139G}. Most cases found a noticeable gain in constraining power and a significant complementarity with 2pt statistics, 
% most often 
typically at the cost of modeling the statistics directly from numerical simulations. 

Amongst the list of challenges faced by these new methods, the modeling of systematics currently stands out as a potential limiting factor. Indeed, the next generation of photometric data, provided by the Vera C. Rubin Observatory\footnote{Vera C. Rubin Observatory:  \url{rubinobservatory.org/}}, the ESA {\it Euclid} telescope\footnote{\textit{Euclid}:  \url{www.euclid-ec.org/}} or the Nancy Grace Roman Space telescope\footnote{Nancy Grace Roman Space Telescope:  \url{roman.gsfc.nasa.gov/}} will by systematics-limited, driving the community to drastically improve the understanding and modeling astrophysical or instrumental effect that contaminate the cosmological signal. In particular, baryonic feedback, caused by powerful astrophysical phenomena such as stellar winds and Active Galactic Nuclei (AGN),  has been recognized as a key source of uncertainty \citep[see][for a review]{Chisari_baryonsReview}. 
The impact of these phenomena on cosmological scales is best studied with hydrodynamical simulations \citep[e.g.,][]{kappaMTNG}. However, the physical prescriptions used to model AGN feedback differ from one simulation to another, and as a result, convergence on their predicted effects has not been achieved yet.
% However, the physics involved is complex enough that we have not yet reached convergence on their effect. 
In addition, hydrodynamical simulations are expensive to run, which limits our ability to explore the joint astrophysical and cosmological parameter space. 

An alternative and flexible approach to incorporate baryonic effects into simulations is through the use of baryonification, an effective technique that transforms the output from gravity-only simulations (often referred to as Dark matter-only simulation, hence labeled  DMO hereafter)  into a modified version that approximates the full matter distribution in the presence of baryonic feedback. Introduced by \cite{Schneider2015}, the Baryonic Correction Model (BCM) modifies the positions of particles in snapshots from gravity-only $N$-body simulations to recover the profile of various baryonic components in matter halos (made of both dark matter and baryonic matter), calibrated against full hydrodynamical simulations. This BCM model has been improved and refined \citep{Schneider_2019, Arico_2021, SchneiderBCM2025}, and some of its free parameters have been constrained directly \citep{Grandis2024}. Speed and flexibility are the main advantages of this method, which avoids all expensive computations, while recovering the results from hydrodynamical simulations at the percent level.
This is essential for modeling the impact of baryons on most beyond-2pt statistics. Furthermore, the original BCM model has been adapted to work on simulated density shells when particle data are not available \citep{Kacprzak_2023}. We have adapted this method in our work to process the output of a new high-resolution simulation described in a companion paper \citep{menafernandez2025}, with details provided in  \cref{subsec:Nbody}.

Recent observational studies combining X-ray, kinematic Sunyaev-Zel'dovich (kSZ), and weak lensing measurements have revealed strong evidence for efficient baryonic feedback leading to substantial gas expulsion from galaxy groups and clusters. Analyses using eROSITA X-ray gas fractions and ACT kSZ data \citep{kovac2025, Siegel2025} consistently suggest stronger feedback than that implemented in most hydrodynamical simulations calibrated to pre-eROSITA data. These findings point to significant suppression of the matter power spectrum exceeding the percent level at  $k>0.3\ h\ {\rm Mpc}^{-1}$ and reaching 20 - 25 \% at $k=5\ h\ {\rm Mpc}^{-1}$. They imply that enhanced AGN activity may be responsible for the large-scale gas depletion and corresponding modification of the matter distribution. Moreover, these joint analyses demonstrate that combining multiwavelength data enables robust constraints on feedback-related parameters such as gas ejection efficiency and AGN heating strength, offering a pathway to break degeneracies between baryonic physics and cosmological parameters. Together, they provide a self-consistent picture of baryonic feedback across mass and length scales, emphasizing the importance of integrating such constraints into cosmological modeling. They also directly constrain the BCM parameters \citep{Grandis2024}, which allows for the improved understanding of feedback processes.
A few steps have been taken to model the baryonic feedback in higher-order statistics for the next-generation surveys: \cite{Burger_2026}\ present a neural-network emulator for baryonic effects on the matter bispectrum, trained on BACCO gravity-only simulations augmented with baryonification and validated against FLAMINGO hydrodynamical simulation, achieving $\leq 2\%$ accuracy 
% (68th percentile) 
across most triangle configurations over $k \in [0.01,\,20]\,h\,\mathrm{Mpc}^{-1}$. 
Complementarily, \citep{Halder2022} developed a response-function model for the shear I3PCF 
% ($\zeta_{\pm}$) 
that is accurate on sub-arcmin scales (where standard bispectrum fits start to fail at a few tenths of an arcmin), reproduces measurements from simulated shear maps, and incorporates baryonic feedback primarily through its impact on the nonlinear matter power spectrum. These strategies will be pivotal for modeling baryonic-feedback to $4\times 3$PCFs and corresponding analyes in galaxy surveys. 
% \cite{HalderBarreira2022_arXiv2201.05607,HalderBarreira2022_arXiv2201.05607pdf}

This paper presents a natural extension of the $3\times2$pt method, where 
% the galaxy clustering (g) and lensing (G) 
the galaxy density field (g) and the weak lensing shear field (G) are instead analysed with 3PCFs, leading to four different combinations: ggg, ggG, gGG and GGG. The resulting $4\times3$PCFs measurements can capture additional information and, despite their increased length of the data vectors, merit our full attention for a number of reasons. First, although the full theoretical modeling of these statistics in configuration space is % not complete yet,
computationally challenging, some key steps have been taken, mostly in Fourier space (i.e., $k$-space), and involve the matter bispectrum ($B(k_1,k_2,k_3)$), which can be modeled with fitting functions \citep[e.g., \textsc{BiHalofit}, described in][]{Bihalofit} or with Effective Field Theory \citep{BispectrumEFT}. Second, measurement tools (e.g.,\textsc{TreeCorr}\footnote{\textsc{TreeCorr}: \url{rmjarvis.github.io/TreeCorr}}, \textsc{Triumvirate}\footnote{\textsc{Triumvirate}: \url{triumvirate.readthedocs.io/en}}) for these methods are efficient and available publicly. In the latest release of \textsc{TreeCorr} (Version 5.1.3), it supports the computation of 3pt cross-correlations between different spin quantities (e.g., galaxy position with spin-0 and the galaxy shape with spin-2). This newly added feature allows for measuring the ggG and the gGG type correlations,  hence the full $4\times3$PCFs, for the first time.

While rapid progress is being made on the theoretical side, including, for example, fast modeling of cosmic shear 3PCF through multipole expansion of the bispectrum \citep{sugiyama2024fastmodelingshearthreepoint}, a substantial amount of work is required to model the full $4\times3$PCFs.
Additionally, no analytical model has been developed to include the impact of baryonic feedback on these statistics. 
% \avi{Hence, we rely on the baryonified simulation to obtain the $4\times3$pt data vector and compare the same with the DMO only case to explore the impact of baryonic feedback modeled using parametrd baryonification technique.}
Hence, we rely on the DMO simulation and on the BCM to obtain $4\times3$PCFs, and explore the impact of baryonic feedback based on the observed differences.
% we baryonify a high-resolution $N$-body simulation and explore the dependence of the $4\times3$pt data vector on the key baryonification parameters. 
We perform this investigation in the context of the upcoming data releases of the Vera C. Rubin Observatory's Legacy Survey of Space and Time (LSST), however, our results are general and can be easily extended to different surveys.

This paper is structured as follows. We review the theoretical background for $3\times2$PCFs and $4\times3$PCFs
% our different estimators 
in \cref{sec:background}. \cref{sec:simulations} introduces the simulations and the BCM methods. Our results are presented and discussed in \cref{sec:results}, while \cref{sec:conclusions} contains our final conclusions and opens up on possible future directions. Additionally, we provide the relations between the 3PCFs and the corresponding bispectrum in \cref{App:theory_3PCF}. The complete expressions of the estimators are shown in \cref{App:estimators}. In \cref{App:Treecorr_benchmark}, we outline the accuracy requirements for measuring the 3PCFs and compare the computation time needed for varying numbers of objects in the galaxy catalog. We also present full data vectors for $3\times2$PCFs and $4\times3$PCFs for relevant combinations of tomographic bins with baryonic feedback in \cref{App:Data_vectors}.

\section{Background}\label{sec:background}
\par\noindent
\crefalias{section}{section}

%\JHD{Needs some context here. Introduce lensing and clustering, and use the gG notation... We could also use $\gamma$-2PCF and $\gamma$-3PCF notation}

Galaxy surveys provide access to two distinct fields that trace the matter distribution: the galaxy density field (g) and the weak lensing shear field (G)\footnote{ The intrinsic alignment (IA) of galaxies is a known contaminant to the observed shear, but we do not consider the impact of IA in this study. For a detailed analysis on the impact of IA on higher order statistics, we refer to \citep{harnoisderaps_2025}.}. These two fields give rise to three types of measurable two-point correlation functions ($3\times2$PCFs), and four types of three-point correlation functions ($4\times3$PCFs), which altogether encode complementary information about the underlying cosmological structures. % Each of which captures how mass is clustered in the Universe. 
In this section, we review the mathematical background of these correlations. % and present how these are estimated from our simulations. 

Galaxy positions are described by the spin-0 over-density field $\delta_{\rm g}$, whereas galaxy shapes are described by spin-2 ellipticity fields, which can be defined along some reference direction $(\boldsymbol{\theta})$ as \( \varepsilon(\boldsymbol{\theta}) = \varepsilon_{\rm t}(\boldsymbol{\theta}) + \mathrm{i} \cdot \varepsilon_\times (\boldsymbol{\theta}) \). Here, $\varepsilon_{\rm t}$ and $\varepsilon_\times$ are the ellipticity components oriented perpendicular and at $45^{\circ}$  compared to the reference direction. 
In practice, $\boldsymbol{\theta}$ is taken as the vector connecting the pair of source galaxies, and its magnitude is denoted by $\vartheta$. 
%In this section, we discuss the theory of the observable 2pt and 3pt correlation functions in a weak lensing survey. 

Note that in this study, we employ the same set of galaxies as both lenses and sources, enabling a self-consistent analysis of the lensing and clustering signals within a unified sample. This, however, is not a requirement for our methods and conclusions to hold.

% \subsection{Cosmic shear}

% \begin{figure*}
%     \centering
%     % First row
%     \begin{subfigure}[t]{0.48\linewidth}
%         \centering
%         \includegraphics[width=\linewidth]{figures/nshells_51_imap_08_dmb_map.pdf}
%         \caption{dmb}
%         \label{fig:dmb}
%     \end{subfigure}
%     \hfill
%     \begin{subfigure}[t]{0.48\linewidth}
%         \centering
%         \includegraphics[width=\linewidth]{figures/nshells_51_imap_08_dmo_map.pdf}
%         \caption{dmo}
%         \label{fig:dmo}
%     \end{subfigure}

%     \vspace{1em} % Space between rows

%     % Second row
%     \begin{subfigure}[t]{0.48\linewidth}
%         \centering
%         \includegraphics[width=\linewidth]{figures/nshells_51_imap_08_diff_map.pdf}
%         \caption{difference (dmb-dmo)}
%         \label{fig:diff}
%     \end{subfigure}

%     \caption{Projected maps for \texttt{imap} =08 (a) dmb, (b) dmo, and (c) difference.}
%     \label{fig:map_comparison}
% \end{figure*}

\subsection{$3\times2$PCFs -- theory}\label{subsec:2pt_theory}
\par\noindent

This section presents the $3\times2$pt statistics, along with the complete expressions to model those both in harmonic and configuration space. A detailed description of the estimators for these statistics can be found in \cref{App:estimators_2PCF}. 
% We measure all 2PCFs, presented in this work, with the public software \textsc{TreeCorr} \citep{Jarvis_2004}, with internal configuration parameters detailed in \cref{sec:results}. 

% \subsubsection{Matter power spectra $C^{\delta\delta}_{\ell}$}
% \par\noindent

% \subsubsection{Convergence power spectra $C^{\kappa\kappa}_{\ell}$}
% \par\noindent

%%-----------------------------------------------------------
\subsubsection*{Shear--shear 2PCF $(\xi_\pm(\vartheta))$}
\par\noindent

The shear--shear 2PCF, or shape--shape (GG) correlation, is a statistic that quantifies the amount of coherent distortion imparted on the shapes of distant galaxies by the foreground large-scale structure (LSS) due to weak gravitational lensing.
% These correlations arise because light from background galaxies is deflected by the same intervening matter distributed in the large-scale structure of the Universe. }
% Specifically, it captures the degree of alignment between galaxy ellipticities as a function of their angular separation on the sky. 
The two components of the shear--shear 2PCF are defined by \citep{Bartelmann_2001, Kilbinger_2015}
\begin{eqnarray}\label{eq:1}
    \xi_{+}^{ij} (\vartheta) &\equiv& \langle \varepsilon^i \varepsilon^{*j} \rangle (\vartheta) = \langle \varepsilon_{\rm t}^i \varepsilon_{\rm t}^j\rangle (\vartheta) + \langle \varepsilon_\times^i \varepsilon_\times^j \rangle (\vartheta) \equiv {\varepsilon^{ij}_{\rm tt} + \varepsilon^{ij}_{\times\times}},
    \\ \label{eq:2}
    \xi_{-}^{ij} (\vartheta) &\equiv& \langle \varepsilon^i \varepsilon^j \rangle (\vartheta) = \langle \varepsilon_{\rm t}^i \varepsilon_{\rm t}^j\rangle (\vartheta) - \langle \varepsilon_\times^i \varepsilon_\times^j \rangle (\vartheta) \equiv {\varepsilon^{ij}_{\rm tt} - \varepsilon^{ij}_{\times\times}},
\end{eqnarray}
where the angle brackets denote the ensemble average over all possible pairs of galaxies, $(i,j)$ refer to the tomographic bins, and the right-hand equivalence is to use shorthand notations of these correlation components.  We omit the imaginary components above, since they vanish for a parity-invariant shear field. We assume uniform shape measurement weights for all galaxies, but in practice, these would vary per object. Note that `$^*$' denotes complex conjugation.  
% A detailed description of all our estimators can be found in \cref{App:B1.3}.

The observed shear--shear 2PCFs can be theoretically modeled as 
\begin{eqnarray}\label{eq:3}
    \xi_{\pm}^{ij}(\vartheta) = \frac{1}{2\pi} \int \ell J_{0/4}(\ell \vartheta) C_{ij}^{\rm GG}(\ell) d\ell, 
\end{eqnarray}
where $J_{\alpha}\ (\alpha=0,4)$ is the $\alpha^{\rm th}$-order Bessel function of the first kind. The shear--shear angular power spectrum $C_{ij}^{\rm GG}(\ell)$ is related to the matter power spectrum $P_{\rm mm}(k,z)$, under the Limber approximation, as 
\begin{eqnarray}\label{eq:4}
    C_{ij}^{\rm GG}(\ell) = \int_0^{\chi_{\rm h}} \frac{g_i(\chi)g_j(\chi)}{\chi^2} P_{\rm mm}\left(k=\frac{\ell+\frac{1}{2}}{\chi},z\right) d\chi, 
\end{eqnarray}
where $\chi_{\rm h}$ is the comoving horizon distance; we assume a flat universe such that the transverse angular diameter distance $f_{K}(\chi) = \chi$. The lensing kernel is given by 
\begin{eqnarray}\label{eq:5}
    g_{i}(\chi) = \frac{3}{2} \frac{H_0^2 \Omega_{\rm m}}{c^2} \frac{\chi}{a(\chi)} \int_0^{\chi_{\rm h}} n_i(\chi') \frac{\chi'-\chi}{\chi} d\chi',
    \label{eq:kernel}
\end{eqnarray}
where $n_i(z)$ is the normalized redshift distribution of the source galaxies, which follows $n_i(\chi)d\chi = n_i(z)dz$, and $a(\chi)$ is the expansion scale factor at a comoving distance $\chi$ away from the observer.  $H_0$, $\Omega_{\rm m}$, and $c$ correspond to the present-day value of the Hubble constant, the matter energy density parameter and the speed of light in vacuum, respectively.

%%-----------------------------------------------------------
\subsubsection*{Galaxy--shear 2PCF $(\gamma_{\rm t}(\vartheta))$}
\par\noindent

The galaxy--shear 2PCF or position--shape
% galaxy-shear  
(gG)\footnote{Also known as or galaxy--galaxy lensing (GGL) correlation.} correlation, is a statistic that measures the average tangential shear profile of background (source) galaxies around foreground (lens) galaxies, effectively measuring the projected mass density profile around the lenses.
% Such a correlation arises as the distribution of foreground (lens) galaxies and the distortion of background (source) galaxy shapes due to weak gravitational lensing are both caused by the large scale structure of the universe.}
The tangential shear profile is defined as 
\begin{eqnarray}\label{eq:6}
    \gamma_{\rm t}^{ij} (\vartheta) \equiv \langle \delta_{\rm{g}}^i \varepsilon^j \rangle (\vartheta) = \langle \delta_{\rm{g}}^i \varepsilon_{\rm t}^j \rangle (\vartheta) \equiv 
    % \varepsilon^{ij}_{\rm gt}
    {\delta_{\rm{g}}^{i} \varepsilon^{j}_{\rm t}},
\end{eqnarray}
understanding that the imaginary components vanish due to parity invariance. %, and the right-hand equivalence is to use shorthand notations of these correlation component. 
% The GGL estimator is defined as
% \begin{eqnarray}\label{eq:7}
%     \widehat{\gamma}_{\rm t}^{ij} (\vartheta) =  \langle (n_{\rm g}^i - n^i_{\rm R})\varepsilon^{j} \rangle , 
% \end{eqnarray} 
% where the tangential component of the ellipticities is measured in the $j^{\rm th}$ bin, relative to the lens galaxies $ (n_{\rm g}^{i})$ first, and then relative to random points $(n^i_{\rm R})$ corresponding to the $i^{\rm th}$ bin. Note that when assigned to a grid, the $n_{\rm g}$ galaxies are used to construct the galaxy over-density field $\delta_{\rm g}$ introduced earlier. 

%\begin{eqnarray}\label{eq:15}
%    \Hat{\gamma}_{\rm t}(\vartheta) = S(D-R),
%\end{eqnarray} 
%where $S_{\rm t}D$ is the sum over all position-source pairs with separations $\vartheta$ of the tangential component of shear in the data catalog, and can be expressed as  
%\begin{eqnarray}\label{eq:16}
%    SD =  \frac{\sum_{ij} w_i w_j  n^D_i \epsilon_{j} } {\sum_{ij} w_i w_j }. 
%\end{eqnarray}
%where the tangential component of the ellipticity of the galaxy at location $j\ (\epsilon_{j})$ is measured relative to the galaxy at location $i\ (n_i^D)$.}

This galaxy--shear 2PCF can be modeled from the galaxy--shear cross-spectrum $C_{ij}^{\rm gG}(\ell)$ as:
\begin{eqnarray}\label{eq:7}
    \gamma_{\rm t}^{ij}(\vartheta) = \frac{1}{2\pi} \int \ell J_{2}(\ell \vartheta) C_{ij}^{\rm gG}(\ell) d\ell, 
\end{eqnarray}
where $J_{2}$ is the $2^{\rm nd}$-order Bessel function of first kind. The  $C_{ij}^{\rm gG}(\ell)$ term is related to the matter power spectrum $P_{\rm mm}(k,z)$ under the Limber approximation as 
\begin{eqnarray}\label{eq:8}
    C_{ij}^{\rm gG}(\ell) = \int_0^{\chi_{\rm h}} \frac{n_i(\chi) g_j(\chi)}{\chi^2} b_i^{\rm lin}P_{\rm mm}\left(k=\frac{\ell+\frac{1}{2}}{\chi},z\right) d\chi, 
\end{eqnarray}
where we assume a linear galaxy bias, i.e., the galaxy overdensity field is linearly related to the matter overdensity field and can be written as $\delta_{\rm g} \approx b^{\rm lin} \delta_{\rm m}$.
Therefore, the matter--galaxy 3D cross-correlation power spectrum can be approximated as $P_{\rm gm} = b^{\rm lin} P_{\rm mm}$, and $b_i^{\rm lin}$ is the linear bias parameter for the $i^{\rm th}$ tomographic bin. The lensing kernel $g_i(\chi)$ is given by \autoref{eq:kernel}. 

\subsubsection*{Galaxy--galaxy 2PCF ($w(\vartheta)$)}
\par\noindent

The galaxy--galaxy 2PCF or position--position (gg)\footnote{Also known as galaxy clustering 2PCF.} correlation, measures the excess probability of finding a galaxy pair at a given distance, compared to a random distribution.
The galaxy--galaxy 2PCF is defined as 
\begin{eqnarray}\label{eq:9}
    w^{ij} (\vartheta) \equiv \langle \delta_{\rm{g}}^i \delta_{\rm{g}}^j \rangle (\vartheta) \equiv \delta_{\rm gg}^{ij}.
\end{eqnarray}
% and is measured with the Landy-Szalay estimator \citep{LS}
% \begin{eqnarray}\label{eq:11}
%        \widehat{w}^{ij}(\vartheta) = \frac{D^iD^j - D^iR^j - R^iD^j + R^iR^j}{R^iR^j}.
% \end{eqnarray} 
% Here, $DD$ and $RR$ are the sum over all object pairs with separation $\vartheta$ from the data catalog and the random catalog, respectively.  Similarly, $DR$ and $RD$ are the sum over all data-random pairs with separation $\vartheta$. In the absence of weights, the $DD$ term in the tomographic setting can be expressed as 
% \begin{eqnarray}\label{eq:12}
%    %DD = \frac{\sum_{ij} w_i w_j  n^D_i n^D_j } {\sum_{ij} w_i w_j },
%    {D^iD^j} = \sum n_{\rm g}^i n_{\rm g}^j \, ,
% \end{eqnarray}
%  and the $DR, RD$ and $RR$ terms can be expressed similarly by replacing $n_{\rm g}$ by $n_{\rm R}$ appropriately.  %\avi{ A detailed description of this estimators can be found in \cref{App:B1.1}.}

Modeling the galaxy--galaxy 2PCFs is  achieved via 
\begin{eqnarray}\label{eq:10}
    w^{ij}(\vartheta) = \frac{1}{2\pi} \int \ell J_{0}(\ell \vartheta) C_{ij}^{\rm gg}(\ell) d\ell, 
\end{eqnarray}
where $J_{0}$ is the $0^{\rm th}$-order Bessel function of first kind. 
Under linear galaxy bias approximation, 
% the galaxy overdensity field can be written as $\delta_{\rm g} \approx b^{\rm lin} \delta_{\rm m}$. 
the galaxy--galaxy angular power spectrum $C_{ij}^{\rm gg}(\ell)$ can be related to the matter power spectrum $P_{\rm mm}(k,z)$  as 
\begin{eqnarray}\label{eq:11}
    C_{ij}^{\rm gg}(\ell) = \int_0^{\chi_{\rm h}} \frac{n_i(\chi) n_j(\chi)}{\chi^2} b_i^{\rm lin}b_j^{\rm lin} P_{\rm mm}\left(k=\frac{\ell+\frac{1}{2}}{\chi},z\right) d\chi, 
\end{eqnarray}
where the galaxy--galaxy 3D power spectrum is approximated as $P_{\rm gg}(k, z) = \left(b^{\rm lin}\right)^2P_{\rm mm}(k, z)$. %\avi{$b_i^{\rm lin}$ and $b_j^{\rm lin}$ are the linear bias parameters for the $i^{\rm th}$ and $j^{\rm th}$ tomographic bins, respectively.}

We emphasize that a non-linear galaxy bias with higher order terms would be required to model the $\gamma_{\rm t}^{ij}(\vartheta)$ and $w^{ij}(\vartheta)$ correlations measured from data on small scales. For this work, however, we restrict ourselves to a linear galaxy bias, as justified in \cref{subsec:Nbody}.

\subsection{$4\times3$PCFs -- theory}\label{subsec:3pt_theory}
\par\noindent

This section presents the $4\times3$PCFs, and builds on the notation presented previously. As mentioned in the introduction, theoretical predictions exist for these statistics, but are of limited accuracy and computationally challenging; hence, we do not discuss them in the main text. We provide a comprehensive summary of these statistics in \cref{App:theory_3PCF}. 
A major challenge is that 3PCFs  span a broad range of triangle configurations \citep[e.g., see Figure 4 of][]{Heydenreich_2023}. Among these, equilateral, isosceles, and squeezed configurations are some of the most commonly studied in the literature \citep{DESY3_Secco}. Although these all have pros and cons, our work focuses on equilateral triangle configurations, where all 3PCFs can be expressed with a single variable, the side of an equilateral triangle in terms of the angular separation $\vartheta_{\rm eq}\ (\approx \vartheta_1\approx \vartheta_2\approx \vartheta_3)$, and many statistics have a vanishing imaginary part. A detailed description of the estimators for these statistics can be found in \cref{App:estimators_3PCF}.

\subsubsection*{Shear--shear--shear 3PCF $(\Gamma_{(\alpha)}(\vartheta_1, \vartheta_2, \vartheta_3))$}
\par\noindent

Shear--shear--shear 3PCF, or shape--shape--shape (GGG) correlation, is the natural extension of the shear--shear 2PCF and comes with an increased complexity. Whereas $\xi_\pm(\vartheta)$ can be defined without any reference direction, no tri-linear scalar can be formed with three two-component quantities alone % but it is quite complicated to define 3PCFs, since with three two-component quantities alone, no tri-linear scalar can be formed 
\citep{Schneider_2003, Schneider_2005a}. 

Though the choice of orientation for shear projection is less obvious (e.g., the orthocenter of the triangle or the side direction), ``natural components'' for the shear--shear--shear 3PCF have been developed by \cite{Schneider_2003}, %. These %natural components 
with rotation properties analogous to $\xi_\pm(\vartheta)$. Following their notation, %\cite{Schneider_2003}, 
the four natural components for a triplet of galaxies from three different tomographic bins ($i, j, k$) located at angular positions $\boldsymbol{\theta}_1, \boldsymbol{\theta}_2$, and $\boldsymbol{\theta}_3$, respectively, are defined as  
\begin{eqnarray}\nonumber 
    \Gamma^{ijk}_{(0)}(\vartheta_1, \vartheta_2, \vartheta_3) &\equiv& \langle \varepsilon^i(\vartheta_1)\varepsilon^j(\vartheta_2)\varepsilon^k(\vartheta_3)\rangle\\ \label{eq:Gamma_0}
    &\equiv& \varepsilon^{ijk}_{\rm ttt} - \varepsilon^{ijk}_{\rm  t\times\times} - \varepsilon^{ijk}_{\times \rm  t \times} - \varepsilon^{ijk}_{\times\times \rm t}  + \mathrm{i}\cdot \big[ \varepsilon^{ijk}_{\rm tt\times}+ \varepsilon^{ijk}_{\rm  t\times t} + \varepsilon^{ijk}_{\rm \times tt} - \varepsilon^{ijk}_{\times\times\times } \big], \\ \nonumber
    \Gamma^{ijk}_{(1)}(\vartheta_1, \vartheta_2, \vartheta_3) &\equiv& \langle \varepsilon^{*i}(\vartheta_1)\varepsilon^j(\vartheta_2)\varepsilon^k(\vartheta_3)\rangle \\ \label{eq:Gamma_1}
    &\equiv& \varepsilon^{ijk}_{\rm ttt} - \varepsilon^{ijk}_{\rm t\times\times} + \varepsilon^{ijk}_{\rm \times t \times} + \varepsilon^{ijk}_{\times\times \rm  t} + \mathrm{i}\cdot \big[ \varepsilon^{ijk}_{\rm tt\times} + \varepsilon^{ijk}_{\rm t\times t} - \varepsilon^{ijk}_{\rm \times tt} + \varepsilon^{ijk}_{\times\times\times } \big],\\ \nonumber
    \Gamma^{ijk}_{(2)}(\vartheta_1, \vartheta_2, \vartheta_3) &\equiv& \langle \varepsilon^i(\vartheta_1)\varepsilon^{*j}(\vartheta_2)\varepsilon^k(\vartheta_3)\rangle \\ \label{eq:Gamma_2}
    &\equiv& \varepsilon^{ijk}_{\rm ttt} + \varepsilon^{ijk}_{\rm t\times\times} - \varepsilon^{ijk}_{\times \rm t \times} + \varepsilon^{ijk}_{\times\times \rm t} + \mathrm{i}\cdot \big[ \varepsilon^{ijk}_{\rm tt\times} - \varepsilon^{ijk}_{\rm t\times t} + \varepsilon^{ijk}_{\times \rm tt} + \varepsilon^{ijk}_{\times\times\times } \big],\\  \nonumber
    \Gamma^{ijk}_{(3)}(\vartheta_1, \vartheta_2, \vartheta_3) &\equiv& \langle \varepsilon^i(\vartheta_1)\varepsilon^j(\vartheta_2)\varepsilon^{*k}(\vartheta_3)\rangle \\ \label{eq:Gamma_3}
    &\equiv& \varepsilon^{ijk}_{\rm ttt} + \varepsilon^{ijk}_{\rm t\times\times} + \varepsilon^{ijk}_{\times \rm t \times} - \varepsilon^{ijk}_{\times\times \rm t} + \mathrm{i}\cdot \big[- \varepsilon^{ijk}_{\rm tt\times} + \varepsilon^{ijk}_{\rm t\times t} + \varepsilon^{ijk}_{\times \rm tt} + \varepsilon^{ijk}_{\times\times\times } \big].
\end{eqnarray}

Analytical expressions that connect the convergence--convergence--convergence bispectrum ($B_{\kappa\kappa\kappa}$) to these natural components can be found in \cref{App:sss_3PCF}.

%\JHD{Can you provide an intuition for what these natural components correspond to, or probe? Also, I would add "We will use these natural components in this paper." }

\subsubsection*{Galaxy--shear--shear 3PCF $(G_\pm(\vartheta_1, \vartheta_2, \vartheta_3))$}
%\avi{It has been used in previous studies e.g. - \textit{Measuring galaxy-galaxy-galaxy-lensing with higher precision and accuracy}, \citep{Linke_2020}}
\par\noindent

Galaxy--shear--shear 3PCF
% (GGLGL) \mj{[MJ: I don't think GGLGL is a helpful acronym.  Just call this galaxy-shear-shear or gGG.  That's enough names for this combination.]} 
or the position--shape--shape (gGG)\footnote{Also known as galaxy--galaxy lensing--galaxy lensing as first developed in \citep{Schneider_2005b}.} correlation captures the correlation between the position of a single lens galaxy and the tangential shears of a pair of source galaxies 
% (and is therefore also referred to as gGG correlation), 
effectively measuring the projected mass distribution around a lens galaxy.

For a foreground lens galaxy at $\boldsymbol{\theta}_1$ (bin $i$) and two background source galaxies located at $\boldsymbol{\theta}_2$ (bin $j$) and $\boldsymbol{\theta}_3$ (bin $k$), natural components of the gGG correlation can be defined using the convention of \cite{Schneider_2005b} as
\begin{eqnarray} \label{eq:G_+}
    G_+^{ijk}(\vartheta_1, \vartheta_2, \vartheta_3) &\equiv& \langle \delta^i_{\mathrm{g}}(\vartheta_1)\varepsilon^j(\vartheta_2)\varepsilon^{*k}(\vartheta_3)\rangle \equiv \delta^i_{\rm g}\varepsilon^{jk}_{\rm tt} + \delta^i_{\rm g}\varepsilon^{jk}_{\times\times} + \mathrm{i}\cdot \big[ \delta^i_{\rm g}\varepsilon^{jk}_{\times \rm t} - \delta^i_{\rm g}\varepsilon^{jk}_{\rm t\times}  \big],  \\  \label{eq:G_-}
    G_-^{ijk}(\vartheta_1, \vartheta_2, \vartheta_3) &\equiv& \langle \delta^i_{\mathrm{g}}(\vartheta_1)\varepsilon^j(\vartheta_2)\varepsilon^k(\vartheta_3)\rangle \equiv \delta^i_{\rm g}\varepsilon^{jk}_{\rm tt} - \delta^i_{\rm g}\varepsilon^{jk}_{\times\times} + \mathrm{i}\cdot \big[ \delta^i_{\rm g}\varepsilon^{jk}_{\times \rm t} + \delta^i_{\rm g}\varepsilon^{jk}_{\rm t\times} \big]. 
\end{eqnarray}
%where the right-hand equivalence is to use shorthand notations of these correlation components.
%\JHD{Can you provide an intuition for what these probe? Is $G_-$ also measuring smaller scales compared to $G_+$, just like to $\xi_\pm$? Maybe also explain here that this combination of gGG is particularly sensitive to XYZ, that is difficult to extract from GGL.}

This statistic is sensitive to the non-Gaussian features of both the matter density field and shear field and provides complementary information to $\gamma_{\rm t}^{ij}(\vartheta)$. The relation between these components and the galaxy--convergence--convergence bispectrum ($B_{{\rm g}\kappa\kappa}$) can be found in \cref{App:gss_3PCF}. To the best of our knowledge, these components are measured for the first time in this paper through simulations. Depending on the ordering of the lens and source tomographic bins, we can construct three observables (gGG, GgG, GGg), and the estimators for all three types of correlations\footnote{Note that \textsc{TreeCorr} uses a different notation: galaxy tracers, or other objects used in number count statistics, are referred to as `N', whereas we use `g'.} are described in \cref{App:estimator_gss}.
% \begin{eqnarray}\label{eq:}
%    \widehat{G}_{\pm}^{ijk}(\vartheta_1, \vartheta_2, \vartheta_3) = \avi{S_{\pm}^jS_{\pm}^k(D^i-R^i)},
% \end{eqnarray}
% \begin{eqnarray}\label{eq:21}
%    \widehat{G}_{\pm}^{ijk}(\vartheta_1, \vartheta_2, \vartheta_3) = 
%    \begin{cases}
%        S_{\pm}^iS_{\pm}^j(D^k-R^k); & \text{GGg config.} \\
%        S_{\pm}^i(D^j-R^j)S_{\pm}^k; & \text{GgG config.} \\
%        (D^i-R^i)S_{\pm}^jS_{\pm}^k; & \text{gGG config.}
%    \end{cases}
% \end{eqnarray}
% where $DS_{\pm}S_{\pm}$ is the position-shape-shape correlation averaged over all triplets of galaxies. In our convention, the two background galaxies are located at $\vartheta_2$ and $\vartheta_3$ relative to the foreground galaxy from the data (or random) catalog at $\vartheta_1$. \textsc{TreeCorr} can compute all three configurations gGG, GgG, and GGg. 
In this paper, we utilize the former only and we choose the lens bins to be at a lower redshift compared to the source bins (i.e., $i < j, k$).  %A detailed description of these estimators can be found in \cref{App:B2.3}.
%$S_{\pm}S_{\pm}D$ and can be expressed as 
%\begin{eqnarray}
%    S_{+}S_{+}D = \frac{\sum_{ijk} w_i w_j w_k  n_i^D \epsilon_j \epsilon_k^*} {\sum_{ijk} w_i w_j w_k}, \\
%    S_{-}S_{-}D = \frac{\sum_{ijk} w_i w_j w_k  n_i^D \epsilon_j \epsilon_k} {\sum_{ijk} w_i w_j w_k}
%\end{eqnarray}
%where the ellipticities $ \epsilon_j, \epsilon_k$ are measured for galaxies at $j,\ k$, and the position of the third galaxy at location $i$ is $(n_i^D)$.} 

\subsubsection*{Galaxy--galaxy--shear 3PCF $(\mathcal{G}(\vartheta_1, \vartheta_2, \vartheta_3))$}
\par\noindent

Galaxy--galaxy--shear 3PCF or the position--position--shape (ggG)\footnote{Also known as galaxy--galaxy--galaxy lensing as first developed in \citep{Schneider_2005b}.} correlation, captures the correlation between the positions of a pair of lens galaxies and the tangential shear distortion of a background source galaxy. Specifically, it %quantifies the correlation between the tangential shear of a background source galaxy around a pair of lens galaxies in the foreground, effectively 
measures the projected mass profile around two lens galaxies, and hence can differentiate regions of different foreground clustering environments. 

For a pair of foreground lens galaxies located at $\boldsymbol{\theta}_1$ (bin $i$), and $\boldsymbol{\theta}_2$ (bin $j$), and a background source galaxy at $\boldsymbol{\theta}_3$ (bin $k$) the ggG correlation can be defined using the convention of \cite{Schneider_2005b} as
\begin{eqnarray}\label{eq:G}
    \mathcal{G}^{ijk}(\vartheta_1, \vartheta_2, \vartheta_3) &\equiv& \langle \delta^i_{\mathrm{g}}(\vartheta_1) \delta^j_{\mathrm{g}}(\vartheta_2) \varepsilon^k(\vartheta_3)\rangle 
    \equiv \delta^{ij}_{\rm gg}\varepsilon^k_{\rm t} + \mathrm{i}\cdot \delta^{ij}_{\rm gg}\varepsilon^k_{\times}.
\end{eqnarray}

This statistic is also sensitive to the non-Gaussian features of both the matter density field and shear field and provides complementary information to $\gamma_{\rm t}^{ij}(\vartheta)$. The relation between these components and the galaxy--galaxy--convergence bispectrum ($B_{{\rm gg}\kappa}$) can be found in \cref{App:ggs_3PCF}. Depending on the ordering of the lens and source tomographic bins, we can construct three observables (ggG, gGg, Ggg), and the estimators for all three types are described in \cref{App:estimator_ggs}. 
\textsc{TreeCorr} can compute all of these configurations. In this paper, we utilize the ggG configuration and we choose the lens bins to be at a lower redshift compared to the source bins (i.e., $i = j< k$). Again, this paper presents the first-ever measurement from simulations of this quantity.

%\JHD{ 'whose  shape is being measured' sounds incorrect, you are not measuring the shape, it is already in the catalog. You correlate the shapes...}
% \JHD{which are directly computed by \textsc{TreeCorr}. As for other statistics involving clustering data, one must isolate the signal from that obtained from pure random positions, hence we define our estimator as}

% \begin{eqnarray}\label{eq:23}
%    \widehat{\mathcal{G}}^{ijk}(\vartheta_1, \vartheta_2, \vartheta_3) = 
%    \begin{cases}
%        S^i(D^j-R^j)(D^k-R^k); & \text{Ggg config.} \\
%        (D^i-R^i)S^j(D^k-R^k); & \text{gGg config.} \\
%        (D^i-R^i)(D^j-R^j)S^k; & \text{ggG config.}
%    \end{cases}
% \end{eqnarray}

% \begin{eqnarray}\label{eq:}
%    \widehat{\mathcal{G}}^{ijk}(\vartheta_1, \vartheta_2, \vartheta_3) = S^i(D^jD^k-D^jR^k-R^jD^k+R^jR^k),
% \end{eqnarray}
% where $DDS$ is the measured correlation between the shape of the background galaxy (located at $\vartheta_3$) and the two foreground galaxies (at $\vartheta_1$ and $\vartheta_2$).  $DRS, RDS$ and $RRS$ can be explained similarly, replacing the foreground galaxy positions by those from our random catalogs appropriately. 

%A detailed description of these estimators can be found in \cref{App:B2.2}
% The relation between this correlation and the galaxy-galaxy-convergence bispectrum ($B_{{\rm gg}\kappa }$) can be found in \cref{App:A3}. 

\subsubsection*{Galaxy--galaxy--galaxy 3PCF $(\zeta(\vartheta_1, \vartheta_2, \vartheta_3))$}
\par\noindent

Galaxy--galaxy--galaxy (clustering) 3PCF or position--position--position (ggg) correlation measures the excess probability, compared to a random distribution, of finding a triplet of galaxies separated by given distances relative to each other. 
This correlation %for a triplet of galaxies, located at $\boldsymbol{\vartheta}_1$, $\boldsymbol{\vartheta}_2$, and $\boldsymbol{\vartheta}_3$, 
can be defined as
\begin{eqnarray}\label{eq:zeta}
    \zeta^{ijk}(\vartheta_1, \vartheta_2, \vartheta_3) \equiv \langle \delta^i_{\mathrm{g}}(\vartheta_1) \delta^j_{\mathrm{g}}(\vartheta_2) \delta^k_{\mathrm{g}}(\vartheta_3) \rangle \equiv \delta^{ijk}_{\rm ggg},
\end{eqnarray}
and is extracted from data through the Landy--Szalay estimator \citep{LS} and described in \cref{App:estimator_ggg}.
% \begin{eqnarray}\label{eq:25}
%        \widehat{\zeta}^{ijk}(\vartheta_1, \vartheta_2, \vartheta_3) =       \frac{(D^i-R^i)(D^j-R^j)(D^k-R^k)}{R^iR^jR^k},
% \end{eqnarray} 
% which is the \cite{LS} equivalent for ggg.
%\avi{where a detailed description of this estimator can be found in \cref{App:B2.1}}
% where we have omitted the tomographic indices on the right-hand side to simplify the notation. 
This statistic is connected to the galaxy--galaxy--galaxy (clustering) bispectrum ($B_{\rm ggg}$, see \cref{App:ggg_3PCF}) and is sensitive to non-Gaussian features in the foreground matter density field.

We emphasize that a non-linear galaxy bias with higher-order terms is indeed important to model bispectra involving the galaxy position measured from actual data, even at the tree level. However, we choose to focus on a linear bias setup to simplify the interpretation. Since we focus exclusively on the baryonified response relative to the DMO scenario, we assume that galaxy bias cancels out at first order. Moreover, we neglect any residual impact of bias, as it is expected to be subdominant compared to feedback effects.
%\JHD{I feel it would be natural to merge the estimator section with what we have here.}
% \subsubsection{Integrated three-point correlation functions}
% \par\noindent

\section{Simulation}
\par\noindent
\label{sec:simulations}

In this section, we introduce the $N$-body simulation that is used in this study, the Baryonic Correction Model, and the shell baryonification technique. 

\subsection{$N$-body, light-cones and catalogs}
\label{subsec:Nbody}
\par\noindent

% The statistics presented in the previous section are measured in simulated galaxy catalogs 
To measure the statistics presented in the last section, we create simulated galaxy catalogs that are representative of about 5000 deg$^2$ of Rubin data after ten years of observations, described in \cite{SRD_V1_LSST}, and hereafter referred to as LSST-Y10. This section summarizes the procedure for generating these mock catalogs.

The gravity-only $N$-body simulation at the core of this work was produced with the Hardware/Hybrid Accelerated Cosmology Code (HACC), which is a high-performance simulation framework developed for large-scale cosmological studies \citep{habib2013hacc}. This simulation follows the non-linear trajectories of $2048^3$ matter particles in a box of size $L_{\rm box} = 600$ Mpc/$h$, and assumes a flat $\Lambda$CDM cosmology with parameters: $\Omega_{\rm c} = 0.22$, $\Omega_{\rm b} = 0.0448$, $h = 0.71$, $n_{\rm s} = 0.963$, $\sigma_8 = 0.8$ and $m_\nu = 0$. It is evolved from redshift 200 to redshift 0, and a total of 500 snapshots are stored, linearly spaced in scale factor\footnote{400 from the total of 500 snapshots are between redshifts 4 and 0, which is the redshift range of interest for our analysis.}. halos are also identified and used in the baryonification process, which we detail in \cref{subsec:shell_BCM}.

We build past lightcones by discretizing the volume around an observer into concentric shells, computing the projected density of particles in each of those shells, and storing them on {\sc Healpix} maps \citep{Healpix, Healpy}\footnote{{\sc Healpix}: \url{http://healpix.sf.net}} with an {\sc Nside} of 8192. This resolution corresponds to a physical scale of 0.43 arcmin or 26 arcsec. The lightcone construction process, including redshift slicing, ray tracing, and computation of the convergence and shear fields under the Born approximation, is described in \cite{menafernandez2025}. All these different steps are performed using the publicly available code \pollux\footnote{{\tt pollux}: \url{https://github.com/LSSTDESC/pollux}}, a new package developed to create weak lensing and large-scale structure simulations from $N$-body particle data. Our lightcones cover an octant of the sky and are constructed using 51 redshift snapshots between redshifts 4 and 0 (out of the total of 400 available). A source plane is placed at the upper redshift edge of each shell, where the convergence and shear maps are computed.

As shown in \cite{menafernandez2025}, our simulation resolves scales down to $k=7\ h$Mpc$^{-1}$ to $z\lesssim 2$, which are needed to describe the small-angle weak-lensing signal. In particular, we reach a $\sim 3\%$ agreement with the Mira-Titan IV emulator of the matter power spectrum  \citep{MTEmu}. In $\ell$-space, the simulation is converged to better than 2\% for $\ell<8000$, again for sources below $z=2$. This results in sub-arcmin convergence for $w(\vartheta)$, while scales above  1-2 arcmin are fully resolved for $\gamma_{\rm t}(\vartheta)$ and $\xi_\pm(\vartheta)$. It is also shown in \cite{menafernandez2025} that the statistics are converged for 51 shells, and that 400 shells are not needed at LSST precision.

\begin{figure}
    \centering
    \includegraphics[width=0.6\linewidth]{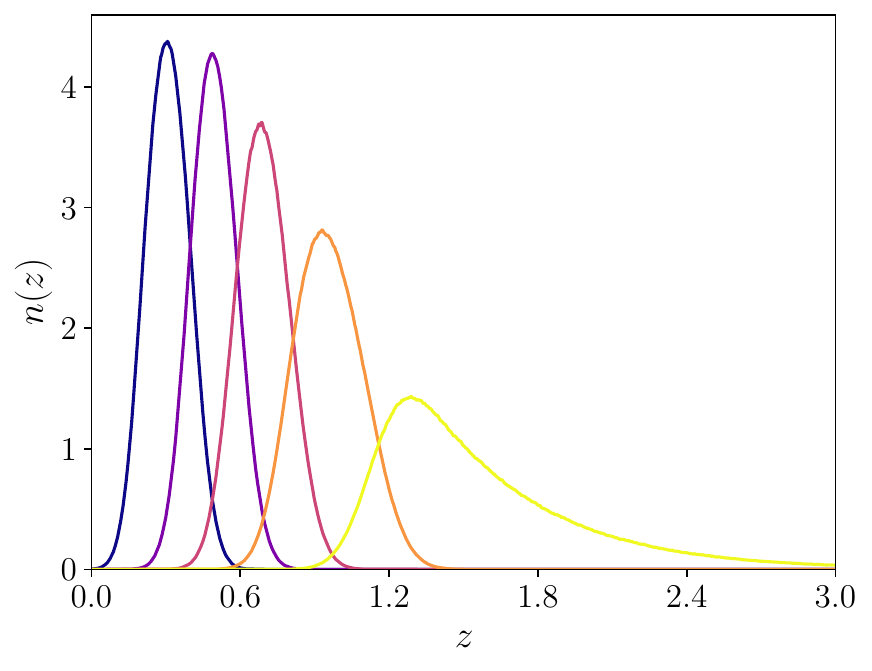}
    \caption{Redshift distributions for the five different tomographic bins used in this work. These are taken from the LSST Science Requirement Document \citep{SRD_V1_LSST}. }
    \label{fig:nz_SRD}
\end{figure}

Galaxy mock catalogs are generated to reproduce the LSST-Y10 tomographic redshift distributions, which are shown in \autoref{fig:nz_SRD}, with a total number density of 30 gal$/{\rm arcmin}^2$. The galaxy positions are obtained by Poisson sampling the density field of each shell, assuming a linear galaxy bias, $b$, with a value of 1 for each of the five tomographic bins. 
In the non-linear regime, a perturbative expansion or a non-linear bias prescription is essential for correctly capturing the galaxy clustering. However, the primary goal of this study is to quantify the impact of baryonic feedback on all the 2PCFs and 3PCFs by examining the ratios between their baryonified and DMO counterparts. Since our analysis focuses on these ratios, any residual effects from imperfect bias modeling are expected to largely cancel out (at least at the leading order), rendering the influence of bias mis-specification subdominant. A comprehensive and precise investigation of this bias prescription will be presented in a subsequent follow-up paper.

We finally linearly interpolate the convergence and shear fields at the three-dimensional position of each galaxy.
For random catalogs, used in the estimator of clustering statistics, we generate positions uniformly distributed on the curved sky over the same footprint. Note that these random catalogs contain 10 times more data points than the corresponding data catalog.

We use this procedure both on the original and on the baryonified shells, whose construction is described next. We emphasize that, as a consequence, the galaxy catalogs differ both in their lensing and in their clustering properties when varying the BCM parameters, and are therefore self-consistent.

\subsection{Baryonic correction model}
\par\noindent
\label{subsec:BCM}

% \textcolor{blue}{Describe the physics, what the parameters correspond to. Equation to describe the displacement field given a measured halo profile and a target matter profile. Show equation where we see the effect of $M_{\rm c}$ and $\theta_{\rm ej}$.
% Make table with fiducial BCM parameters and other explorations.
% }

We follow the Baryonic Correction Model (BCM), developed by \cite{Schneider_2019}, to propagate the effect of baryonic feedback on the total matter distribution. This section presents a review of the method, while in \cref{subsec:shell_BCM} we detail its implementation at the mass-shell level. The BCM considers the matter density field as a collection of halos where baryon feedback is important, superimposed on a large-scale cosmic web where baryons trace dark matter. The profile of each halo is influenced by both the 1-halo and 2-halo components of matter clustering. In the gravity-only (DMO) case, the total matter density field is expressed as
\begin{eqnarray}\label{eq:rho_dmo}
\rho_{\mathrm{DMO}}(r) = \rho_{\mathrm{NFW}}(r) + \rho_{\rm 2h}(r),
\end{eqnarray}
where $\rho_{\mathrm{NFW}}(r)$ and $\rho_{\rm 2h}(r)$ represent a generalized Navarro-Frenk-White (NFW) profile \citep{NFW} and the 2-halo term, respectively. This profile depends on the cosmological parameters of the simulation, the halo mass (more precisely on the parameter $M_{200}$), and its concentration ($c_{200}$). Both of these halo parameters are measured within a radius where the density is 200 times the critical density of the universe. The two-halo term is measured from the background matter density, halo bias and the linear halo correlation function: $\rho_{\rm 2h}(r) = (b_{\rm h}\xi_{\rm lin}(r)+1)\Omega_{\rm m}\rho_{\rm crit}$ \citep[][see their equation 2.7]{Schneider_2019}.

In the baryonified case (DMB), the 2-halo term is unchanged, while the 1-halo term is decomposed into three components: the collisionless matter (clm), the gas (gas), and the central galaxy (cga). The collisionless component $\rho_{\mathrm{clm}}$ is primarily composed of dark matter but also includes satellite galaxies and intracluster stars. Therefore, the matter density profile for the DMB case is given by
\begin{eqnarray}\label{eq:rho_dmb}
\rho_{\mathrm{DMB}}(r) = \rho_{\mathrm{clm}}(r) + \rho_{\mathrm{gas}}(r) + \rho_{\mathrm{cga}}(r) + \rho_{\rm 2h}(r).
\end{eqnarray}
% The baryonification process transforms the dmo density field $\rho_{\mathrm{dmo}}$ into the baryonified field $\rho_{\mathrm{dmb}}$ using the integrated mass profile:
The density profiles can be transformed into the corresponding integrated mass profiles  
\begin{eqnarray}\label{eq:mass_profile}
M_{\rm X}(r) = \int_0^r s^2 \rho_{\rm X}(s) \, ds; \ \ \ \ {\rm X} =  \{\rm DMO,\ DMB\}.
\end{eqnarray}
Both $M_{\rm DMO}(r)$ and $M_{\rm DMB}(r)$ are bijective functions and can be inverted to obtain $r_{\rm DMO}(M)$ and $r_{\rm DMB}(M)$, respectively. This leads to the key idea behind BCM: it applies physically motivated particle displacements to gravity-only simulations, generating full three-dimensional corrected density fields that mimic the impact of baryons. Indeed, the displacement function is defined as the mapping in distance $r$ such that the original profile transforms into the target profile, namely:
\begin{eqnarray}\label{eq:displacement_func}
d(r_{\mathrm{DMO}};\, M,c_{200}) = r_{\mathrm{DMB}}(M,c_{200}) - r_{\mathrm{DMO}}(M,c_{200}),
\end{eqnarray}
which depends on the halo mass and concentration, as well as on BCM parameters that describe the target profile $\rho_{\rm DMB}(r)$.

\begin{table}[tbp]
\centering
\begin{tabular}{|c|c|c|c|c|c|c|}
\hline
$M_{\rm c}$ & \multicolumn{5}{c|}{$\theta_{\rm ej}$} \\
\cline{2-6}
$(\times 10^{13} M_{\odot})$& 2 & 3 & 4 & 5 & 6 \\
\hline
2.5 &$\cdot$ & $\cdot$ & $\checkmark$ & $\cdot$ & $\cdot$\\
5.0 &$\cdot$ & $\cdot$ & $\checkmark$ & $\cdot$ & $\cdot$ \\
10.0 &$\checkmark$ & $\checkmark$ & $\checkmark$ & $\checkmark$ & $\checkmark$ \\
20.0 &$\cdot$ & $\cdot$ & $\checkmark$ & $\cdot$ & $\cdot$ \\
40.0 &$\cdot$ & $\cdot$ & $\checkmark$ & $\cdot$ & $\cdot$ \\
% $\alpha$ & $\beta$ & $\alpha$ and $\beta$\\
\hline
\end{tabular}
\caption{\label{tab:1} Parameter space for the 2 BCM parameters explored in this study, which are varied one at a time while the other parameter is fixed at its fiducial value. $M_{\rm c}$ governs the amount of gas expelled beyond the halo boundary, and $\theta_{\rm ej}$ encodes the maximal ejection radius relative to the halo boundary. Our fiducial BCM model corresponds to $M_{\rm c} = 10\times 10^{13} M_{\odot}$, and $\theta_{\rm ej}=4$.}
\end{table}

The original baryonification model of \cite{Schneider_2019} includes 14 free parameters that govern the behavior of $\rho_{\mathrm{clm}}$, $\rho_{\mathrm{gas}}$, $\rho_{\mathrm{cga}}$, and $\rho_{\rm 2h}$. Specifically, four parameters control the gas profile, five for the stellar component, three for the dark matter, and two for the 2-halo term. In the initial study, only five of these parameters were varied. Among those, the parameter $M_{\rm c}$, which regulates the slope of the gas profile in the outskirts of the halo and the parameter $\theta_{\rm ej}$ that determines the ejection radius of the expelled gas outside the halo, were seen to have the most significant impact on lensing observables \citep{Schneider_2019}. 

The slope of the gas profile ($\beta$), which consists of two free model parameters, can be expressed as  
\begin{eqnarray}\label{eq:slope}
    \beta(M_{200}) = 3 - \bigg(\frac{M_{\rm c}}{M_{200}}\bigg)^\mu,
\end{eqnarray}
where $M_{\rm c}$ and $\mu$ are the parameters related to the slope of the gas profile. $M_{\rm c}$ defines the characteristic mass scale where the slope becomes shallower than $3$, and $\mu$ defines how fast the slope becomes shallower towards small halo masses. 
The slope is allowed to have both positive and negative values but is bound from above, i.e., $\beta \leq 3$. This means that the gas profile can be shallower than the NFW profile but never steeper.

The ejection radius of the gas can be parameterized by 
\begin{eqnarray}\label{eq:ejection_radius}
    r_{\rm ej} = \theta_{\rm ej} r_{200},
\end{eqnarray}
where $\theta_{\rm ej}$ specifies the maximum radius of gas ejection relative to the halo boundary specified by the parameter $r_{200}$. $\theta_{\rm ej}$ is always greater than 1 for consistency purposes.

% \begin{table}[tbp]
% \centering
% \begin{tabular}{|c|c|}
% \hline
% $M_{\rm c} (\times 10^{13} M_{\odot})$ & $\theta_{\rm ej}$ \\
% \hline
% 2.5 & 2 \\
% 5.0 & 3 \\
% 10.0 & 4 \\
% 20.0 & 5 \\
% 40.0 & 6 \\
% % $\alpha$ & $\beta$ & $\alpha$ and $\beta$\\
% \hline
% \end{tabular}
% \caption{\label{tab:1} Parameter space for the 2 BCM parameters explored in this study, \JHD{which are varied one at a time while the other parameter is fixed at its fiducial value. Our fiducial BCM model corresponds to $M_{\rm c} = 10\times 10^{13} M_{\odot}$, and $\theta_{\rm ej}=4$.}}
% \end{table}

The parameter space that we explore contains 5 values for each of the parameters $M_{\rm c}$ and $\theta_{\rm ej}$ as shown in \cref{tab:1}. 
% The fiducial case corresponds to $M_{\rm c, fid}=1\times10^{14} M_\odot$ and $\theta_{\rm ej, fid} = 4$ 
The fiducial case corresponds to $(M_{\rm c, fid}, \theta_{\rm ej, fid}) = (1\times10^{14} M_\odot,  4)$. 
We assume no redshift dependence for these BCM parameters here, however this is another promising avenue, as shown in \cite{Arico_2021}. 

% The model used in F22 focused on varying $M_{\rm c}$ and its redshift evolution, modeled as a power law:

% \begin{equation}
% M_{\rm c} = M_{\rm c}^0 (1 + z)^{\nu},
% \end{equation}

% where $M_{\rm c}^0$ and $\nu$ are treated as free parameters. All other parameters were fixed to the best-guess model (B-avrg), as detailed in Table 2 of \cite{Schneider_2019}. These fixed values are motivated by observational constraints from X-ray gas fractions and hydrostatic mass bias studies~\cite{Vikhlinin2006, Pratt2009, Planck2013, Mantz2016}.

%\JHD{[This section is missing a comparison between the BCM model used here and other BCM parameterisations, why we think our model is good enough, what are its limitations, etc.]}
%\avi{JHD, could you please modify that --
%BCM applies physically motivated particle displacements to dark-matter-only (DMO) simulations, generating full three-dimensional corrected density fields that mimic the impact of baryons, suitable for a wide range of statistical analyses. BCM parameters are directly linked to observable quantities such as gas fractions and thermal Sunyaev–Zel'dovich (tSZ) profiles. Compared to simpler halo-model-based approaches (e.g., HMCode), BCM offers improved accuracy for non-Gaussian and multi-probe analyses while maintaining low computational cost. 
One should keep in mind that BCM models rely on parametric gas and stellar profiles, whose degeneracies and calibration limitations, particularly at high redshift or under extreme feedback scenarios, can impact accuracy. The model performs best on scales and feedback regimes similar to those used during its tuning. %In contrast, newer emulators like BACCO provide broader coverage of cosmological and feedback scenarios but lack BCM’s explicit field-level physical consistency. 
A notable advancement of the BCM framework is the baryonic correction approach by \cite{Arico_2021}, which builds on similar principles and extends the calibration to simultaneously match both the matter power spectrum and bispectrum (including equilateral and squeezed configurations), achieving approximately 1\% accuracy for two- and three-point statistics. This demonstrates that BCM-style methods can robustly capture baryonic effects across higher-order clustering statistics. The current work targets projected statistics, both in lensing and in clustering, which involve a wide redshift integration. This alleviates some of the residual differences between our model and that of \cite{Arico_2021}, as well as the differences between 3D and 2D baryonification methods. 

As shown in  \cite{Arico_2021}, at LSST precision, 2-3 parameters are sufficient to describe accurately the full high-dimensional BCM space (due to degeneracies) at the level of the power spectrum, which motivates our choice to vary only $M_{\rm c}$ and $\theta_{\rm ej}$, however it is unclear whether this holds for $4\times3$PCFs 
% probes 
and will need to be assessed in the future.

\begin{figure*}
    \centering
    \begin{minipage}[t]{0.32\textwidth}
        \centering
        \includegraphics[width=\textwidth]{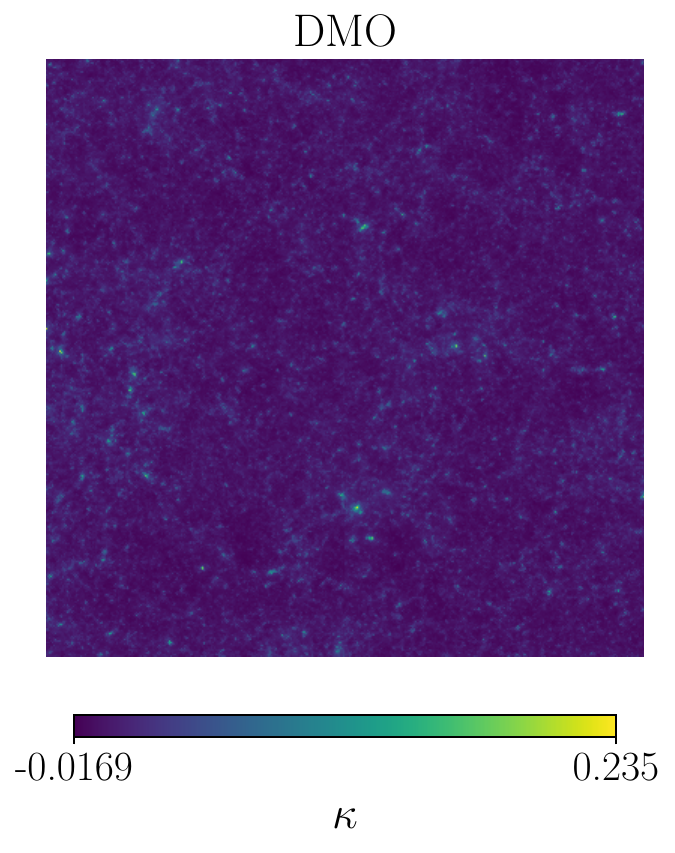}
    \end{minipage}
    \hfill
    \begin{minipage}[t]{0.32\textwidth}
        \centering
        \includegraphics[width=\textwidth]{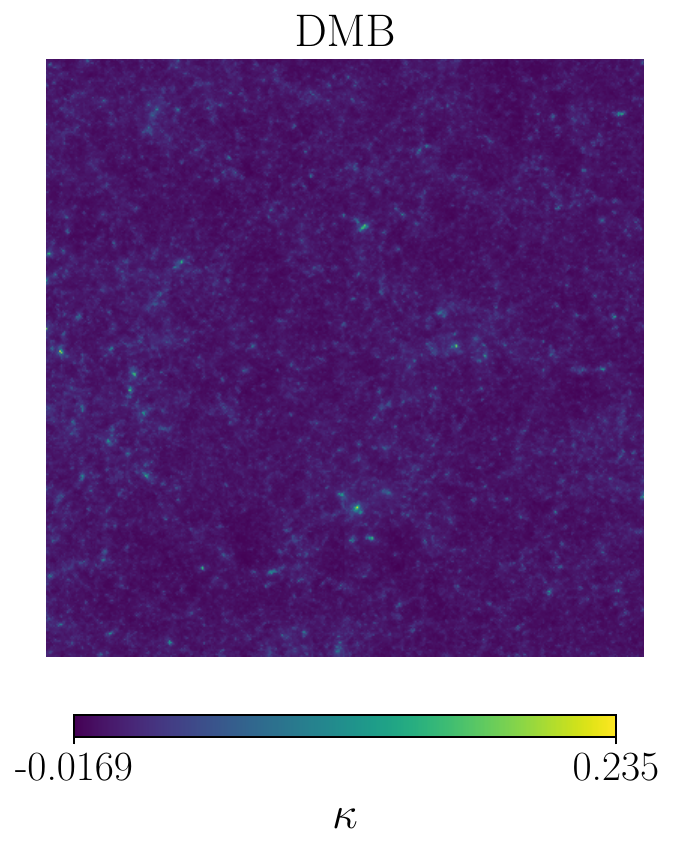}
    \end{minipage}
    \hfill
    \begin{minipage}[t]{0.313\textwidth}
        \centering
        \includegraphics[width=\textwidth]{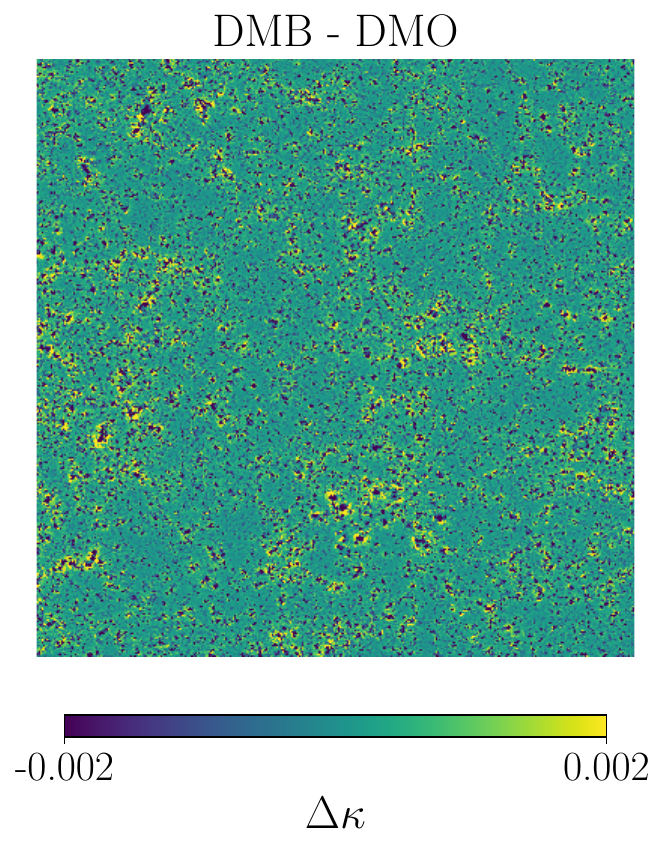}
    \end{minipage}
    \caption{Cartesian projection of the convergence field in a region that is 10 deg on the side for the third tomographic bin for one of our simulations.
    $\kappa$ maps are shown for gravity only (DMO) and dark matter with baryonic feedback (DMB) cases in the left and the middle panels, respectively. 
    % \textbf{Left panel:} gravity only (DMO), showing $\kappa$. \textbf{Middle panel:} dark matter with baryonic feedback (DMB), showing $\kappa$. 
    The right panel represents the difference between the two, $\Delta\kappa$. Halos causing the majority of the convergence seen in the left panel are baryonified: their mass profile is lowered towards their center, as material is pushed to the outskirt. These regions of increased density surrounding lowered density can be seen in the $\Delta\kappa$ map as dark spots centered on pale clouds. 
\label{fig:map_density_shell}}
\end{figure*}

\subsection{Shell baryonification}
\label{subsec:shell_BCM}
\par\noindent

% \textcolor{blue}{Refer to cosmo-grid and explain the baryonification method}

We follow the shell-baryonification method %\avi{and the code} 
developed by \cite{Fluri_2022, Kacprzak_2023} and released publicly along with the \textsc{CosmoGridV1} simulations\footnote{\textsc{CosmoGridV1}: \url{http://www.cosmogrid.ai/}}, and baryonify the output of the DMO (\textsc{HACC}) simulations introduced in \cref{subsec:Nbody}. 

% In the first step, halo catalogs from the dmo \textsc{HACC} simulations are constructed using a Friends-of-Friends (FoF) halo finder with a linking length set to 20\% of the mean particle separation. catalogs are generated at each simulation time step. All FoF halos which are selected contain at least 100 particles within the virial radius, defined as $r \equiv r_{200}$. 

As described in  \cite{heitmann2024newworldssimulationslargescale},  halo catalogs from the \textsc{HACC} simulations are generated using the Friends-of-Friends (FoF) halo-finding algorithm with a linking length of $b = 0.168$ times the mean inter-particle separation. Only halos containing at least 500 particles are selected for further analysis.
For each of these halos, the mass overdensity is computed in concentric spheres centered on the minimum of the Newtonian potential. %The radius at which the enclosed density drops below 200 times the critical density, defined as $\rho_{\rm c} = 3H^2/8\pi G$, is identified as the radius of the spherical overdensity (SO) halo\JHD{$r_{200}$}. The corresponding enclosed mass is denoted as $M_{\rm 200}$. For more details, we refer to \cite{heitmann2024newworldssimulationslargescale}. 

Each halo is fitted with a standard Navarro-Frenk-White (NFW) profile using logarithmically binned mass shells to determine its concentration parameter $c_{200}$ and mass $M_{200}$. Lightcone halos are then selected from the full catalog, following the same boundaries and replication scheme as used in the original lightcone construction. We use {\sc colossus}\footnote{\textsc{colossus}: \url{bdiemer.bitbucket.io/colossus/}} to convert between FoF quantities measured from our halo finder and the virial quantities required by our BCM model.
% halos located within 20 Mpc of a shell boundary were assigned to both adjacent shells, allowing them to influence particles in both regions.

Following the \textsc{UFalcon} convention used in \cite{Kacprzak_2023}, it is assumed that all halos and particles within a shell reside at the shell's mean redshift $z_{\rm m}$. We then define the projected mass within a
cylinder of radius $r$ as 
\begin{eqnarray}\label{eq:32}
M^{\rm p}_{\rm X}(r) = 2\pi \int_0^r s \int_0^{z_{\text{max}}} \rho_{\rm X}(s, z) \, dz \, ds,
\end{eqnarray}
whose dominant contribution comes from the halo on which the cylinder is centered. We set the upper limit of the integration to $z_{\text{max}}$ corresponds to the redshift for $50r$ (i.e., $z_{\text{max}}=z(50r)$). Although the projected mass diverges as $z_{\text{max}}$ increases due to the 2-halo term, the projected displacement function, which is given by the relation
\begin{eqnarray}\label{eq:33}
d_{\rm p}(r_{\text{DMO}};\ M^{\rm p},c_{200}) = r_{\text{DMB}}(M^{\rm p},c_{200}) - r_{\text{DMO}}(M^{\rm p},c_{200}),
\end{eqnarray}
remains finite as the divergent contributions cancel out.

The shell baryonification is implemented by displacing 
% the pixels 
the density field of the high-resolution shells in the \textsc{HACC} simulation and interpolating on the original grid. Assuming a locally flat sky, we displace all pixels within a radius of $50r$ from each halo using the projected displacement function. 
New values are assigned to the {\sc Healpix} pixel positions via linear interpolation based on the displaced coordinates. 
Finally, the resolution of the baryonified shells is the same as the original density maps, namely $\textsc{Nside} = 8192$. 
% Finally, the resolution of the shells has been kept the same $\textsc{nside} = 8192$, effectively applying a smoothing kernel. 

The high-resolution convergence map computed for the third tomographic redshift bin from the DMO shells is shown in \autoref{fig:map_density_shell}, alongside the difference map defined as DMB - DMO. It is clear from these that the largest differences occur in regions of high convergence, which are dominated by the presence of large halos. The high resolution of these maps allows us to study the impact of different parameter choices on small-scale non-Gaussian statistics.

We generate baryonified mass density for all model parameters listed in \autoref{tab:1} and produce mock galaxy catalogs including galaxy shapes and positions, 
% lensing and clustering catalogs 
based on the methods presented in \cref{subsec:Nbody}. We focus our attention on understanding the impact of these parameter choices on the 2PCFs and 3PCFs, which are described in \cref{sec:background}, and are presented next. %We note that the shell-based baryonification can be applied for any desired parameter configuration using the raw \textsc{nside=8192} maps and halo catalogs.

% where for pairs of galaxies with separation $r_p$ and $\Pi$. 

\section{Results}
\par\noindent
\label{sec:results}
This section presents the impact of baryonification on our galaxy clustering and weak lensing statistics, as measured from our mock galaxy catalogs.

\subsection{Measurements}\label{subsec:measurements}
\par\noindent

We compute all correlation functions with  \textsc{TreeCorr}\footnote{\textsc{TreeCorr}: \url{https://github.com/rmjarvis/TreeCorr}} \citep{Jarvis_2004}. \textsc{TreeCorr} has been used extensively in the literature for 2PCFs, and we use them here to compute the $3\times2$PCFs data vectors in 15 bins over the range $[0.5 < \vartheta < 120]$ arcmin, with a \texttt{bin\_slop} parameter set to 0.01\footnote{This is a high-accuracy mode which can in principle be relaxed, especially for 3PCF measurements that are noisier in nature. However, we wanted to test and benchmark the code in a realistic analysis setting in \cref{App:Treecorr_benchmark}, hence this choice of \texttt{bin\_slop}.}. In practice, additional scale cuts can be applied in data analyses to exclude angular separation where the modeling or the measurements are uncertain or severely contaminated by additional survey systematics, which we defer to future work. 

The mixed-spin 3PCFs are a recent feature addition to \textsc{TreeCorr} (new in version 5.1.3), which allows us to correlate galaxy positions (described by spin-0 density field) 
% maps 
with galaxy shapes (described by spin-2 ellipticity fields).  Specifically, \textsc{TreeCorr} extends the work of \cite{Porth24}, who developed a fast multipole-based algorithm for computing the shear--shear--shear 3PCF, to cases where quantities of arbitrary spin can appear on each point of the triangle. 

The triangles formed by the three points in the 3PCF are specified by two of their sides and the angle between them, called the SAS (Side-Angle-Side) configuration in \textsc{TreeCorr}.  The sides are each binned logarithmically, so the binning is referred to as \texttt{LogSAS}. We use 50 sky regions to compute the jackknife error bars. The shared vertex of the two sides is labeled as point P1; the other two points are labeled as P2 and P3. The sides extending from P1 to these points are respectively called $d_2$ and $d_3$, while the angle between them is denoted by $\phi$, measured in radians.
The orientation is defined such that the angle $\phi$ lies between $0$ and $\pi$, sweeping counter-clockwise from $d_2$ to $d_3$.

Unlike the SSS (Side-Side-Side) configuration, where each triangle is uniquely assigned to a single bin, this SAS-based definition centers each triangle at P1. As a result, in auto-correlation analyses, each triangle is counted three times--once for each vertex acting as the center. In cross-correlation analyses, the point ordering is more specific: Objects from the first catalog are placed at the central vertex (P1), those from the second catalog at P2 (opposite $d_2$, i.e., at the end of $d_3$), and those from the third catalog at P3 (opposite $d_3$, i.e., at the end of $d_2$).

% We use 15 logarithmic angular bins from 0.5 to 5 degrees. The error bars are estimated using the jackknife method. 

% The disadvantage of these estimators is that their computational complexity scales with $\mathcal{O}(N^3_{\rm gal})$, which is quite challenging to execute even for a moderate amount of galaxies (i.e., $N_{\rm gal} \geq 10^6$). To avoid this complexity, \texttt{TREECORR} constructs a hierarchical ball tree out of the galaxy sample and computes the correlation functions from the tree \cite{Heydenreich_2023}. 

The original catalogs contain $N_{\rm gal}\sim 10^8$ galaxies in each tomographic bin; since our simulated data are noise-free, we can significantly downsample our catalogs 
without loss of accuracy. Therefore, to speed up the computation of the correlations, we choose to randomly downsample all the catalogs by a factor of 10 such that the resulting 3 gal/arcmin$^2$ roughly resembles the galaxy density and noise levels expected at the level of DES-Y3\footnote{To mimic the LSST-Y1 level galaxy density, the original catalog should instead be downsampled by a factor of 3, which is not our intention here.}. We verified that the jackknife error bars obtained for both the original and downsampled catalog are comparable (see \cref{App:Treecorr_benchmark}).
Note that the random catalogs, which are generated for each tomographic bin during the computation of position correlations, are 10 times larger compared to the original catalog. However, the results presented here are insensitive to this choice, as the downsampling is performed randomly and therefore entails no loss of generality; it merely leads to marginally noisier measurements.
% However, the results presented here are not affected by this choice because the downsampling is random, and resulting in no loss of generality; this would only lead to slightly noisier measurements. %As the downsampled catalog roughly mimics the galaxy density of LSST-Y1, the data vectors and the level of noise presented in this paper are representative of what we can expect from LSST-Y1 data. 
This downsampling allows us to compute the 3PCFs for an ensemble of $\sim 10^7$ galaxies in about 2-3 minutes for a single node on NERSC's Pelumutter system. The computational time to process the catalogs and measure the correlations increases with the number of galaxies as $N$ log $N$, as detailed in \cref{App:Treecorr_benchmark}.
Of course, downsampling will not be performed on real data for LSST-Y10, hence some of the conclusions based on sensitivity presented in this work will need to be re-evaluated in the future. %future analyses will require more computing resources to measure these correlations. 

% \footnote{\mj{[MJ: I think the asymptotic behavior of the TreeCorr 3PCF algorithm is O(N logN), so I wouldn't expect it to be massively slower for 10x more objects.  Ofc, you might not be that close to the asympototic behavior yet at $10^7$, but still, I would have thought it would be pretty reasonable for $10^8$ objects.  You might need to raise bin\_slop a little though.  0.01 is pretty small for 3PCF!  0.1 should be fine and not introduce much noise to the measurement.]}}

 % It takes $~30-35$ minutes for $N_{\rm gal}\backsim 10^7$ while it takes $6-7$ hours for $N_{\rm gal}\backsim 10^8$ for the same setup.  

% Initially, the randoms were generated $3.3$ times larger compared to the DESI tracers \cite{Lange_2024}. The matched catalogs are being generated by choosing the common objects from the shape samples from DES, HSC and the position tracers (BGS) of DESI. In this process, we loose a large fraction of galaxies from both samples, and the final catalog contains only a small subset of the both samples. As we do not downsample the number of randoms, 
% the number of objects available in the random catalog ($N_{\rm randoms} \backsim 10^7-10^8$) is 10-20 times larger compared to the objects in the matched catalog.  
% Computing the 3PCFs with randoms is highly expensive, so we choose not to consider the random subtraction from these correlations. We emphasize that the inclusion of randoms is not negotiable for the detection of ggI and gII correlations. 

In total, we create ten galaxy catalogs, one from the DMO maps and nine from baryonified maps corresponding
to the model variations 
%to a fiducial and four for each of the two baryonic parameters 
presented in \autoref{tab:1}.  As mentioned earlier, our analysis focuses on the two most influential parameters: the characteristic mass scale $M_{\rm c}$ and the maximum radius of gas ejection compared to halo boundary $\theta_{\rm ej}$; we leave the exploration of other BCM parameters for future studies. In the next sections, we quantify the impact of baryonic feedback on $3\times2$PCFs and $4\times3$PCFs.

\begin{figure*}
    \centering
    \begin{subfigure}[b]{\linewidth}
        \centering
        \includegraphics[width=\linewidth]{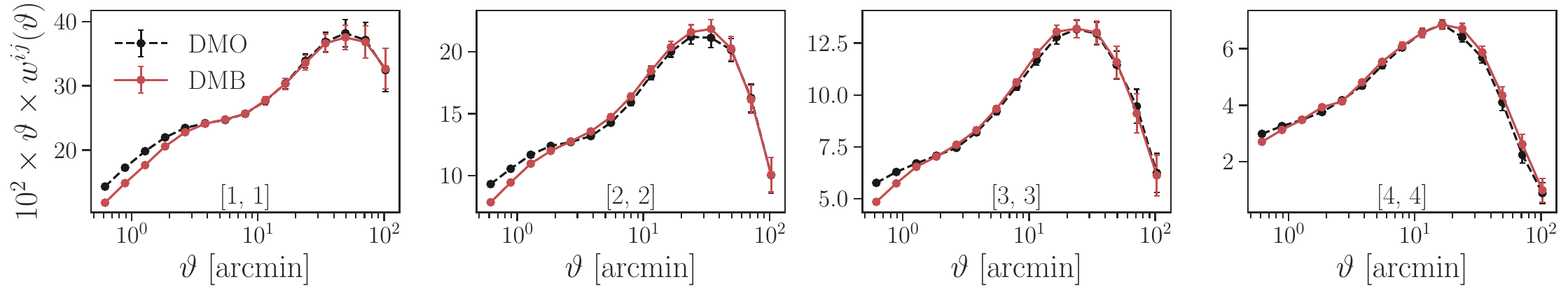}
        \caption{Galaxy--galaxy 2PCF ($w^{ij}(\vartheta)$).}
        \label{fig:w_dmb_dmo_dv}
    \end{subfigure}
    % \vspace{0.5cm} % Optional spacing between subfigures
    \begin{subfigure}[b]{\linewidth}
        \centering
        \includegraphics[width=\linewidth]{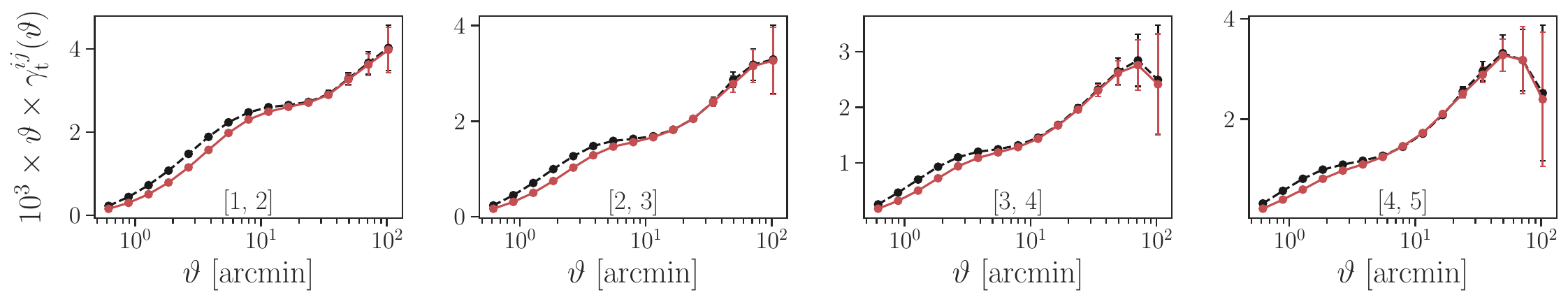}
        \caption{Galaxy--shear 2PCF ($\gamma^{ij}_{\rm t}(\vartheta)$).}
        \label{fig:gammat_dmb_dmo_dv}        
    \end{subfigure}

    \begin{subfigure}[c]{\linewidth}
        \centering
        \includegraphics[width=\linewidth]{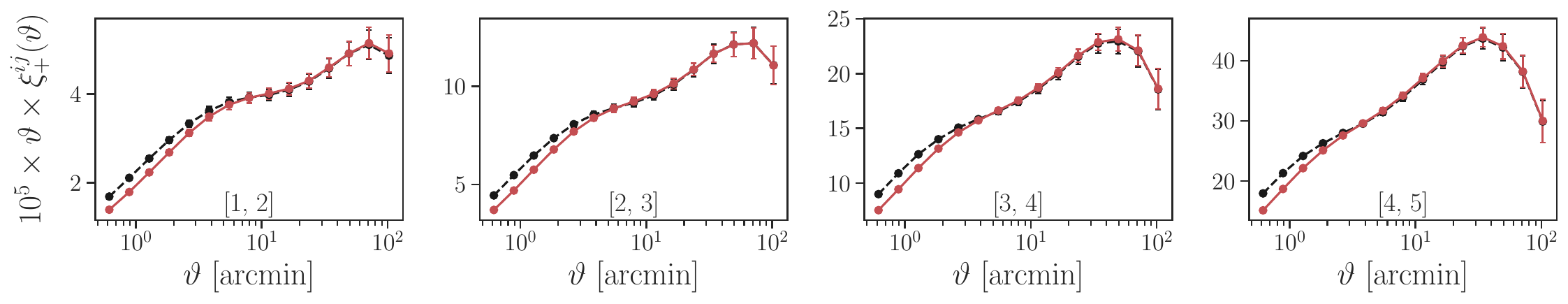}
        \caption{Shear--shear 2PCF ($\xi_+^{ij}(\vartheta)$).}
        \label{fig:xip_dmb_dmo_dv}        
    \end{subfigure}

    \begin{subfigure}[c]{\linewidth}
        \centering
        \includegraphics[width=\linewidth]{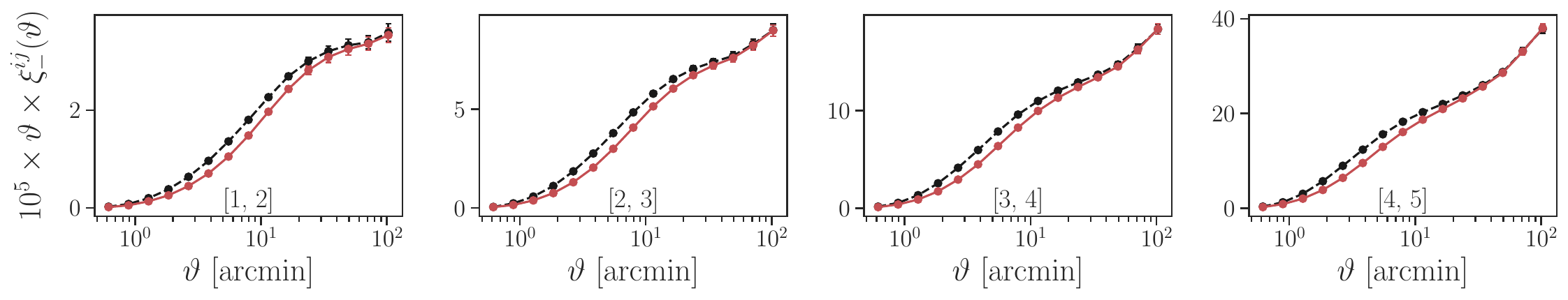}
        \caption{Shear--shear 2PCF ($\xi_-^{ij}(\vartheta)$).}
        \label{fig:xim_dmb_dmo_dv}        
    \end{subfigure}
    
    \caption{All three types of 2PCFs are shown for a selected combination of tomographic bins. The data vectors shown in black and red correspond to the DMO and DMB cases, respectively. Data vectors for all combinations of tomographic bins and baryonic feedback models are shown in \cref{App:Data_vectors}. }
    \label{fig:w_gammat_xipm_dmb_dmo}
\end{figure*}

\begin{figure*}
    \centering
    \begin{subfigure}[b]{\linewidth}
        \centering
        \includegraphics[width=\linewidth]{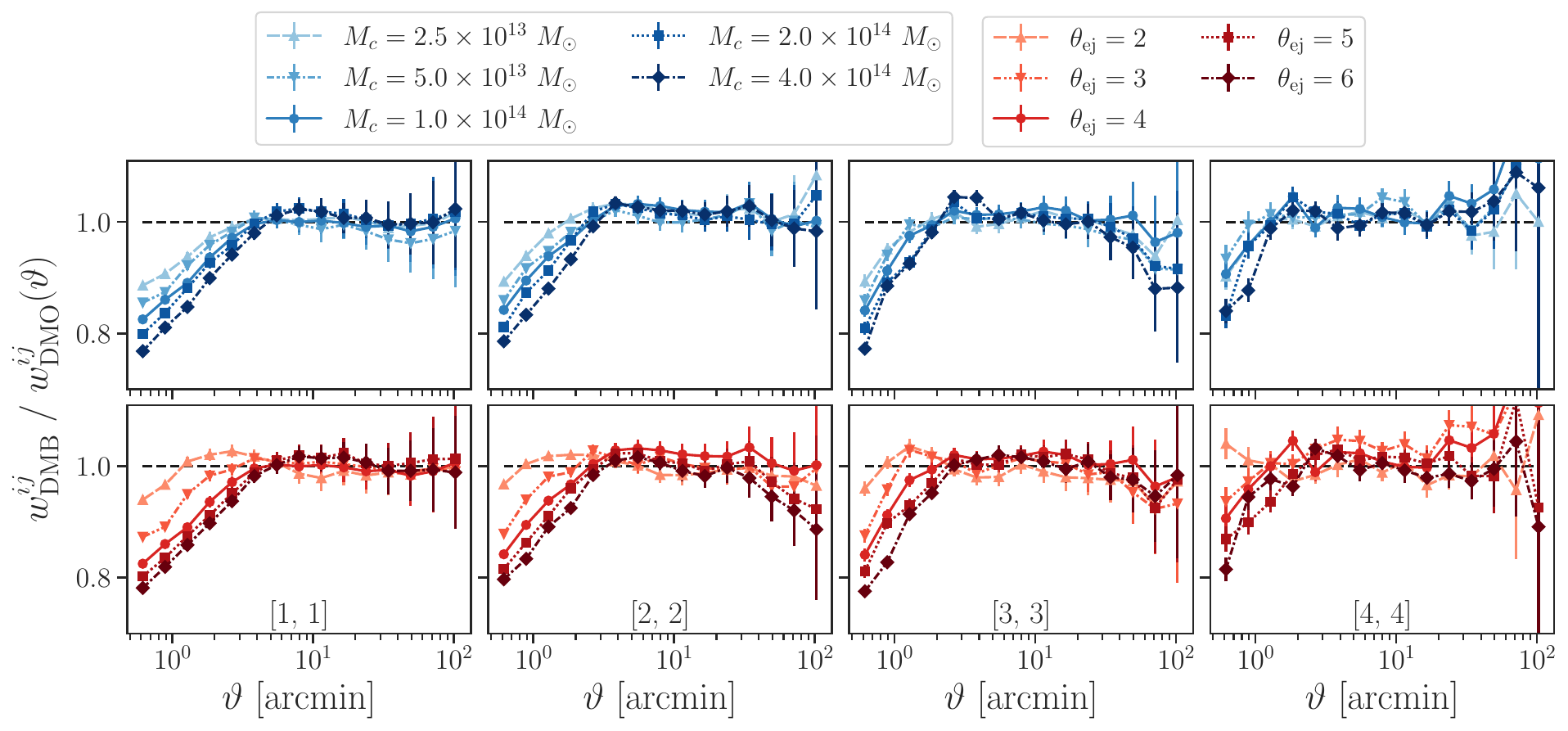}
        \caption{Baryonic suppression effect on galaxy--galaxy 2PCF ($w^{ij}(\vartheta)$) for auto-correlating tomographic bins.}
        \label{fig:w_dmb_over_dmo}
    \end{subfigure}
    
    \vspace{0.5cm} % Optional spacing between subfigures

    \begin{subfigure}[b]{\linewidth}
        \centering
        \includegraphics[width=\linewidth]{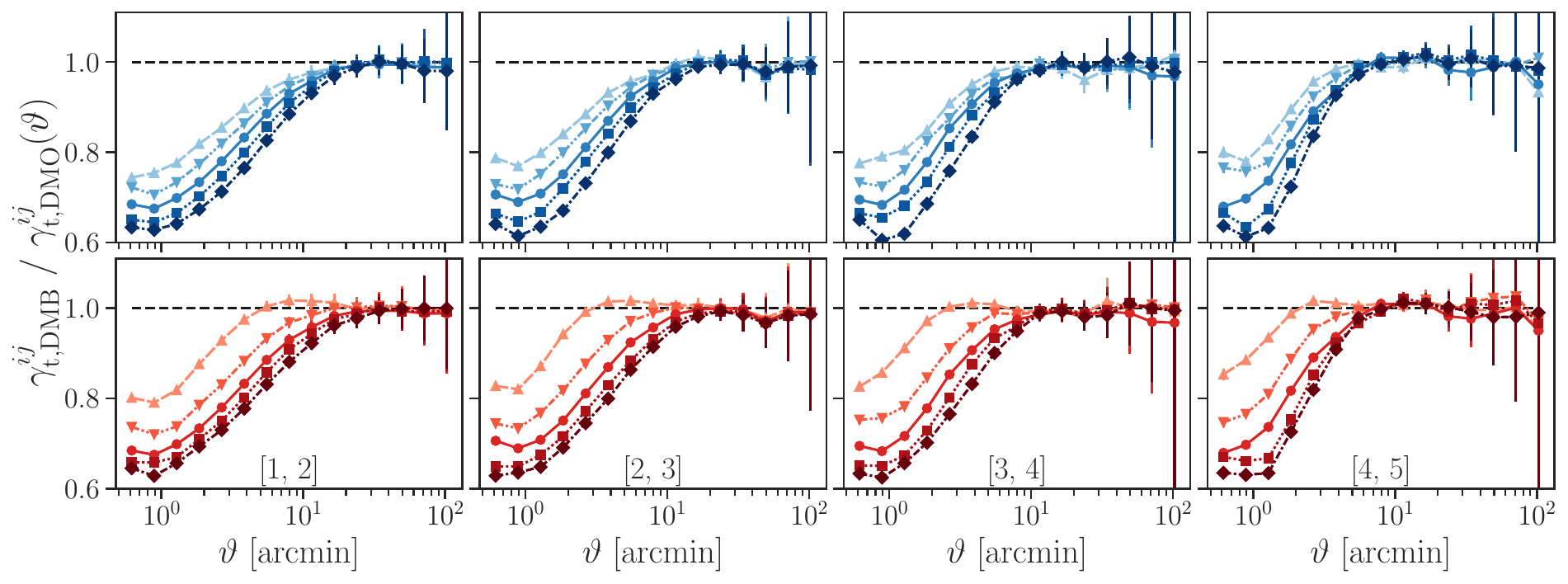}
        \caption{Baryonic suppression effect on the galaxy--shear 2PCF ($\gamma_{\rm t}^{ij}(\vartheta)$) for adjacent tomographic bins.}
        \label{fig:gammat_dmb_over_dmo}
    \end{subfigure}
    
    \caption{Baryonic suppression effects on galaxy--galaxy and galaxy--shear 2PCF for the two most impactful BCM parameters: characteristic mass scale $M_{\rm c}$ and parameter that determines the maximum ejection radius of the expelled gas outside the halo $\theta_{\rm ej}$. The fiducial case corresponds to $(M_{\rm c, fid}, \theta_{\rm ej, fid}) = (1\times10^{14} M_\odot, 4)$, and is shown with a solid line in all panels. Darker color shades represent higher values of the parameter under consideration, while lighter shades indicate lower values, as shown in the legend. Tomographic bin combinations are mentioned within parentheses in each plot.}
    \label{fig:w_gammat_dmb_over_dmo}
\end{figure*}

\begin{figure*}
    \centering
    \begin{subfigure}[b]{\linewidth}
        \centering
        \includegraphics[width=\linewidth]{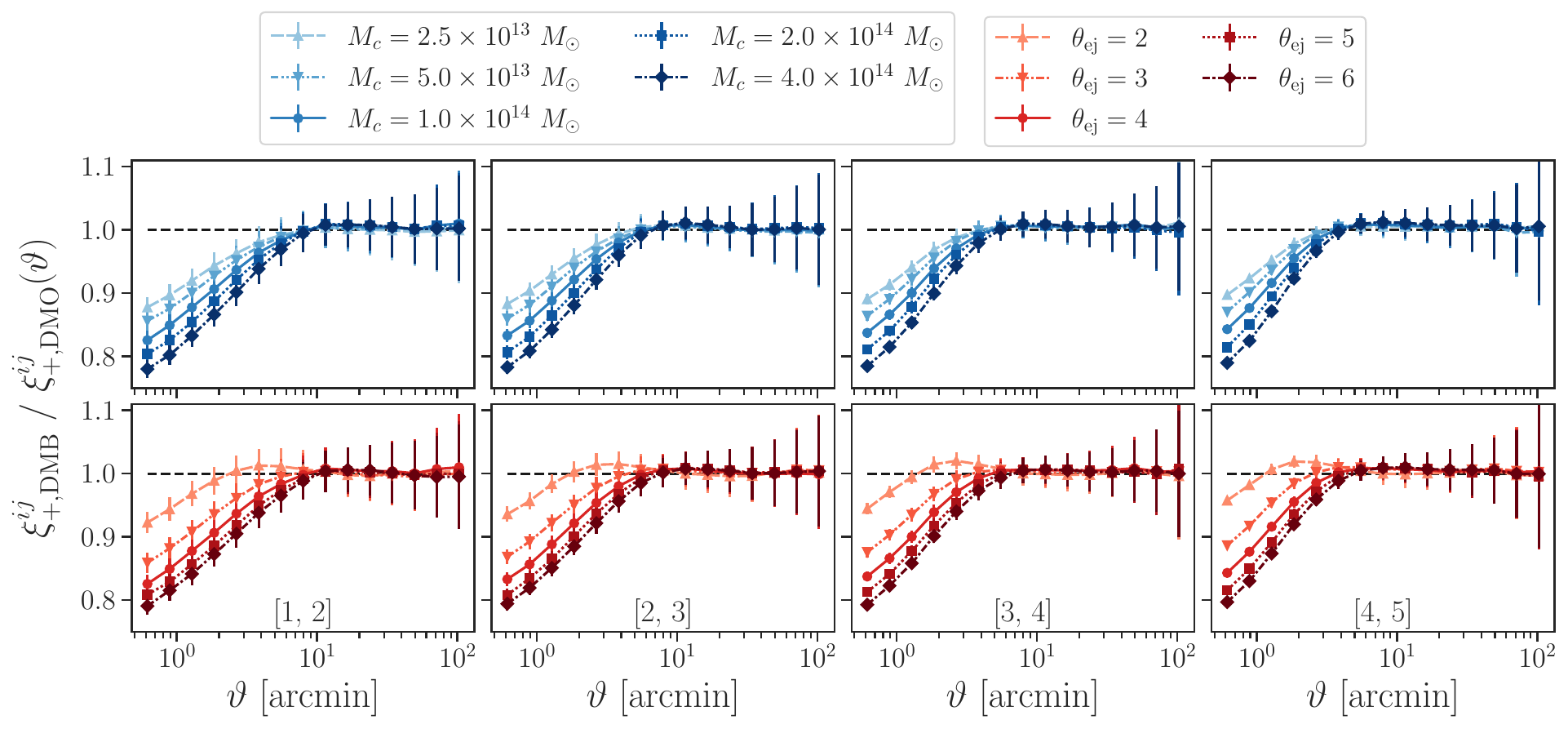}
        \caption{Baryonic suppression effect on $\xi_{+}^{ij}(\vartheta)$ for adjacent tomographic bins.}
        \label{fig:xip_dmb_over_dmo}
    \end{subfigure}
    
    \vspace{0.5cm} % Optional spacing between subfigures

    \begin{subfigure}[b]{\linewidth}
        \centering
        \includegraphics[width=\linewidth]{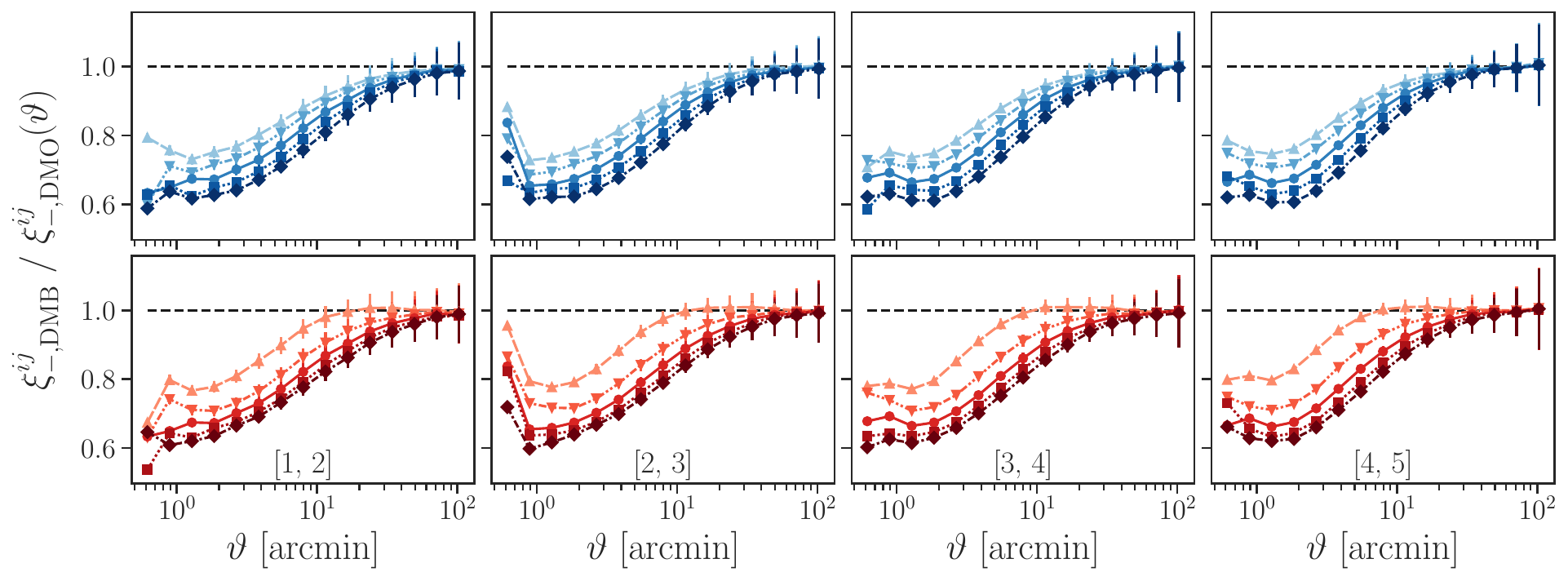}
        \caption{Baryonic suppression effect on the $\xi_{-}^{ij}(\vartheta)$ for adjacent tomographic bins.}
        \label{fig:xim_dmb_over_dmo}
    \end{subfigure}
    
    \caption{Same as \cref{fig:w_gammat_dmb_over_dmo}, but for shear--shear 2PCF. 
    % $\xi_{+}^{ij}(\vartheta)$ (a) and $\xi_{-}^{ij}(\vartheta)$ (b).
    }

    \label{fig:xip_xim_dmb_over_dmo}
\end{figure*}

\subsection{$3\times2$PCFs -- results}\label{subsec:3x2pt_measurments}
\par\noindent

We first investigate the impact of baryonic suppression on the $3\times2$PCFs, namely the galaxy--galaxy clustering ($w(\vartheta)$), the galaxy--shear ($\gamma_{\rm t}(\vartheta)$) and the shear--shear ($\xi_{\pm}(\vartheta)$) correlations. The data vectors are displayed for selected pairs of tomographic bins in \cref{fig:w_gammat_xipm_dmb_dmo},  showing in black and red the DMO and DMB cases, respectively. Data vectors for all tomographic-bin combinations and each baryonic feedback model are provided in \cref{App:Data_vectors} for completeness. The impact of baryonic physics is visible by eye over a broad range of scales and redshifts. To highlight differences between our 9 BCM models, we next 
% The data vectors for all relevant tomographic bin combinations and for all choices of BCM parameters are shown in \cref{App:D}. %These have been reported in the literature \citep{DESY3_BCM} for slightly different BCM models and we find excellent qualitative agreement. 
present our results in the form of ratios between the measurements from simulations including baryonic physics  to those from the gravity-only simulation. The results for $w(\vartheta)$, $\gamma_{\rm t}(\vartheta)$, $\xi_{+}(\vartheta)$ and $\xi_{-}(\vartheta)$ are shown in \autoref{fig:w_dmb_over_dmo}, \autoref{fig:gammat_dmb_over_dmo}, \autoref{fig:xip_dmb_over_dmo}, and \autoref{fig:xim_dmb_over_dmo}, respectively. These figures show, for a selection of tomographic bin pairs, the consequences of varying individually the two BCM parameters $M_{\rm c}$ and $\theta_{\rm ej}$,  in the upper and lower rows, respectively. We show results for $w(\vartheta)$ only for auto-correlation between tomographic bins, as the cross-correlations are highly suppressed and typically not measured. Results for $\gamma_{\rm t}(\vartheta)$ and $\xi_{\pm}(\vartheta)$ are shown for cross-correlations between adjacent tomographic bins.

As expected, increasing the value of these two parameters amounts to making stronger feedback mechanisms, which translate here into ratios that depart from unity at increasing angular scales and dive deeper. The total suppression reaches approximately 5--20\% for $w(\vartheta)$, 20--35\% for $\gamma_{\rm t}(\vartheta)$ and 10--40\% for $\xi_\pm(\vartheta)$; these numbers are affected by the physical scales probed by the various statistics and are larger when the sensitivity to small scales is increased. This also reflects in the redshift evolution of the ratios, which all show a greater departure from unity in the left panels (lower redshift) compared to those on the right. 

% \textcolor{purple}{To guide the intuition, 1Mpc at the mean redshift of tomographic bin 1 corresponds to 4 arcmin.}

The smallest scales are probed by $\xi_-(\vartheta)$ and $\gamma_{\rm t}(\vartheta)$, which both show that the baryonic suppression reaches a minimum and begins an upturn in most DMB models, in line with the BCM and hydrodynamical simulations where the stellar component tends to cool and condense well within the 1-halo regime, effectively increasing the density relative to the DMO model. The scales mostly unaffected by the baryonic effects for the $3\times2$PCFs are shown in \autoref{tab:2}. 
As one can see, clustering probes are protected over a wider range of angular separations compared to lensing, as the latter is sensitive to the matter between the source and the observer, which projects to larger angles on the sky. The scale cuts proposed in this study for $3\times2$PCFs are in agreement with \cite{DESY3_Amon, DESY3_Secco, KiDSLegacy_Wright, HSCY3_Cl, HSCY3_2PCF}.

% In Figure \ref{fig:w_dmb_over_dmo}, we observe a suppression of approximately \(10\%-20\%\) and \(5\%-20\%\) in \(w(\theta)\) for auto-correlating bins at low redshift, corresponding to the specified ranges of \(M_{\rm c}\) and \(\theta_{\rm ej}\), respectively, below the scale 1 arc minute. In contrast, the impact of baryonic feedback on bins at high redshift is comparatively mild.
% In Figure \ref{fig:gammat_dmb_over_dmo}, we observe a suppression of approximately \(20\%-40\%\) in \(\gamma_{\rm t}(\theta)\) for cross-correlating (adjacent) bins at all redshifts, corresponding to the specified ranges of \(M_{\rm c}\) and \(\theta_{\rm ej}\), respectively. 
% \ref{fig:xip_dmb_over_dmo}, we observe a suppression of approximately \(10\%-20\%\) in \(\xi_{+}(\theta)\) for cross-correlating (adjacent) bins at all redshifts, corresponding to the specified ranges of \(M_{\rm c}\) and \(\theta_{\rm ej}\), respectively. 
% \ref{fig:xim_dmb_over_dmo}, we observe a suppression of approximately \(20\%-40\%\) in \(\xi_{-}(\theta)\) for cross-correlating (adjacent) bins at all redshifts, corresponding to the specified ranges of \(M_{\rm c}\) and \(\theta_{\rm ej}\), respectively. 

\begin{figure*}
    \centering
    \begin{subfigure}[b]{\linewidth}
        \centering
        \includegraphics[width=\linewidth]{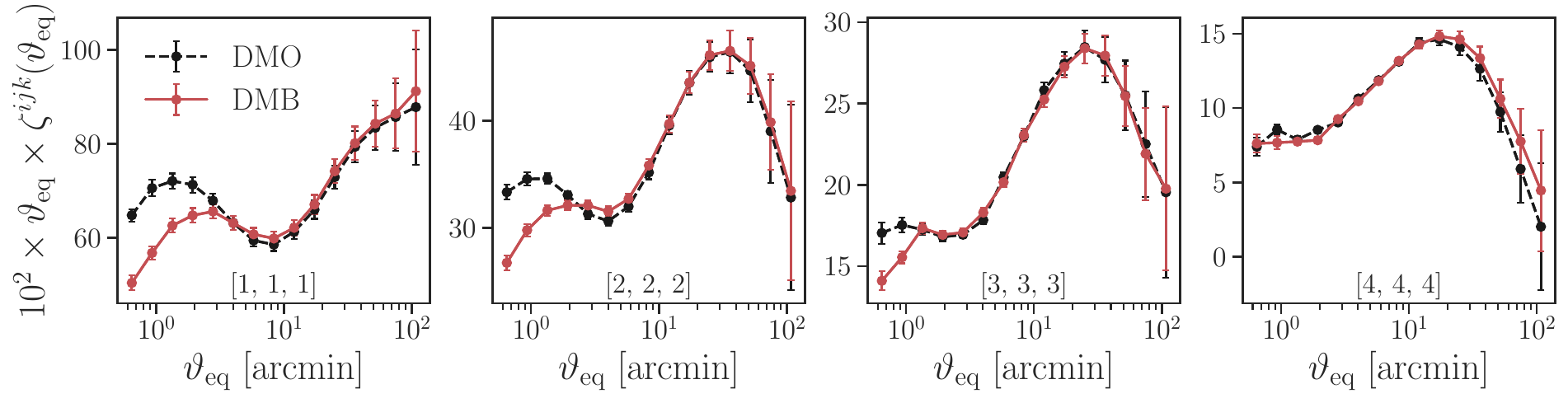}
        \caption{Galaxy--galaxy--galaxy 3PCF ($\zeta^{ijk}(\vartheta_{\rm eq})$, auto-correlations only) for the first four tomographic bins.}
        \label{fig:zeta_dmb_dmo_dv}
    \end{subfigure}
    % \vspace{0.5cm} % Optional spacing between subfigures
    \begin{subfigure}[b]{\linewidth}
        \centering
        \includegraphics[width=\linewidth]{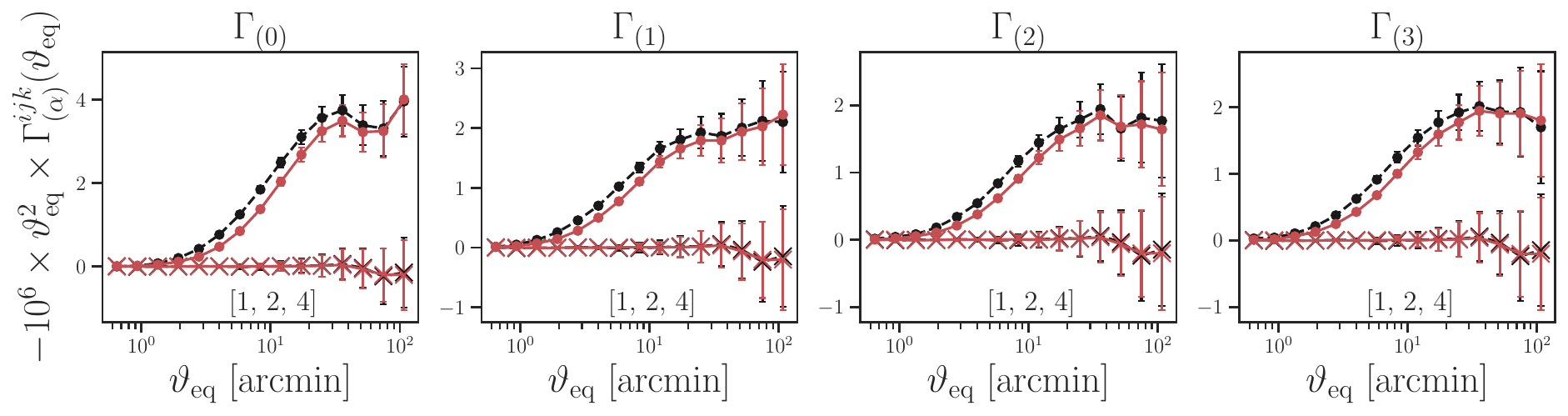}
        \caption{Natural components of shear--shear--shear 3PCF ($\Gamma^{ijk}_{(\alpha)}(\vartheta_{\rm eq})$) for tomographic bin combination [1, 2, 4]. Circular and cross data points represent the real and imaginary components. Each column corresponds to the 4 natural components ($\alpha = [0, 3]$). }
        \label{fig:Gamma_dmb_dmo_dv}        
    \end{subfigure}
     
    \caption{The two types of 3PCFs, arising from the correlations of identical spin fields, namely galaxy positions (spin-0) and galaxy shapes (spin‑2), are shown for a selected set of tomographic bin combinations. These measurements are for equilateral triangle configuration only, and are presented as a function of the length side ($\vartheta_{\rm eq}$). Black and red symbols correspond to the DMO and DMB cases, respectively. } 
    \label{fig:zeta_Gamma_dmb_dmo}
\end{figure*}

\begin{figure*}
    \centering
    \begin{subfigure}[b]{0.7\linewidth}
        \centering
        \includegraphics[width=\linewidth]{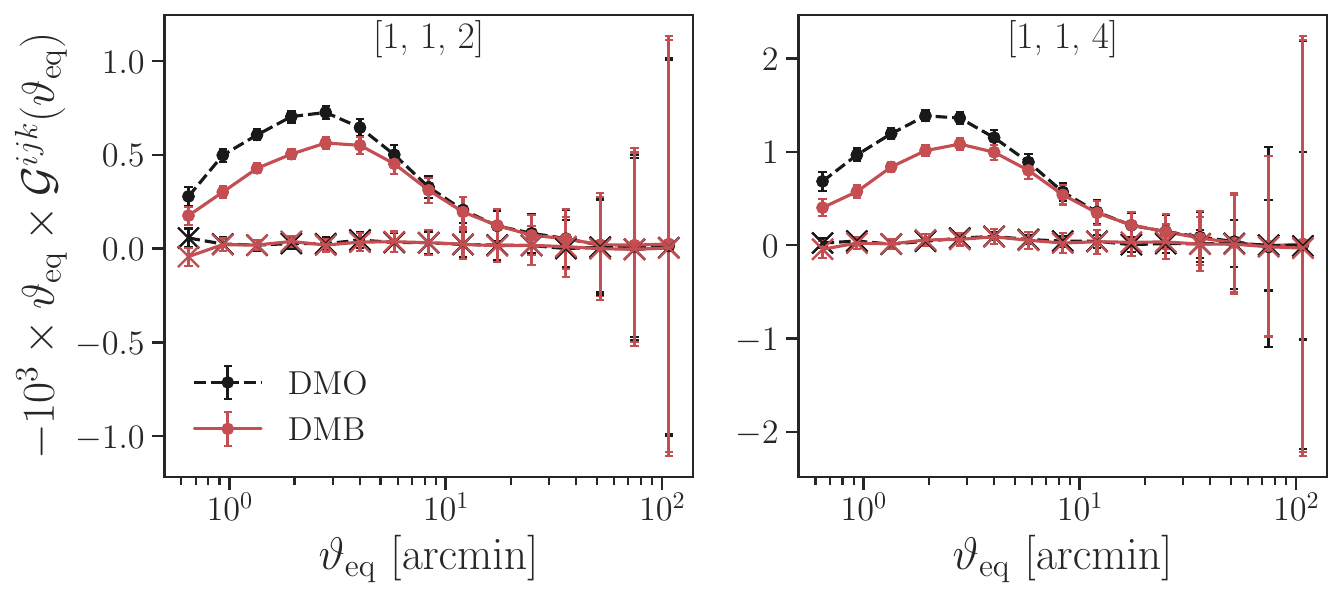}
        \caption{Galaxy--galaxy--shear 3PCF ($\mathcal{G}^{ijk}(\vartheta_{\rm eq})$) for the tomographic bin combinations [1, 1, 2] and [1, 1, 4].}
        \label{fig:ggG_dmb_dmo_dv}
    \end{subfigure}
    \hspace{0.5cm} % Optional spacing between subfigures
    \begin{subfigure}[b]{0.7\linewidth}
        \centering
        \includegraphics[width=\linewidth]{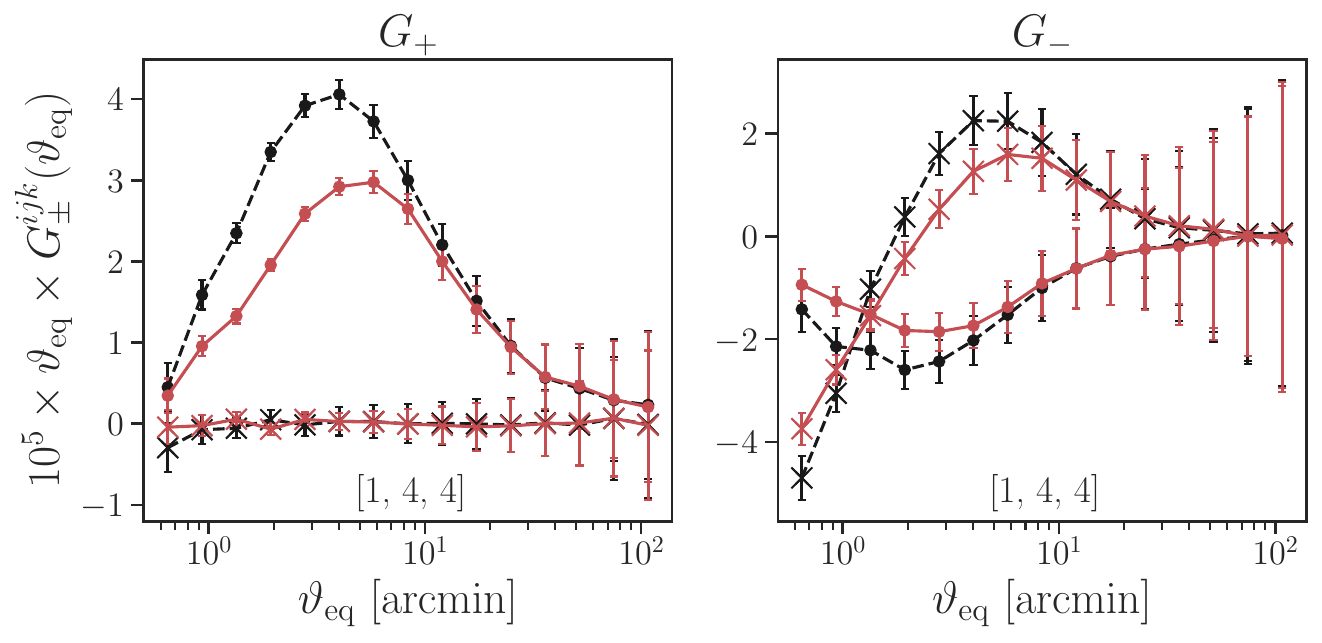}
        \caption{Natural components of galaxy--shear--shear 3PCF ($G^{ijk}_{\pm}(\vartheta_{\rm eq})$ for tomographic bin combination [1, 4, 4]. %Each column corresponds to the 2 natural components ($\alpha = [0, 1]$). 
        }
        \label{fig:gGG_dmb_dmo_dv}        
    \end{subfigure}
     
    \caption{The two types of 3PCFs arising from the cross-correlations of different spin fields, namely galaxy positions (spin‑0) and galaxy shapes (spin‑2), are shown for a selected set of tomographic bin combinations. These measurements correspond to the equilateral triangle configuration, and they are presented as a function of the equal-length side $(\vartheta_{\rm eq})$. As for \cref{fig:Gamma_dmb_dmo_dv}, filled circles and crosses represent the real and imaginary components. The data vectors shown in black (solid line) and red (dashed) correspond to the DMO and DMB cases, respectively. } 
    \label{fig:ggG_gGG_dmb_dmo}
\end{figure*}

\begin{figure*}
    \centering
    \begin{subfigure}[b]{\linewidth}
        \centering
        \includegraphics[width=\linewidth]{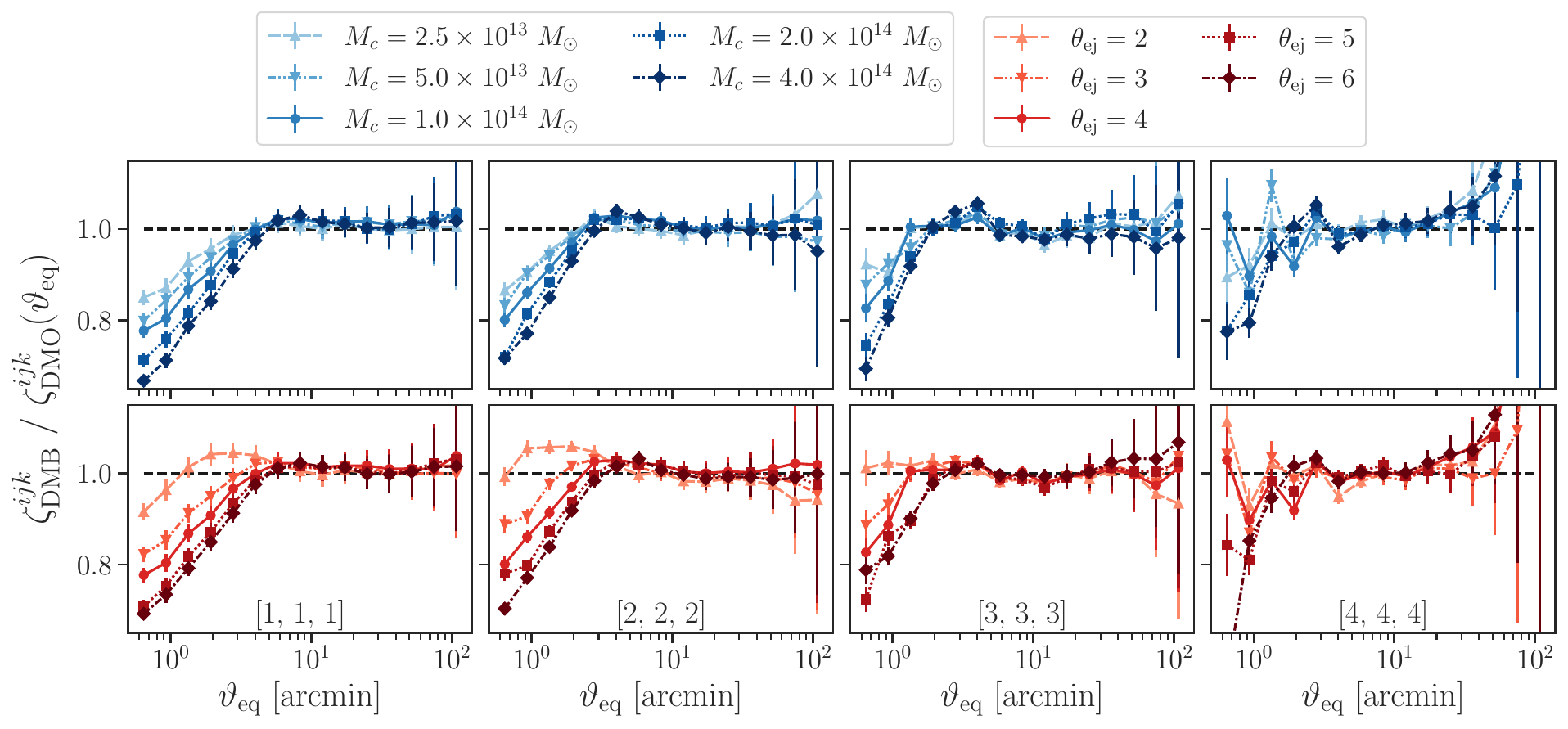}
        \caption{Baryonic suppression effect on galaxy--galaxy--galaxy 3PCF ($\zeta^{ijk}(\vartheta_{\rm eq})$) for the first four auto-correlating tomographic bins.}
        \label{fig:zeta_dmb_over_dmo}
    \end{subfigure}
    
    \vspace{0.5cm} % Optional spacing between subfigures

    \begin{subfigure}[b]{\linewidth}
        \centering
        \includegraphics[width=\linewidth]{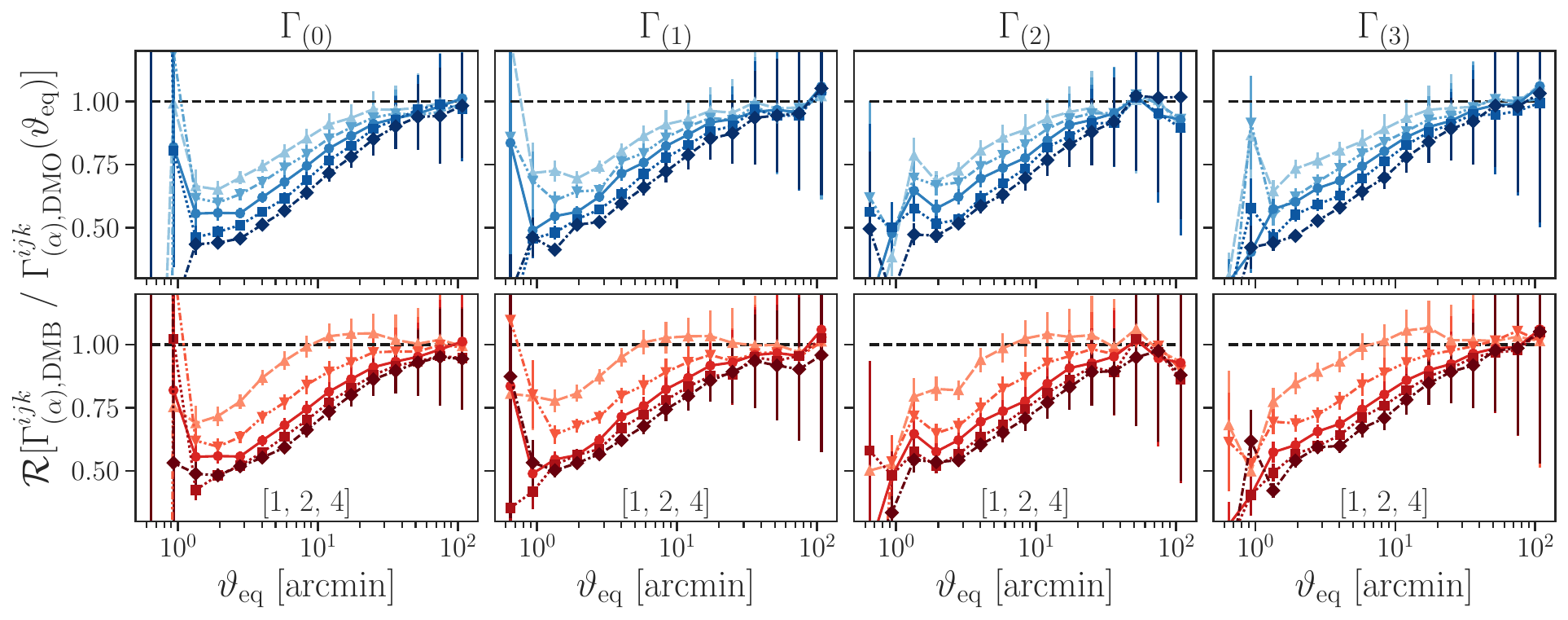}
        \caption{Baryonic suppression effect on the real part of the four natural components of shear--shear--shear 3PCF ($\Gamma_{(\alpha)}^{ijk}(\vartheta_{\rm eq})$) for tomographic bin combination [1, 2, 4]. As shown in Appendix \ref{App:Data_vectors}, the imaginary part vanishes for this statistic.} %Each column corresponds to the 4 natural components ($\alpha = [0, 3]$).
        \label{fig:Gamma_dmb_over_dmo}
    \end{subfigure}
    
    \caption{Same as \cref{fig:w_gammat_dmb_over_dmo}, but for galaxy--galaxy--galaxy 3PCF (a), and real part of the four natural components of shear--shear--shear 3PCF (b).}
    \label{fig:zeta_Gamma_dmb_over_dmo}
\end{figure*}

\begin{figure*}
    \centering
    \begin{subfigure}[b]{0.75\linewidth}
        \centering
        \includegraphics[width=\linewidth]{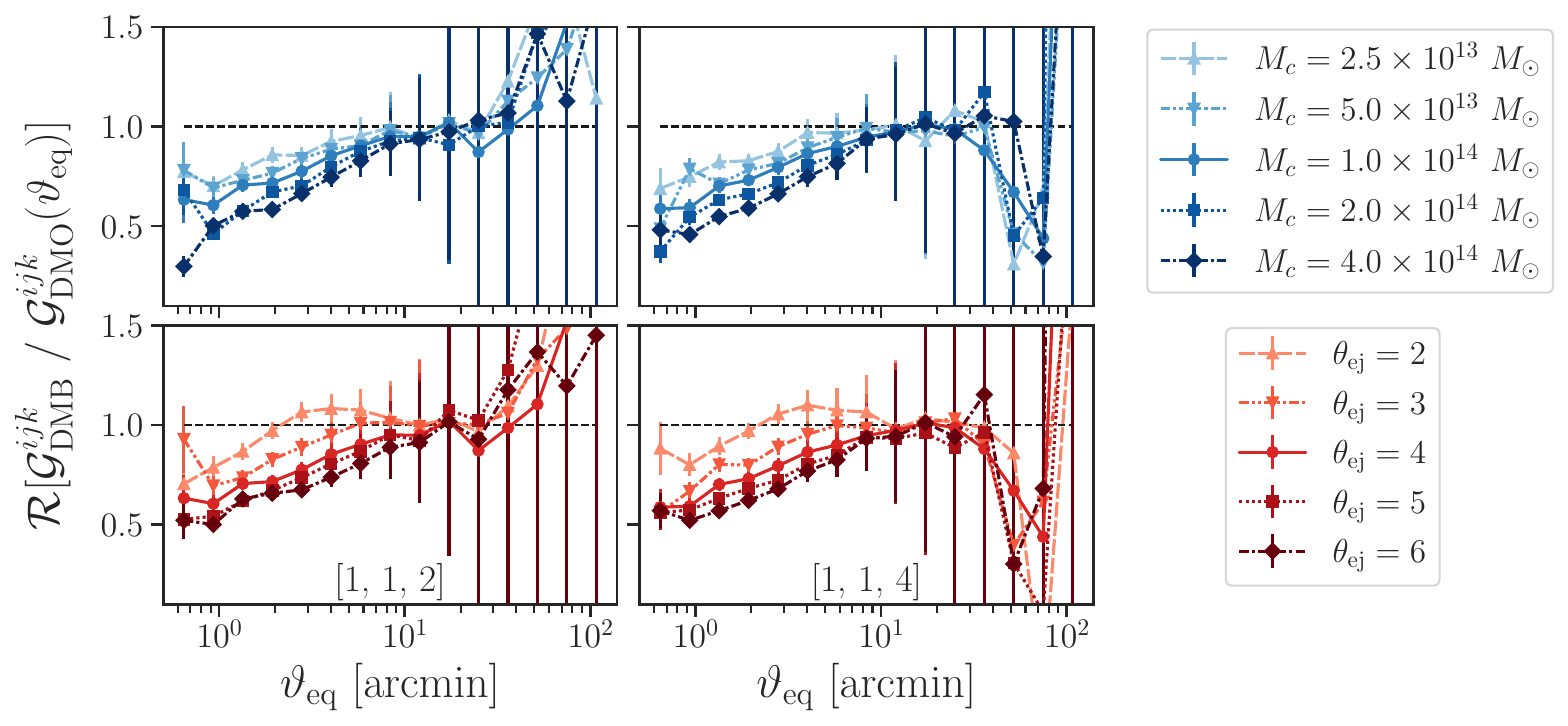}
        \caption{Baryonic suppression effect on the real part of the galaxy--galaxy--shear 3PCF $(\mathcal{G}^{ijk}(\vartheta_{\rm eq}))$ for tomographic bin combinations [1, 1, 2], and [1, 1, 4].}
        \label{fig:ggG_dmb_over_dmo}
    \end{subfigure}
    
    % \vspace{0.5cm} % Optional spacing between subfigures

    \begin{subfigure}[b]{0.75\linewidth}
        \centering
        \includegraphics[width=\linewidth]{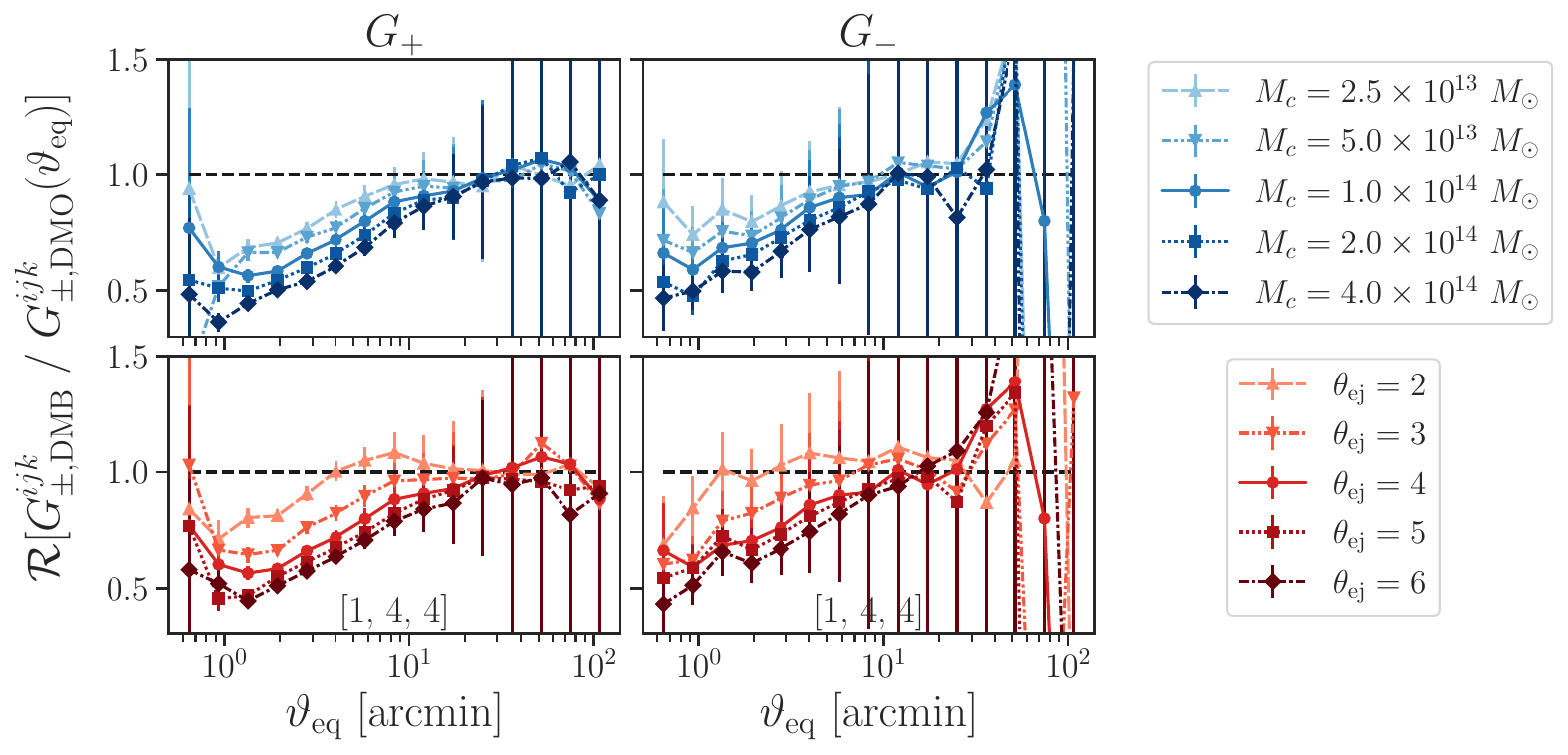}
        \caption{Baryonic suppression effect on the real part of the natural components of the galaxy--shear--shear 3PCF $(G_\pm^{ijk}(\vartheta_{\rm eq}))$ for tomographic bin combination [1, 4, 4]. The left and right columns correspond to $G_+$ and $G_-$, respectively. }
        \label{fig:gGG_dmb_over_dmo}
    \end{subfigure}
    
    \caption{Same as \cref{fig:w_gammat_dmb_over_dmo}, but for the real part of galaxy--galaxy--shear 3PCF (a) and the two natural components of galaxy--shear--shear 3PCF (b).}
    \label{fig:ggGL_gGG_dmb_over_dmo}
\end{figure*}

\subsection{$4\times3$PCFs -- results}\label{subsec:4x3pt_measurments}
\par\noindent

We first remark that baryonic feedback acts directly on the matter density field $\delta(x)$, hence we expect statistics that are more sensitive to higher powers of  $\delta(x)$ (such as 3PCFs) to be more affected, and therefore reach lower values (corresponding to higher suppression) in their DMB/DMO ratios. %This is discussed in the next section.
%As mentioned above, the impact of baryon feedback is expected to be larger for three-point statistics than for two-point statistics, which 
This is indeed confirmed for all parts of our $4\times3$PCF data vectors. The outputs from \textsc{TreeCorr} contain all types of triangle configurations, but in this study, we focus on the baryonic effects in the equilateral triangle configuration. 
The full shapes of the data vectors for relevant combinations of tomographic bin triplets are presented in \cref{App:Data_vectors}, and we once again focus here on the ratio between the nine different DMB cases and the DMO case, for a few selections of tomographic bin triplets. 

The data vectors for the ggg and the natural components of GGG 3PCFs are shown in \cref{fig:zeta_dmb_dmo_dv} and \cref{fig:Gamma_dmb_dmo_dv}, respectively, clearly contrasting the DMO and DMB models. The former statistic is shown for auto-tomographic bins only, as the signal is highly suppressed otherwise. We also note that the impact of baryons is weaker in the highest redshift bin, since the physical scales affected project to angular scales smaller than those probed by $\zeta^{444}(\vartheta_{\rm eq})$. The natural components of GGG 3PCF are shown for the redshift bin combination [1, 2, 4] as a representative example, although other redshift combinations have an overall similar shape. As observed, $\Gamma_{(\alpha)}^{124}$ is affected by baryons at almost all angular scales due to the lensing projection.  

Next, we show the first measurement of the ggG and gGG statistics ($\mathcal{G}(\vartheta_{\rm eq})$ and $G_{\pm}(\vartheta_{\rm eq})$, respectively) in \cref{fig:ggG_dmb_dmo_dv} and \cref{fig:gGG_dmb_dmo_dv}, respectively. We choose the tomographic bin combinations [1, 1, 2] and [1, 1, 4] for the former to ensure a strong signal, in which the two galaxy positions are in the same tomographic bin and therefore auto-correlate, combined with a lensing component well in the background and therefore probing the matter--galaxy cross-correlation efficiently. This choice can be relaxed for the $G_\pm(\vartheta_{\rm eq})$ statistics, where the signal is strong provided that the lens sample is in the foreground. We therefore opt to show the $G_\pm(\vartheta_{\rm eq})$ results for the [1, 4, 4] combination.  
A measurement where lenses are behind sources can be used to probe magnification, as in \cite{Duncan_magnification} or intrinsic alignment of galaxies \citep{IA_lens_flipped}, which we leave for future work. Data vectors for more tomographic bin combinations of these $4\times3$PCFs with all baryonic suppression models are shown in \cref{fig:ggg_dmb_dmo_allbin}, \cref{fig:ggG_dmb_dmo_allbin}, and \cref{fig:gGG_dmb_dmo_allbin}, and \cref{fig:Gamma_dmb_dmo_allbin}, respectively.
%In all these figures, similar to 2PCFs, data vectors in black and red correspond to DMO and DMB cases, respectively. 
% Data vectors for all tomographic-bin combinations and each baryonic feedback model are provided in \cref{App:Data_vectors} for completeness.

\cref{fig:zeta_dmb_over_dmo} presents the ratio $\zeta_{\rm DMB}/\zeta_{\rm DMO}(\vartheta_{\rm eq})$, where we observe a suppression that exceeds 30\% at the smallest scales. Comparing this with the ratio obtained for $w(\vartheta)$, the signal has a similar shape and an amplitude that is larger by
%  $10-15\%$,
%a factor of $\sim$3/2, aligned with our interpretation that the ratio roughly scales with the order of the field. This is a rule of thumb and should not be over-interpreted, but holds well at low redshift: 
up to 50\%. %for the smallest 5 angular bins, the baryonic suppression in $\zeta^{111}$, averaged over the 5 $M_{\rm c}$ models, is $1.40 \pm
%0.09$ times larger than that for $w^{11}$; when averaged over the $\theta_{\rm ej}$ models instead, this ratio is $1.52 \pm 0.20$. 
The scale-mixing caused by lensing blurs this relation in the other statistics, but an enhancement in the suppression is generally seen with respect to that of $w(\vartheta)$, regardless of the statistics and redshift, sometimes 90\% larger. This can be seen by comparing the data reported in \cref{tab:2} and \cref{tab:3}.

The redshift evolution is also clearly visible from \cref{fig:zeta_dmb_over_dmo}, as the ratio is close to unity for separations larger than 1 arcmin at higher redshifts, whereas it departs significantly at lower redshifts. We also notice a small bump before the drop, which is most prominent in the $\theta_{\rm ej} = 2$ model. Hints of this could be seen in the $w(\vartheta)$ measurements, but with a lower significance. The feature here is unambiguous and is caused by the excess clustering formed at the outskirts of the halos, where the hot gas is ejected (and therefore varies with $\theta_{\rm ej}$ but not $M_{\rm c}$). 

We present in \cref{fig:Gamma_dmb_over_dmo} the ratios of the real part of four natural components of the shear--shear--shear 3PCF for DMB and DMO cases, $\Gamma_{(0-3)}(\vartheta_{\rm eq})$, where the imaginary part vanishes for equilateral triangles. For other triangle configurations, the imaginary parts could serve as additional data vectors alongside the real parts. 
% We focus on one 
For the representative tomographic bin combination [1, 2, 4], the ratios are similar in shape and amplitude for all components. However, $\Gamma_{(0)}(\vartheta_{\rm eq})$ has the highest signal-to-noise and thus has a greater detection power.

The baryonic suppression on the real component of the ggG and gGG statistics %, $\mathcal{G}_{\rm DMB}/\mathcal{G}_{\rm DMO}(\vartheta_{\rm eq})$ and $G_{\pm, \rm DMB}/G_{\pm, \rm DMO}(\vartheta_{\rm eq})$, 
are shown in \cref{fig:ggG_dmb_over_dmo} and \cref{fig:gGG_dmb_over_dmo}, respectively.
The ratios measured in these statistics are similar to those of $\zeta(\vartheta_{\rm eq})$, except that the bump is suppressed. 
We notice that the sample variance is smaller for $G_+(\vartheta_{\rm eq})$ than for $G_-(\vartheta_{\rm eq})$, and that the suppression is stronger and reaches saturation in the latter.
% former. 
% In fact, $G_+(\vartheta_{\rm eq})$  probes scales similar to $\xi_+(\vartheta)$, while $G_-(\vartheta_{\rm eq})$ has more resemblance with $\xi_-(\vartheta)$.  
These three statistics reach 25-55\% suppression, depending on the BCM model. We show here only the real part, noting that the imaginary part of $\mathcal{G}(\vartheta_{\rm eq})$ and $G_+(\vartheta_{\rm eq})$ vanish, however imaginary part of $G_-(\vartheta_{\rm eq})$ does not vanish for equilateral triangles; see \cref{fig:gGG_dmb_dmo_dv}.

% \avi{The data vectors for the four natural components corresponding to the shear--shear--shear 3PCF for the DMO and the DMB cases are shown in \cref{fig:Gamma_dmb_dmo_dv} for the tomographic bin combination [1, 2, 4].}

% \subsection{Scale affected by baryons}
% \par\noindent

The scales that are mostly unaffected by baryonic effects for the $4\times3$PCFs are shown in \autoref{tab:3}. As for the 2PCFs, probes with a higher number of lensing fields are more sensitive to the matter distribution contained between the galaxy sample and the observer, which is at a lower redshift, and therefore are more sensitive the impact of baryons at fixed angular separations.

To guide the intuition about the physical scales where baryonic feedback is important for galaxy clustering 2PCF and 3PCF, it is reasonable to evaluate those at the mean redshift of the tomographic bin. The angular scales below which baryonic feedback is significant for $w^{11}(\vartheta)$, and $\zeta^{111} (\vartheta_{\rm eq})$ are 4 arcminutes, which corresponds to a physical scale of about 1.1 Mpc for the first tomographic bin, when deprojected with the mean bin redshift of 0.3. Note that it is only possible to convert angles into physical scales for a fixed redshift; however, in this study we are primarily interested in 3PCFs involving cross-correlations between broad redshift bins with even broader lensing kernels, which makes non-trivial the determination of a dominant physical scale.
% 1Mpc at the mean redshift of tomographic bin 1 corresponds to 4 arcmin.

%It is important to note that this heuristic scaling does not imply identical contributions from
It is important to realize that modifications to the density field do not impact equally shear and density fields. %The galaxy density field ($\delta_{\rm g}$) is a scalar quantity, while the shear field ($\varepsilon$) is a spin-2 quantity derived from the projected gravitational potential, which integrates the matter distribution along the line of sight. Consequently, 
Clustering statistics such as $w(\vartheta)$ and $\zeta(\vartheta_{\rm eq})$ respond directly to changes in the matter overdensity field $\delta(x)$, whereas shear-based statistics ($\xi_{\pm}(\vartheta)$, $\Gamma_{(\alpha)}(\vartheta_{\rm eq})$) are influenced through lensing kernels and projection effects that smooth and redistribute the impact of baryonic feedback across scales. 
%The approximate factor of $3/2$ reflects the difference in field order--quadratic for two-point and cubic for three-point correlations--rather than an assumption of equal sensitivity. 
In practice, the ratio between 3pt baryonic suppression and 2pt baryonic suppression varies across probes and scales, as seen in 
% \cref{fig:w_gammat_dmb_over_dmo}, \cref{fig:Shear_3pcf_dmb_over_dmo}
% Figures~5--7 and Tables~2--3, 
\cref{tab:2}, and \cref{tab:3} where suppression ratios range from near unity to values well above 50\%, depending on the mix of density and shear fields, projection effects, and feedback parameters. 
% See Appendix \ref{App:3/2} for more details on this comparison.

\begin{table*}[tbp]
\centering
\begin{tabular}{|c|c|c|c|}
\hline
\textbf{Correlation} & \multicolumn{2}{c|}{\textbf{Baryonic Suppression below 1 arcmin}} &
 \textbf{Scale cuts} ($\vartheta_{5\%}$)\\
\cline{2-3}
 & $M_{\rm c}$ & $\theta_{\rm ej}$ & (arcmin)\\
\hline
$w(\vartheta)$ & $\sim10\%-20\%$ & $\sim5\%-20\%$ & $\sim 2$\\
$\gamma_{\rm t}(\vartheta)$ & $\sim20\%-40\%$ & $\sim20\%-40\%$ & $\sim 10$\\
$\xi_+(\vartheta)$ & $\sim10\%-20\%$ & $\sim10\%-20\%$ & $\sim 4$\\
$\xi_{-}(\vartheta)$ & $\sim20\%-40\%$ & $\sim20\%-40\%$ & $\sim 30$\\
\hline
\end{tabular}
\caption{\label{tab:2} Estimated baryonic suppression effects for the $3\times2$PCFs below 1 arcminute scale for the minimum and maximum values of $M_{\rm c}$ and $\theta_{\rm ej}$ are shown in Columns 2 and 3, respectively. As a reference for comparison, column 4 represents the scales above which the statistics are less than 5\% affected by baryonic feedback for the fiducial DMB model. 
% The values reported here are for the fiducial DMB model, and for a few tomographic bin pairs.
}
\end{table*}

\begin{table*}[tbp]
\centering
\begin{tabular}{|c|c|c|c|}
\hline
\textbf{Correlation} & \multicolumn{2}{c|}{\textbf{Baryonic Suppression below 5 arcmin}} & \textbf{Scale cuts}($\vartheta_{5\%}$)\\
\cline{2-3}
 & $M_{\rm c}$ & $\theta_{\rm ej}$ & (arcmin)\\
\hline
$\zeta(\vartheta)$ & $\sim15\%-35\%$ & $\sim5\%-35\%$ & $\sim 3$\\
$\mathcal{G}(\vartheta)$ & $\sim25\%-60\%$ & $\sim20\%-50\%$ & $\sim 8$\\
$G_+(\vartheta)$ & $\sim20\%-50\%$ & $\sim15\%-45\%$ & $\sim 7$\\
$G_-(\vartheta)$ & $\sim20\%-60\%$ & $\sim20\%-60\%$ & $\sim 15$\\
$\Gamma_{(\alpha)}(\vartheta)$ & $\sim20\%-60\%$ & $\sim20\%-60\%$ & $\sim 40$\\
\hline
\end{tabular}
\caption{\label{tab:3} Estimated relative strength of baryonic suppression effects for the $4\times3$PCFs below 5 arcminute scale for the minimum and maximum values of $M_{\rm c}$ and $\theta_{\rm ej}$ are shown in Columns 2 and 3, respectively. As in \cref{tab:2}, column 4 represents the scales above which $4\times3$PCFs are less than 5\% affected by baryonic feedback for the fiducial DMB model. 
% The values reported here are for the fiducial DMB model, and for a few tomographic bin triplets.
}
\end{table*}

\subsection{Probing BCM parameters in the future}\label{subsec:probing_bcm_params}
\par\noindent

Forward modeling the impact of baryons on weak lensing statistics allows the inclusion of smaller angular scales in cosmological analyses, which improves the general precision on parameter constraints. The capacity of the $4\times3$PCFs at measuring departures from the DMO model could be complementary to that of $3\times2$PCFs, hence could be used to obtain interesting priors on the baryonic parameters $M_{\rm c}$ and $\theta_{\rm ej}$ as in \cite{Grandis2024}. A simple way to quantify our ability to measure these parameters is to extract the angular scale at which the DMO model can be rejected from the baryonified data. It corresponds to the cut needed to protect the analysis against significant biases when baryonified data is analyzed with gravity-only theory. To make a fair comparison between 2PCFs and 3PCFs, whose data vector sizes are very different, we consider here only a single combination of tomographic bins, but could easily extend this to the full data vector that includes all combinations of tomographic bins. We carry out this hypothesis rejection exercise by computing the $p$-values for a single representative combination of tomographic bins, using the diagonal part of the jackknife error in these calculations, and a threshold of $p=0.01$. We note that our error propagation is not optimal and would benefit from using a full matrix, however our jackknife sample has only 50 patches, which produces a noisy matrix whose inversion is noisy and biased. We plan to work with a larger ensemble of mock data in the future to improve on the covariance estimation, and compare this with analytical calculations, which are tractable.

The scale cuts derived from hypothesis rejection are not only relevant for protecting analyses against baryonic biases but also have direct implications for cosmological inference. By identifying angular ranges where baryonic effects remain below 
a given threshold, we ensure that the retained scales are dominated by gravitational physics and thus provide unbiased constraints on cosmological parameters such as the matter density and the amplitude of matter fluctuations. Conversely, excluding small scales reduces statistical power, which can weaken constraints if not compensated for by complementary probes. 
This trade-off highlights the importance of incorporating accurate baryonic models: doing so would allow us to safely include smaller scales, thereby tightening cosmological constraints while simultaneously constraining feedback parameters. Future joint analyses of $3\times2$PCFs and $4\times3$PCFs will exploit this synergy to break degeneracies between cosmology and baryonic physics.

Our results are reported in \cref{tab:4}, where smaller values indicate statistics for which the DMO model is close to the baryonified data over a large range of scales, pointing to probes less sensitive to BCM parameters. We first observe that the clustering 2PCFs and 3PCFs, $w(\vartheta)$ and $\zeta(\vartheta_{\rm eq})$, have similar cut angles for all 9 models considered here, varying by only one angular bin. As expected, stronger models ($M_{\rm c}=4\times10^{14}M_\odot$ and $\theta_{\rm ej}=6$) require a larger angular cut if they are to remain consistent with DMO.
Second, the lensing 3PCFs are bracketed 
by $\xi_+(\vartheta)$ and $\xi_-(\vartheta)$, with a greatest sensitivity to baryons coming from $\xi_-$, followed by the first natural component $\Gamma_{(0)}(\vartheta_{\rm eq})$. Finally, the cross-correlation terms $G_\pm(\vartheta_{\rm eq})$ and $\mathcal{G}(\vartheta_{\rm eq})$ are all closer to DMO than the galaxy--galaxy lensing term $\gamma_{\rm t}(\vartheta)$. These results suggest that 3PCFs are not necessarily better or worse at measuring baryonic feedback; their signal is more affected, but the overall precision is lower (i.e., 3PCFs have larger error bars than 2PCFs, 
% as seen in  \cref{fig:w_dmb_over_dmo} and \cref{fig:zeta_dmb_over_dmo} for galaxy clustering 2PCF and 3PCF, respectively). 
The final answer will, of course, emerge from a joint $3\times2$PCFs + $4\times3$PCFs MCMC analysis where cosmology and BCM parameters will be jointly varied. This requires analytical modeling of the full $4\times3$PCFs data vector presented here, and the covariance matrix, either from theory or from simulation-based emulation, neither of which is fully available yet.

\begin{table*}[tbp]
\centering
\begin{tabular}{|c|c|c|c|c|c|}
\hline
\textbf{Correlation} & Fiducial & $M_{\rm c} = 2.5 \times 10^{13}  M_{\odot}$ & $M_{\rm c} = 4 \times 10^{14} M_{\odot}$ & $\theta_{\rm ej} = 2$ & $\theta_{\rm ej} = 6$\\
\hline \hline
$w^{11}(\vartheta)$ & 1.84''  &  1.28" & 2.66" & 0.89" & 2.66"\\
$\zeta^{111}(\vartheta)$ & 1.93" & 0.93" & 1.93" & 0.65" & 1.93" \\
\hline
$\xi_+^{12, 14}(\vartheta)$ & 1.84"  &  0.89" & 1.84" & 0.62" & 1.84"\\
$\xi_-^{12, 14}(\vartheta)$ & 16.52"  &  11.46" & 23.81" & 5.52" & 23.81" \\
$\mathcal{R}[\Gamma_{(0)}^{124}(\vartheta)]$ & 12.03"  &  5.79" & 17.34" & 2.79" & 17.34"\\
$\mathcal{R}[\Gamma_{(1-3)}^{124}(\vartheta)]$ & 5.80"  &  4.02" & 8.35" & 1.93" & 8.35"\\
\hline
$\gamma_{\rm t}^{14}(\vartheta)$ & 7.96"  &  5.52" & 11.47" & 2.66" & 11.47" \\
$\mathcal{R}[\mathcal{G}^{114}(\vartheta)]$ & 1.94"  & 1.34" & 2.79" & $<$ 0.65" & 2.79"\\
$\mathcal{R}[G_+^{144}(\vartheta)]$ & $<$0.65" & $<$0.65"  & 0.93" & $<$ 0.65" & $<$ 0.65"\\
$\mathcal{R}[G_-^{144}(\vartheta)]$ & 4.02"  &  2.79" & 5.79" & 1.94" & 5.79" \\
\hline
\end{tabular}
\caption{\label{tab:4} Angular scales below which the DMO model can be ruled out at 99\% confidence by each probe, at fixed cosmology. Comparisons between the 2pt and 3pt correlations for similar probes are shown for specific tomographic bin combinations. Again, the fiducial case corresponds to $(M_{\rm c, fid}, \theta_{\rm ej, fid}) = (1\times10^{14} M_\odot, 4)$.}
\end{table*}

\section{Discussions and Conclusions}
\label{sec:conclusions}
\par\noindent

In this work, we studied the impact of baryonic feedback on both the $3\times2$PCFs and $4\times3$PCFs using the baryonic correction model and shell-based baryonification method of \cite{Kacprzak_2023}. We present some of these measurements for the first time, and argue that $4\times3$PCFs can be used as a powerful cosmological probe, complementing the conventional $3\times2$PCFs.  %Recall that the non-Gaussian information contained in the matter field strongly motivates an inclusion of these higher-order statistics, which must handle signal contamination by key astrophysical effects such as baryonic feedback}. 
The recent feature added to \textsc{TreeCorr} marks a significant advancement toward enabling studies of the full set of 3PCFs, including cross-correlations between galaxy positions and shapes.
% that enables the computation of cross-correlations between different spin quantities (e.g., galaxy position and shape) in 3PCFs opens up a new horizon in cosmology. 

Moreover, the signature of baryonic physics is strong and distinct in the 3PCFs, which should allow for degeneracy breaking from cosmological parameters in analyses. For example, cosmological parameters (such as the matter density and the amplitude of matter fluctuations) primarily affect the overall amplitude and shape of correlation functions across all scales, whereas baryonic feedback introduces localized, scale-dependent suppression at small scales and distinct features such as the ``bump'' associated with the gas ejection radius ($\theta_{\rm ej}$).  These patterns are not easily mimicked by cosmological variations, making them distinguishable.  $3\times2$PCFs are powerful for constraining cosmology on large scales but are unable to capture higher-order information.
% mostly sensitive to Gaussian features.  
In contrast, $4\times3$PCFs probe higher-order information of non-Gaussian structures such as filaments and clusters and capture small-scale information where baryonic effects dominate. 
Combining 2PCFs and 3PCFs leverages their complementary responses, resulting in robust cosmological constraints and functionality to mitigate baryonic physics systematic effects. In other words, because these probes react differently to cosmology and feedback, a joint analysis can effectively separate their contributions, reducing biases in cosmological inference. This will be explored in future work.

Challenges remain in the modeling of higher-order statistics, especially in achieving an accurate understanding of astrophysical systematics such as the baryonic feedback investigated here, but also the intrinsic alignment of galaxies \citep{SkySim5000IA}, or the magnification bias in the presence of non-linear galaxy bias \citep{non-linear-bias}. In addition to that, 
% modeling 3PCFs from theory requires accurate prescriptions for nonlinear three-point statistics (e.g., beyond current bi-halofit models), 
analyses must contend with high-dimensional data vectors, which necessitate compression techniques, optimal selection of configurations, or reliable analytical covariance matrices.

While the BCM framework provides a computationally efficient and physically motivated alternative to hydrodynamical simulations, it remains an approximation. In particular, the simple form of the model adopted here does not capture the full complexity of baryonic processes, and more sophisticated versions, such as the updated \cite{SchneiderBCM2025} model, or large suites of fully hydrodynamical simulations as those of \cite{vanDaalenSuite}, would offer greater accuracy. Furthermore, extending the modeling from projected 2D statistics to full 3D correlation functions would provide additional fidelity. The long-term scope of our method, however, is not so much to describe the baryonic sector perfectly, but to marginalize over plausible scenarios in order to de-bias constraints on the dark sector. For this, the requirements on the BCM accuracy are relaxed, and since the parameters are associated with physical properties, priors already exist on some of these \citep{Grandis2024}, which help narrow down the high-dimensional parameter volume.

There are several avenues for future progress. Extending the modeling to include a larger number of BCM parameters, exploring joint treatments of baryons with other systematics such as intrinsic alignment of galaxies, and testing the robustness of results against hydrodynamical simulations at the data vector level first, then with full inference, are natural next steps. 
% Beyond two-point and three-point statistics, higher-order or alternative probes may provide further sensitivity to baryonic effects. 
In addition, varying triangular configurations, moving beyond linear galaxy bias assumptions in both mocks and theoretical modeling, and refining theoretical predictions of the matter bispectrum will be crucial for unlocking the full potential of 3PCFs.

Overall, our findings highlight both the opportunities and challenges of using 3PCFs as cosmological probes in the presence of baryonic feedback. With continued theoretical development, improved models, and joint analyses across probes, 3PCFs can play a key role in advancing precision cosmology in the era of high-precision stage-IV surveys.

\section*{Acknowledgements}
\par\noindent

We are deeply grateful to Tomasz Kacprzak for his generous help with developing the interface between the public \textsc{CosmoGridV1} baryonification code and our HACC simulations. 
%We would like to thank XYZ for their thoughtful comments on the manuscript. 
This paper has undergone internal review in the LSST Dark Energy Science Collaboration. % REQUIRED
The internal reviewers were Anik Halder and Dani Leonard, and we thank them for their valuable comments and suggestions. % Optional but recommended
A special thanks also to  Janis Fluri, who developed an earlier version of the baryonification code and made it public. 
We acknowledge Cora Uhleman, Lina Castiblanco Tolosa and Sofía del Pilar Samario Nava for providing useful comments and feedback on the manuscript. 

All authors contributed to the development and writing of this paper. AB and JHD led the analysis; JMF produced the core simulation products specific to this work, based on $N$-body simulations produced by KH, and post-processed by {\sc Pollux}, which is co-developed by CD; MJ developed the {\sc TreeCorr} measurement tool;  
MI contributed to the calculation, validation and interpretation of the 3PCFs for this project.
%and an overlapping project for 3 PCFs IA, as well as contrasting baryonic effects on 2PCFs versus 3PCFs for degeneracies with cosmological parameters.   
%while XYZ.

JHD acknowledges partial support from an STFC Ernest Rutherford Fellowship (project reference ST/S004858/1). The work of KH at Argonne National Laboratory was supported under the U.S. DOE contract DE-AC02-06CH11357.
MI and AB acknowledge support in part by the Department of Energy, Office of Science, under Award Number DE-SC0022184 and also in part by the U.S. National Science Foundation under grant AST2327245.
This research used resources of the Argonne Leadership Computing Facility, which is a DOE Office of Science User Facility supported under Contract DE-AC02-06CH11357. 
%LC and CU were supported by the STFC Astronomy Theory Consolidated Grant ST/W001020/1 from UK Research \& Innovation,  CU was also supported by the European Union (ERC StG, LSS\_BeyondAverage, 101075919). 

The DESC acknowledges ongoing support from the Institut National de Physique Nuclé-aire et de Physique des Particules in France; the Science \& Technology Facilities Council in the United Kingdom; and the
Department of Energy and the LSST Discovery Alliance
in the United States.  DESC uses resources of the IN2P3 
Computing Center (CC-IN2P3--Lyon/Villeurbanne - France) funded by the Centre National de la Recherche Scientifique; the National Energy Research Scientific Computing Center, a DOE Office of Science User 
Facility supported by the Office of Science of the U.S.\ Department of Energy under Contract No.\ DE-AC02-05CH11231; STFC DiRAC HPC Facilities, funded by UK BEIS National E-infrastructure capital grants; and the UK particle physics grid, supported by the GridPP Collaboration.  This work was performed in part under DOE Contract DE-AC02-76SF00515.

Some of the results in this paper have been derived using the \textsc{healpy} and \textsc{healpix} packages. The authors acknowledge the developers and contributors of the open-source software packages that made this work possible, including but not limited to \textsc{NumPy} (Harris et al. 2020) and  \textsc{Matplotlib} (Hunter 2007). These tools were essential in performing the analysis and generating the results presented in this work.

%%%%%%%%%%%%%%%%%%%%%%%%%%%%%%%%%%%%%%%%%%%%%%%%%%
\section*{Data Availability}

 %The baryonified shells are available at NERSC - \url{$CFS/lsst/shared/external/HACC-HOS-Y1/BCM}. The galaxy catalogs are avaialble at - .............

Simulated data products generated in this paper can be made available upon request.

% %%%%%%%%%%%%%%%%%%%% REFERENCES %%%%%%%%%%%%%%%%%%

% The best way to enter references is to use BibTeX:
% \bibliographystyle{unsrt}
% \bibliographystyle{JCAP}
% \bibliographystyle{JHEP}
% \bibliography{citation}{} % if your bibtex file is called example.bib

%%%%%%%%%%%%%%%%%%%% APPENDICES %%%%%%%%%%%%%%%%%%

% \newpage
\appendix
\crefalias{section}{appendix}
\section{Relation between 3PCFs and corresponding bispectra}\label{App:theory_3PCF}
\par\noindent
\crefalias{subsection}{appendix}

\subsection*{Shear--shear--shear 3PCF} \label{App:sss_3PCF}
\par\noindent

The natural components $\Gamma_{(\alpha)}$ ($\alpha=0,1,2,3$), of the shear--shear--shear 3PCF can be expressed in terms of the convergence bispectrum. This formulation was originally developed by \cite{Schneider_2005a} and has since been implemented in numerical integration frameworks by \cite{Heydenreich_2023} and \cite{gomes2025cosmologysecondthirdordershear}. These expressions involve integrals containing Bessel functions $J_2$ and $J_6$, which exhibit strong oscillatory behavior, making direct numerical evaluation challenging. The expression for $\Gamma_{(0)}$ is given by
\begin{eqnarray}\label{eq:A1}
    \Gamma_{(0)}^{ijk}(\vartheta_1,\vartheta_2,\vartheta_3) &=& 2\pi\int_0^{\infty}\frac{\ell_1\,d\ell_1}{(2\pi)^2} \int_0^{\infty}\frac{\ell_2\,d\ell_2}{(2\pi)^2} \int_0^{2\pi}d\varphi\, B^{ijk}_{\kappa\kappa\kappa}(\ell_1,\ell_2,\varphi)\, e^{2i\bar{\beta}} \\ \nonumber
    && \qquad \qquad \qquad \qquad \qquad \qquad \times \Big[e^{i(\phi_1-\phi_2-6\alpha_3)}J_6(A_3) \\ \nonumber
    && \qquad \qquad \qquad \qquad \qquad \qquad \quad 
    +\ e^{i(\phi_2-\phi_3-6\alpha_1)}J_6(A_1) \\ \nonumber
    && \qquad \qquad \qquad \qquad \qquad \qquad \quad 
    +\  e^{i(\phi_3-\phi_1-6\alpha_2)}J_6(A_2)\Big].
\label{Gamma0eq}
\end{eqnarray}

Similarly, the expression for $\Gamma_{(1)}$ is
\begin{eqnarray}
    \Gamma_{(1)}^{ijk}(\vartheta_1,\vartheta_2,\vartheta_3) &=& 2\pi\int_0^{\infty}\frac{\ell_1\,d\ell_1}{(2\pi)^2} \int_0^{\infty}\frac{\ell_2\,d\ell_2}{(2\pi)^2} \int_0^{2\pi}d\varphi\, B^{ijk}_{\kappa\kappa\kappa}(\ell_1,\ell_2,\varphi)\, e^{2i\bar{\beta}}  \\ \nonumber
    && \qquad \qquad \qquad \qquad \qquad \qquad \times \Big[ e^{i(\phi_1- \phi_2 +2\phi_3 +2\bar{\beta} -2\varphi -2\alpha_3)}J_2(A_3) \\ \nonumber
    && \qquad \qquad \qquad \qquad \qquad \qquad \quad 
    +\ e^{i(\phi_3-\phi_2-2\bar{\beta}-2\alpha_1)}J_2(A_1) \\  \nonumber
    && \qquad \qquad \qquad \qquad \qquad \qquad \quad 
    +\ e^{i(\phi_3-\phi_1-2\phi_2+2\bar{\beta}+2\varphi-2\alpha_2)}J_2(A_2) \Big] . \label{Gamma1eq}
\end{eqnarray}

In these expressions, the convergence bispectrum is parameterized by the wavevector magnitudes $\ell_1$, $\ell_2$, and the angle $\varphi$ between them, with $\ell_3 = \sqrt{\ell_1^2 + \ell_2^2 - 2\ell_1\ell_2\cos\varphi}$. The angle $\bar{\beta}$ represents the orientation between $\ell_3$ and the average direction of $\ell_1$ and $\ell_2$. The remaining components, $\Gamma_{(2)}$ and $\Gamma_{(3)}$, are obtained by cyclically permuting the indices in the expression for $\Gamma_{(1)}$.

The auxiliary quantities $A_3$ and $\alpha_3$ are defined through
\begin{eqnarray}\label{eq:A3}
A_3\sin{\alpha_3} &=& (\ell_1x_2 - \ell_2x_1)\sin{\left(\frac{\varphi + \phi_3}{2}\right)}, \\ \label{eq:A4}
A_3\cos{\alpha_3} &=& (\ell_1x_2 + \ell_2x_1)\cos{\left(\frac{\varphi + \phi_3}{2}\right)},
\end{eqnarray}

with $A_1$, $A_2$, $\alpha_1$, and $\alpha_2$ similarly defined via cyclic permutations.

\subsection*{Galaxy--shear--shear 3PCF}\label{App:gss_3PCF}
\par\noindent
\crefalias{section}{appendix}

The components of the galaxy--shear--shear 3PCF ($G_\pm$) can be expressed in terms of the galaxy--convergence--convergence bispectrum. Following the original formulation of \cite{Schneider_2005b}, $G_\pm$ can be expressed as 
\begin{eqnarray}\label{eq:A5}
    G_+^{ijk}(\vartheta_1,\vartheta_2,\phi_3) &=& \int_0^{\infty}\frac{\ell_1\,d\ell_1}{2\pi} \int_0^{\infty}\frac{\ell_2\,d\ell_2}{2\pi} \int_0^{2\pi}\frac{d\psi}{2\pi}\, B_{{\rm g}\kappa\kappa}^{ijk}(\ell_1,\ell_2,\psi)\, e^{-2i(\phi_3-\psi)} J_0(A), \\ \label{eq:A6}
    G_-^{ijk}(\vartheta_1,\vartheta_2,\phi_3) &=& \int_0^{\infty}\frac{\ell_1\,d\ell_1}{2\pi} \int_0^{\infty}\frac{\ell_2\,d\ell_2}{2\pi} \int_0^{2\pi}\frac{d\psi}{2\pi}\, B_{{\rm g}\kappa\kappa}^{ijk}(\ell_1,\ell_2,\psi)\, e^{4i\nu} J_4(A).
\label{G_pm}
\end{eqnarray}

In these expressions, the galaxy--convergence--convergence bispectrum is parameterized by the wavevector magnitudes $\ell_1$, $\ell_2$, and the angle $\psi$ between them. 
% with $\ell_3 = \sqrt{\ell_1^2 + \ell_2^2 - 2\ell_1\ell_2\cos\psi}$. 
The angle $\phi_3$ represents the angle between $\boldsymbol{\vartheta}_1$ and $\boldsymbol{\vartheta}_2$.

The auxiliary quantities $A$ and $\nu$ are defined through
\begin{eqnarray}\label{eq:A7}
    A^2 &=& \ell_1^2\vartheta_1^2 + \ell_1^2\vartheta_1^2 - 2\ell_1\ell_2\vartheta_1\vartheta_2 {\rm cos}(\phi_3-\psi),\\ \label{eq:A8}
    {\rm e}^{2i\nu} &=& \frac{1}{A^2}\big[ 2\ell_1\ell_2\vartheta_1\vartheta_2 + (\ell_1\vartheta_1)^2 {\rm e}^{i(\phi_3-\psi)} + (\ell_2\vartheta_2)^2 {\rm e}^{-i(\phi_3-\psi)}\big],
\end{eqnarray}
where $\boldsymbol{\ell}_1 \cdot \boldsymbol{\vartheta}_1 + \boldsymbol{\ell}_2 \cdot \boldsymbol{\vartheta}_2 = A {\rm cos} (\eta-\nu)$ and $\eta = \eta' - \zeta$. 

% Next, we decompose the angles $\varphi_i$ and $\beta_i$ into their average and difference components. Specifically, we define
% \[
% \varphi_1 = \zeta - \frac{\phi_3}{2}, \quad \varphi_2 = \zeta + \frac{\phi_3}{2}, \quad \beta_1 = \eta' - \frac{\psi}{2}, \quad \beta_2 = \eta' + \frac{\psi}{2},
% \]
% where 

% Using these definitions, the scalar product of the wavevectors and position vectors can be expressed as
% \[
% \boldsymbol{\ell}_1 \cdot \boldsymbol{\theta}_1 + \boldsymbol{\ell}_2 \cdot \boldsymbol{\theta}_2 = A \cos(\eta - \nu),
% \]
% where $\eta = \eta' - \zeta$.

\subsection*{Galaxy--galaxy--shear 3PCF}\label{App:ggs_3PCF}
\par\noindent
\crefalias{subsection}{appendix}

The galaxy--galaxy--shear 3PCF ($\mathcal{G}$) can be expressed in terms of the galaxy--galaxy--convergence bispectrum. Following the original formulation of \cite{Schneider_2005b}, $\mathcal{G}$ can be expressed as 
\begin{eqnarray}
    \mathcal{G}^{ijk}(\vartheta_1,\vartheta_2,\phi_3) &=& \int_0^{\infty}\frac{\ell_1\,d\ell_1}{2\pi} \int_0^{\infty}\frac{\ell_2\,d\ell_2}{2\pi} \int_0^{2\pi}\frac{d\psi}{2\pi}\, B_{{\rm gg}\kappa}^{ijk}(\ell_1,\ell_2,\psi)\ \\ \nonumber
    && \qquad \qquad \qquad \qquad \qquad \quad 
    \times \frac{\big(\ell_1 {\rm e}^{-i\psi/2} + \ell_2 {\rm e}^{-i\psi/2}\big)^2}{|\ell|^2}\ e^{2i\nu} J_2(A),
\label{G}
\end{eqnarray}
where all the variables are described above.

\subsection*{Galaxy--galaxy--galaxy 3PCF}\label{App:ggg_3PCF}
\par\noindent
\crefalias{subsection}{appendix}

The galaxy--galaxy--galaxy 3PCF ($\zeta$) can be expressed in terms of the galaxy clustering bispectrum. Following the original formulation of \cite{arvizu2024}, $\zeta$ can be expressed as 
% \citep{yue2024paircountingbinning}
\begin{eqnarray}
\zeta^{ijk}(\vartheta_1,\vartheta_2,\phi_3) 
% &=& 2\pi\int_0^{\infty}\frac{\ell_1\,d\ell_1}{(2\pi)^2} \int_0^{\infty}\frac{\ell_2\,d\ell_2}{(2\pi)^2} \int_0^{\infty}\frac{\ell_3\,d\ell_3}{(2\pi)^2} \ {\rm exp}[{\rm i}(\ell_1\cdot\vartheta_1 + \ell_2\cdot\vartheta_2 + \ell_3\cdot\vartheta_3)] \ \delta_{\rm D} (\ell_1+\ell_2+\ell_3) B_{\rm ggg}(\ell_1, \ell_2, \ell_3),\\
&=& \int_0^{\infty}\frac{\ell_1\,d\ell_1}{2\pi} \int_0^{\infty}\frac{\ell_2\,d\ell_2}{2\pi} \ J_0(\ell_1 \vartheta_1)\ J_0(\ell_2 \vartheta_2)\ B_{\rm ggg}^{ijk}(\ell_1, \ell_2, \psi).\label{eq:A10}
\end{eqnarray}
% where all the variables are described in the sub\cref{App:A2}.
In this expression, the galaxy clustering bispectrum is parameterized by the wavevector magnitudes $\ell_1$, $\ell_2$, and the angle $\psi$ between them, with $\ell_3 = \sqrt{\ell_1^2 + \ell_2^2 - 2\ell_1\ell_2\cos\psi}$. The angle $\phi_3$ represents the angle between $\boldsymbol{\vartheta}_1$ and $\boldsymbol{\vartheta}_2$.

% \clearpage
\section{Estimators} 
\par\noindent\label{App:estimators}
\crefalias{section}{appendix}

% Treecorr, 2t, 3pt... angular scales and binning, accuracy parameters, randoms, downsampling.
We introduce the estimators for $3\times2$PCFs and $4\times3$PCFs
% estimators in a simplified version, where the weights are all taken to be unity. Here we generalize these equations and define 3PCF estimators in the case 
%where the galaxy catalogs are associated with a weight per galaxy
including galaxy weights here, either associated with lensing or clustering.

\subsection{$3\times2$PCFs -- estimators}\label{App:estimators_2PCF}
\par\noindent
\crefalias{subsection}{appendix}

We adopt the estimators to measure the 2PCFs in a survey by generalizing the usual Landy-Szalay (LS) estimator for the galaxy--galaxy correlation function developed by \cite{Mandelbaum_2006} and \cite{Hirata_2007}. 
We denote the data catalog and the random catalog by $D$ and $R$, respectively.

\subsubsection*{Galaxy--galaxy 2PCF}\label{App:estimator_gg}
\par\noindent
% \crefalias{subsection}{appendix}

The estimator for galaxy--galaxy (clustering) 2PCF is defined as
\begin{eqnarray}\label{eq:B1}
       \widehat{w}^{ij}(\vartheta) = \frac{D^iD^j - D^iR^j - R^iD^j + R^iR^j}{R^iR^j},
\end{eqnarray} 
where $DD$ and $RR$ are the sum over all galaxy pairs with separation $\vartheta$ from the data catalog and the random catalog, respectively.  Similarly, $DR$ and $RD$ are the sum over all galaxy pairs with separation $\vartheta$ between the data catalog and the random catalog. These can be expressed as 
\begin{eqnarray}\label{eq:B2}
   D^iD^j = \frac{\sum_{\rm pairs} w^i_{\rm g} w^j_{\rm g}  n_{\rm g}^i n_{\rm g}^j } {\sum_{\rm pairs} w^i_{\rm g} w^j_{\rm g}};\ \ &&  
   D^iR^j = \frac{\sum_{\rm pairs} w^i_{\rm g} w^j_{\rm r}  n_{\rm g}^i n_{\rm r}^j } {\sum_{\rm pairs} w^i_{\rm g} w^j_{\rm r}}; \\    
   R^iD^j = \frac{\sum_{\rm pairs} w^i_{\rm r} w^j_{\rm g}  n_{\rm r}^i n_{\rm g}^j } {\sum_{\rm pairs} w^i_{\rm r} w^j_{\rm g}}; \ \ && 
   R^iR^j = \frac{\sum_{\rm pairs} w^i_{\rm r} w^j_{\rm r}  n_{\rm r}^i n_{\rm r}^j } {\sum_{\rm pairs} w^i_{\rm r} w^j_{\rm r}},
\end{eqnarray}
where $n^i_{\rm g}$ and $n^i_{\rm r}$ represent the position of galaxy in the $i^{\rm th}$ tomographic bin from the data catalog and random catalog with weights $w^i_{\rm g}$ and $w^i_{\rm r}$, respectively. 

\subsubsection*{Galaxy--shear 2PCF}\label{App:estimator_gs}
\par\noindent

The estimator for the galaxy--shear 2PCF (or GGL) is defined as
\begin{eqnarray}\label{eq:B3}
    \widehat{\gamma}_{\rm t}^{ij}(\vartheta) = (D^i-R^i)S^j,
\end{eqnarray} 
where $DS$ and $RS$ are the sum over all position-source pairs and position-random pairs with separations $\vartheta$ of the tangential component of shear in the data catalog, and can be expressed as  
\begin{eqnarray}\label{eq:B4}
    D^iS^j =  \frac{\sum_{\rm pairs} w^i_{\rm g} w^j_{\rm g}  n_{\rm g}^i \varepsilon^{j} } {\sum_{\rm pairs}  w^i_{\rm g} w^j_{\rm g} }; \ \ 
    R^iS^j =  \frac{\sum_{\rm pairs} w^i_{\rm r} w^j_{\rm g}  n_{\rm r}^i \varepsilon^{j} } {\sum_{\rm pairs}  w^i_{\rm r} w^j_{\rm g} }. 
\end{eqnarray}
where the tangential component of the ellipticity of the galaxy at $j^{\rm th}$ tomographic bin $(\varepsilon^{j})$ is measured relative to the galaxy and randoms at $i^{\rm th}$ tomographic bin $(n^i_{\rm g})$ and $(n^i_{\rm r})$, respectively.

\subsubsection*{Shear--shear 2PCF}\label{App:estimator_ss}
\par\noindent

The estimators for the shear--shear 2PCFs are defined as
\begin{eqnarray}\label{eq:B5}
    \widehat{\xi}_{\pm}^{ij}(\vartheta) = S_\pm^i S_\pm^j,
\end{eqnarray} 
where $S_{\pm}S_{\pm}$ is the sum over all pairs with separations $\vartheta$ of shear in the data catalog, and can be expressed as  
\begin{eqnarray}\label{eq:B6}
    S_{+}^iS_{+}^j =  \frac{\sum_{\rm pairs} w^i_{\rm g} w^j_{\rm g} \varepsilon^{i} \varepsilon^{*j} } {\sum_{\rm pairs} w^i_{\rm g} w^j_{\rm g} }; \ \ 
    S_{-}^iS_{-}^j =  \frac{\sum_{\rm pairs} w^i_{\rm g} w^j_{\rm g} \varepsilon^{i} \varepsilon^{j} } {\sum_{\rm pairs} w^i_{\rm g} w^j_{\rm g} }, 
\end{eqnarray}
where $\varepsilon^{i}$ and $\varepsilon^{j}$ are the measured ellipticities of the galaxies at $i^{\rm th}$ and $j^{\rm th}$ tomographic bins, respectively.

\subsection{$4\times3$PCFs -- estimators}\label{App:estimators_3PCF}
\par\noindent
\crefalias{subsection}{appendix}

We denote the vertices of the triangle located at $\boldsymbol{\theta}_1$ (tomograhic bin $i$), $\boldsymbol{\theta}_2$ (tomograhic bin $j$), and $\boldsymbol{\theta}_3$ (tomograhic bin $k$).

\subsubsection*{Galaxy--galaxy--galaxy 3PCF} \label{App:estimator_ggg}
\par\noindent

The estimator for galaxy--galaxy--galaxy 3PCF is defined as
\begin{eqnarray}\label{eq:B7}
       \widehat{\zeta}^{ijk}(\vartheta_1, \vartheta_2, \vartheta_3) &=& \frac{(D^i-R^i)(D^j-R^j)(D^k-R^k)}{R^iR^jR^k}, \\ \nonumber
       % &=&\frac{D^iD^jD^k - (D^iD^jR^k + D^iR^jD^k + R^iD^jD^k)  + (D^iR^jR^k + R^iD^jR^k + R^iR^jD^k) - R^iR^jR^k}{R^iR^jR^k},
       &=&\frac{D^iD^jD^k - (D^iD^jR^k + \mathrm{2\  perm.})  + (D^iR^jR^k + \mathrm{2\  perm.}) - R^iR^jR^k}{R^iR^jR^k},
\end{eqnarray} 
where $DDD$ and $RRR$ are the sum over all galaxy triplets located at $\boldsymbol{\theta}_1$, $\boldsymbol{\theta}_2$, and $\boldsymbol{\theta}_3$ from the data catalog and the random catalog, respectively. $DDR$ are the sum over all galaxy triplets with a pair of galaxies from the data catalog and the third galaxy from the random catalog. Similarly, $DRR$ are the sum over all galaxy triplets with a pair of galaxies from the random catalog and the third galaxy from the data catalog. 
They can be expressed as 
\begin{eqnarray}\nonumber
   D^iD^jD^k = \frac{\sum_{\rm triplets} w^i_{\rm g} w^j_{\rm g} w^k_{\rm g}  n^i_{\rm g} n^j_{\rm g} n^k_{\rm g} } {\sum_{\rm triplets} w^i_{\rm g} w^j_{\rm g} w^k_{\rm g}};  \ \ &&
   D^iD^jR^k = \frac{\sum_{\rm triplets} w^i_{\rm g} w^j_{\rm g} w^k_{\rm r}  n^i_{\rm g} n^j_{\rm g} n^k_{\rm r} } {\sum_{\rm triplets} w^i_{\rm g} w^j_{\rm g} w^k_{\rm r}}, \\
   D^iR^jR^k = \frac{\sum_{\rm triplets} w^i_{\rm g} w^j_{\rm r} w^k_{\rm r}  n^i_{\rm g} n^j_{\rm g} n^k_{\rm r} } {\sum_{\rm triplets} w^i_{\rm g} w^j_{\rm r} w^k_{\rm r}}; \ \  &&
   R^iR^jR^k = \frac{\sum_{\rm triplets} w^i_{\rm r} w^j_{\rm r} w^k_{\rm r}  n^i_{\rm g} n^j_{\rm r} n^k_{\rm r} } {\sum_{\rm triplets} w^i_{\rm r} w^j_{\rm r} w^k_{\rm r}},
\end{eqnarray}
where $n^i_{\rm g}$ and $n^i_{\rm r}$ represent the position of galaxy in the $i^{\rm th}$ tomographic bin from the data catalog and random catalog with weights $w^i_{\rm g}$ and $w^i_{\rm r}$, respectively.    
($D^iR^jD^k, R^iD^jD^k$) and ($R^iD^jR^k, R^iR^jD^k$) can be expressed  similarly as $D^iD^jR^k$ and $D^iR^jR^k$, respectively by replacing $n_{\rm g}$ by $n_{\rm r}$ appropriately.

\subsubsection*{Galaxy--galaxy--shear 3PCF}\label{App:estimator_ggs}
\par\noindent

For the galaxy--galaxy--shear 3PCF, the estimator should count the triplets of galaxies from the catalog, and accumulate the shear of the third galaxy given two other galaxies whose positions are being measured in each triangle configuration. Depending on the ordering of the lens and source tomographic bins, we can construct three observables (ggG, gGg, Ggg) and the estimators for these three observables are defined as 
\begin{eqnarray}\label{eq:B10}
   \widehat{\mathcal{G}}_{\rm ggG}^{ijk}(\vartheta_1, \vartheta_2, \vartheta_3) &=& (D^i-R^i)(D^j-R^j)S^k\\ \nonumber
   &=& D^iD^jS^k - D^iR^jS^k-R^iD^jS^k + R^iR^jS^k,\\\label{eq:B11}
   \widehat{\mathcal{G}}_{\rm gGg}^{ijk}(\vartheta_1, \vartheta_2, \vartheta_3) &=& (D^i-R^i)S^j(D^k-R^k) \\ \nonumber
   &=& D^iS^jD^k - D^iS^jR^k-R^iS^jD^k + R^iS^jR^k,\\\label{eq:B12}
   \widehat{\mathcal{G}}_{\rm Ggg}^{ijk}(\vartheta_1, \vartheta_2, \vartheta_3) &=& S^i(D^j-R^j)(D^k-R^k) \\ \nonumber
   &=& S^iD^jD^k -S^iD^jR^k-S^iR^jD^k + S^iR^jR^k,
\end{eqnarray}
where $DDS$ and $RRS$ type estimators represent the sum over all triplets of galaxies, where the shear of a background galaxy is correlated with the positions of the two foreground galaxies from the data catalog and random catalog, respectively. $DRS$ type estimators represent the sum over all triplets of galaxies, where the shear of a background galaxy is correlated with the positions of the two foreground galaxies -- one from the data catalog and the other from the random catalog.
They can be expressed as 
\begin{eqnarray}\nonumber
    D^iD^jS^k = \frac{\sum_{\rm triplets} w^i_{\rm g} w^j_{\rm g} w^k_{\rm g}  n^i_{\rm g} n^j_{\rm g} \varepsilon^k} {\sum_{\rm triplets} w^i_{\rm g} w^j_{\rm g} w^k_{\rm g}}; && \ \
    D^iR^jS^k = \frac{\sum_{\rm triplets} w^i_{\rm g} w^j_{\rm r} w^k_{\rm g}  n^i_{\rm g} n^j_{\rm r} \varepsilon^k} {\sum_{\rm triplets} w^i_{\rm g} w^j_{\rm r} w^k_{\rm g}}; \\
    R^iR^jS^k &=& \frac{\sum_{\rm triplets} w^i_{\rm r} w^j_{\rm r} w^k_{\rm g}  n^i_{\rm r} n^j_{\rm r} \varepsilon^k} {\sum_{\rm triplets} w^i_{\rm r} w^j_{\rm r} w^k_{\rm g}}
\end{eqnarray}
where $\varepsilon^k$, which is the ellipticity of the galaxy at $k^{\rm th}$ tomographic bin, is correlated with the positions of the two galaxies from $i^{\rm th}$, and $j^{\rm th}$ tomographic bins. The rest of the estimators can be expressed similarly by replacing $n_{\rm g}$ by $n_{\rm r}$ appropriately.

\subsubsection*{Galaxy--shear--shear 3PCF}\label{App:estimator_gss}
\par\noindent

For the galaxy--shear--shear 3PCF, the estimator should count the triplets of galaxies from the catalog, and accumulate the product of shears of the two galaxies whose shapes are being measured relative to the third galaxy whose position is being measured. Depending on the ordering of the lens and source tomographic bins, we can construct three observables (gGG, GgG, GGg) and the estimators for these three observables are defined as 
\begin{eqnarray}\label{eq:B14}
   \widehat{G}^{ijk}_{\rm gGG, \pm}(\vartheta_1, \vartheta_2, \vartheta_3) = (D^i-R^i) S^j_{\pm}S^k_{\pm} = D^iS^j_{\pm}S^k_{\pm} - R^iS^j_{\pm}S^k_{\pm}, \\\label{eq:B15}
   \widehat{G}^{ijk}_{\rm GgG, \pm}(\vartheta_1, \vartheta_2, \vartheta_3) =  S^i_{\pm}(D^j-R^j)S^k_{\pm} = S^i_{\pm}D^jS^k_{\pm} - S^i_{\pm}R^jS^k_{\pm}, \\\label{eq:B16}
   \widehat{G}^{ijk}_{\rm GGg, \pm}(\vartheta_1, \vartheta_2, \vartheta_3) =  S^i_{\pm}S^j_{\pm}(D^k-R^k) = S^i_{\pm}S^j_{\pm}D^k - S^i_{\pm}S^j_{\pm}R^k,
\end{eqnarray}
where $DS_{\pm}S_{\pm}$ and $RS_{\pm}S_{\pm}$ are the sum over all triplets of galaxies, where the shapes of two background galaxies are being correlated with the position of a foreground galaxy from the data catalog and random catalog, respectively. They can be expressed as 
\begin{eqnarray}\nonumber
    D^iS_{+}^jS_{+}^k = \frac{\sum_{\rm triplets} w^i_{\rm g} w^j_{\rm g} w^k_{\rm g}  n^i_{\rm g} \varepsilon^j \varepsilon^{*k}} {\sum_{\rm triplets} w^i_{\rm g} w^j_{\rm g} w^k_{\rm g} };\ \ &&
    R^iS_{+}^jS_{+}^k = \frac{\sum_{\rm triplets} w^i_{\rm r} w^j_{\rm g} w^k_{\rm g}  n^i_{\rm g} \varepsilon^j \varepsilon^{*k}} {\sum_{\rm triplets} w^i_{\rm r} w^j_{\rm g} w^k_{\rm g} }, \\
    D^iS_{-}^jS_{-}^k = \frac{\sum_{\rm triplets} w^i_{\rm g} w^j_{\rm g} w^k_{\rm g}  n^i_{\rm g} \varepsilon^j \varepsilon^{k}} {\sum_{\rm triplets} w^i_{\rm g} w^j_{\rm g} w^k_{\rm g} };\ \ &&
    R^iS_{-}^jS_{-}^k = \frac{\sum_{\rm triplets} w^i_{\rm r} w^j_{\rm g} w^k_{\rm g}  n^i_{\rm g} \varepsilon^j \varepsilon^{k}} {\sum_{\rm triplets} w^i_{\rm r} w^j_{\rm g} w^k_{\rm g} }, 
    % S_{+}S_{+}D = \frac{\sum_{ijk} w_i w_j w_k  n_i^D \epsilon_j \epsilon_k^*} {\sum_{ijk} w_i w_j w_k}, \\
    % S_{-}S_{-}D = \frac{\sum_{ijk} w_i w_j w_k  n_i^D \epsilon_j \epsilon_k} {\sum_{ijk} w_i w_j w_k}
\end{eqnarray}
where $\varepsilon^j$ and $\varepsilon^k$ are the ellipticities of galaxies at $j^{\rm th}$ and $k^{\rm th}$ tomographic bins, respectively and are correlated with the position of the third galaxy from $i^{\rm th}$ tomographic bin. The rest of the estimators can be expressed similarly by replacing $n_{\rm g}$ by $n_{\rm r}$ appropriately.

\subsubsection*{Shear--shear--shear 3PCF}\label{App:estimator_sss}
\par\noindent

For the shear--shear--shear 3PCF, the estimator should count the triplets of galaxies in the survey and accumulate the product of their shears in tangential and cross-orientations. The estimators for the natural components of the shear--shear--shear 3PCF are
\begin{eqnarray}\label{eq:B19}
     \widehat{\Gamma}_{(0)}^{ijk}(\vartheta_1, \vartheta_2, \vartheta_3)  &=& \frac{\sum_{\rm tripltets} w_{\rm g}^i w_{\rm g}^j w_{\rm g}^k \varepsilon^i \varepsilon^j \varepsilon^k}{\sum_{\rm tripltets} w_{\rm g}^i w_{\rm g}^j w_{\rm g}^k}, \\ \label{eq:B20}
    \widehat{\Gamma}_{(1)}^{ijk}(\vartheta_1, \vartheta_2, \vartheta_3)  &=& \frac{\sum_{\rm tripltets} w_{\rm g}^i w_{\rm g}^j w_{\rm g}^k \varepsilon^{*i} \varepsilon^j \varepsilon^k}{\sum_{\rm tripltets} w_{\rm g}^i w_{\rm g}^j w_{\rm g}^k}, \\ \label{eq:}
    \widehat{\Gamma}_{(2)}^{ijk}(\vartheta_1, \vartheta_2, \vartheta_3)  &=& \frac{\sum_{\rm tripltets} w_{\rm g}^i w_{\rm g}^j w_{\rm g}^k \varepsilon^i \varepsilon^{*j} \varepsilon^k}{\sum_{\rm tripltets} w_{\rm g}^i w_{\rm g}^j w_{\rm g}^k}, \\ \label{eq:}
    \widehat{\Gamma}_{(3)}^{ijk}(\vartheta_1, \vartheta_2, \vartheta_3)  &=& \frac{\sum_{\rm tripltets} w_{\rm g}^i w_{\rm g}^j w_{\rm g}^k \varepsilon^i \varepsilon^j \varepsilon^{*k}}{\sum_{\rm tripltets} w_{\rm g}^i w_{\rm g}^j w_{\rm g}^k},
\end{eqnarray}
where the $\varepsilon^i,\ \varepsilon^j$, and $ \varepsilon^k$ are the measured ellipticities of the galaxies from $i^{\rm th}$, $j^{\rm th}$ and $k^{\rm th}$ tomographic bins, respectively.

\begin{figure*}
    \centering
    \includegraphics[width=0.47\linewidth]{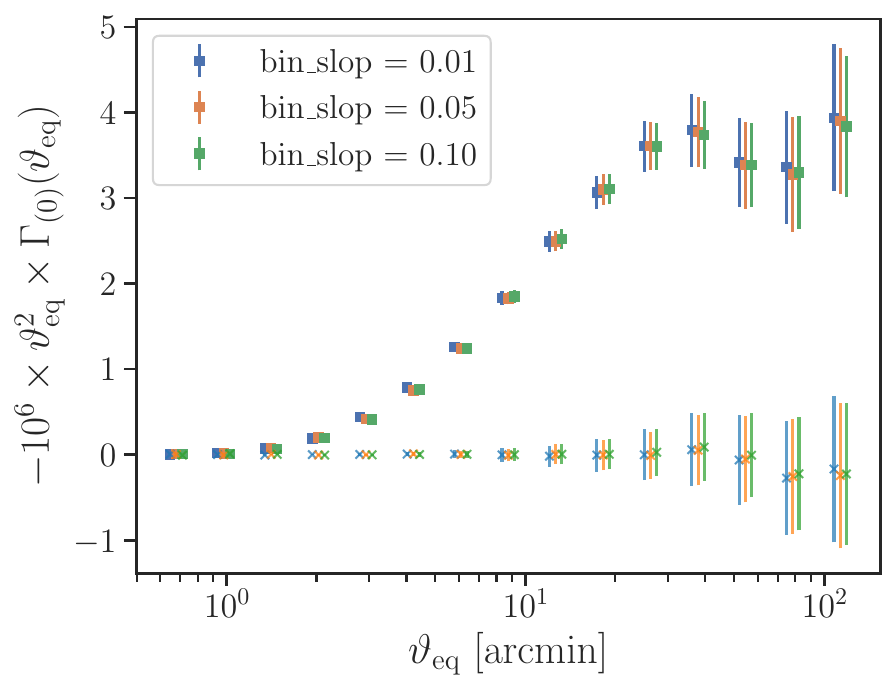}
    \includegraphics[width=0.47\linewidth]{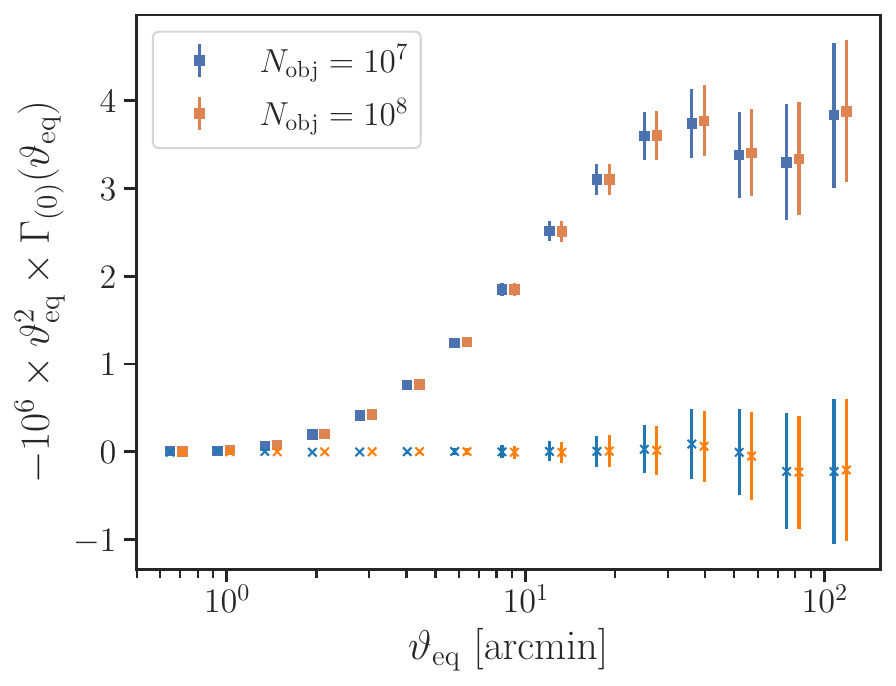}
    \caption{The two-point galaxy clustering correlation function (for the DMO case) for a particular combination of tomographic bins ($i=1, j=2, k=4$) using different choices
    for \texttt{bin\_slop} (left panel) and $N_{\rm{obj}}$ (right panel).  In both cases, the measurements change very little with different choices for
    either parameter. Data points with filled squares and crosses correspond to the real and imaginary components, respectively. The results shown in the rest of this paper use $N_{\rm{obj}} = 10^7$ and \texttt{bin\_slop} = 0.01.}
    \label{fig:accuracy}
\end{figure*}

\section{Computation time of 3PCF with variation of number of objects in catalog and \lowercase{\texttt{bin\_slop}} parameter}
\label{App:Treecorr_benchmark}
\crefalias{section}{appendix}
\par\noindent

Most of the computations presented in this paper used approximately $10^7$ galaxies per tomographic bin. Here we explore how the computation time varies both with
the number of objects being correlated and with a tunable parameter in \textsc{TreeCorr}, called \texttt{bin\_slop}. The naive scaling of a three-point correlation
calculation is $O(N^3)$. However, the tree-based algorithm used by \textsc{TreeCorr} brings this down to a more manageable asymptotic scaling of $O(N \log N)$,
assuming the \texttt{bin\_slop} parameter is sufficiently large.

The effect of the \texttt{bin\_slop} parameter in \textsc{TreeCorr} is essentially to allow the bins into which the correlation function is accumulated to be
slightly fuzzy at the edges. Triangles (or pairs for two-point correlations) that would fall very close to the edge of a bin are allowed to be
placed in the next bin over. For 3PCFs, the \texttt{bin\_slop} parameter governs how close to the edge the triangle has to be before being allowed to be placed
in the next bin. A \texttt{bin\_slop} value of 0.1 means that triangles must be within 0.1 bin width of an edge to be allowed to be placed in the next
bin instead. Typically, roughly equal numbers of triangles slop upwards as slop downwards, so there is usually very little
impact on the accuracy of the computation. But it allows \textsc{TreeCorr} to greatly increase the computational efficiency, since many triangles
may be calculated at once and placed together into a single bin.

\begin{table*}
\centering
\begin{tabular}{|c|c|c|c|}
\hline 
\textbf{No. of objects} & {\textbf{\texttt{bin\_slop}}} & \textbf{Catalog processing time} & \textbf{Computation time}\\
& & \textbf{(sec.)} & \textbf{(sec.)} \\
% \cline{2-3}
%  & $M_{\rm c}$ & $\theta_{\rm ej}$ & (arcmin)\\
\hline 
 & 0.01 & 93.48 & 277.67\\
$\mathcal{O}(10^7)$ & 0.05 & 93.74 & 187.47\\
& 0.10 & 93.22 & 166.10\\
\hline
 & 0.01 & 687.19 & 7941.53\\
$\mathcal{O}(10^8)$ & 0.05 & 696.32 & 2030.61\\ & 0.10 & 695.65 & 1228.06\\
\hline
\end{tabular}
\caption{\label{tab:C1} Computation time for shear 3PCF ($\Gamma_{(\alpha)}^{ijk}$) for a particular combination of tomographic bins ($i=1, j=2, k=4$) while varying the numbers of objects per bin and \texttt{bin\_slop} parameter. We use a single node on \textsc{NERSC} to compute the correlation.}
\end{table*}

\autoref{tab:C1} shows some computation time for a few choices of $N$ and \texttt{bin\_slop} for a particular combination of tomographic
bins ($i=1, j=2, k=4$).  We break the total time into the time it takes \textsc{TreeCorr} to read in and process the input catalog and the time
it takes \textsc{TreeCorr} to compute the 3PCF. When \texttt{bin\_slop} is 0.1, the scaling with $N$ is apparently a bit
faster than $O(N)$ for this particular geometry and choice of triangle shapes being computed. For other configurations, it will vary
somewhat, but it should normally be something similar to $O(N \log N)$, which is the theoretical asymptotic behaviour.
However, when \texttt{bin\_slop} is as low as 0.01, the scaling becomes much worse.  Processing 10 times as many objects takes over 20 times as long.

\autoref{fig:accuracy} compares the measurements for the different configurations given in \autoref{tab:C1}.  As these data do not include
any shape noise, the measurements for $10^7$ and $10^8$ objects per tomographic bin are nearly identical. Furthermore, it is clear that the
numerical noise introduced by \textsc{TreeCorr} via \texttt{bin\_slop} as large as $0.1$ is fairly small. For real data with shape noise, the numerical
noise from \texttt{bin\_slop} $= 0.1$ would be completely negligible for this kind of measurement. 
Of course, it is always worth checking whether a given choice of \texttt{bin\_slop} is sufficiently small by comparing the measurements to
the results from a smaller \texttt{bin\_slop} value (say a factor of 2 smaller), but we anticipate that \texttt{bin\_slop}=0.1 would be a good choice for most use cases. The results shown in other parts of this paper use $N_{\rm{obj}} = 10^7$ and \texttt{bin\_slop} = 0.01. 
The latter is a conservative choice based on our desire to obtain high-accuracy results, which could be relaxed in the future.

%\mj{[MJ: I think the asymptotic behavior of the TreeCorr 3PCF algorithm is O(N logN), so I wouldn't expect it to be massively slower for 10x more objects.  Ofc, you might not be that close to the asympototic behavior yet at $10^7$, but still, I would have thought it would be pretty reasonable for $10^8$ objects.  You might need to raise bin\_slop a little though.  0.01 is pretty small for 3PCF!  0.1 should be fine and not introduce much noise to the measurement.]}

% \clearpage
\section{Data vectors with baryonic suppression}
\label{App:Data_vectors}
\crefalias{section}{appendix}
\par\noindent

We show here the full data vectors for $3\times2$PCFs and $4\times3$PCFs for relevant combinations of tomographic bins. The data points represent measurements obtained from catalogs constructed using the DMO maps. The blue and the red bands illustrate the range spanned by the measurements corresponding to the minimum and maximum values of the parameters $M_{\rm c}$, and $\theta_{\rm ej}$, respectively. The bin combination is shown in parentheses in each plot. 

% \newpage
% \begin{figure*}
%     \centering
%     \includegraphics[width=\linewidth]{figures/gg_dmb_dmo.pdf}
%     \caption{The galaxy-galaxy 2PCF is shown for auto-correlating tomographic bins, where the data points represent measurements obtained from catalogs constructed using the gravity-only (DMO) maps. The blue and the red bands illustrate the range spanned by the measurements corresponding to the minimum and maximum values of the parameters $M_{\rm c}$, and $\theta_{\rm ej}$, respectively. The bin combination is shown in parentheses in each plot.}\label{fig:w_dmb_dmo}
% \end{figure*}

% \begin{figure*}
%     \centering
%     \includegraphics[width=\linewidth]{figures/gamma_t_dmb_dmo.pdf}
%     \caption{The galaxy-shear 2PCF is shown for all combinations of tomographic bins, where the data points represent measurements obtained from catalogs constructed using the gravity-only (DMO) maps. The blue and the red bands illustrate the range spanned by the measurements corresponding to the minimum and maximum values of the parameters $M_{\rm c}$, and $\theta_{\rm ej}$, respectively. The bin combination is shown in parentheses in each plot.}\label{fig:gamma_t_dmb_dmo}
% \end{figure*}

\begin{figure*}
    \centering
    \begin{subfigure}[b]{\linewidth}
        \centering
        \includegraphics[width=\linewidth]{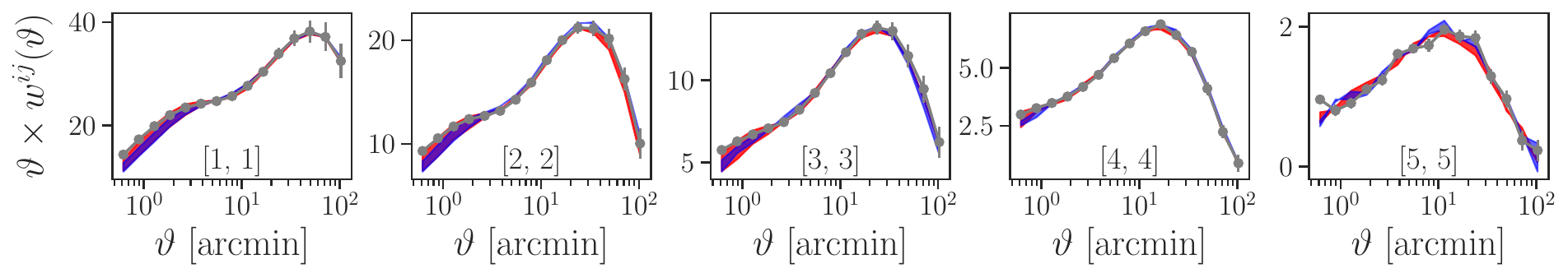}
        \caption{Galaxy-galaxy 2PCF ($w^{ij}(\vartheta)$).}
        \label{fig:w_dmb_dmo}
    \end{subfigure}
    \hfill
    \begin{subfigure}[b]{\linewidth}
        \centering
        \includegraphics[width=\linewidth]{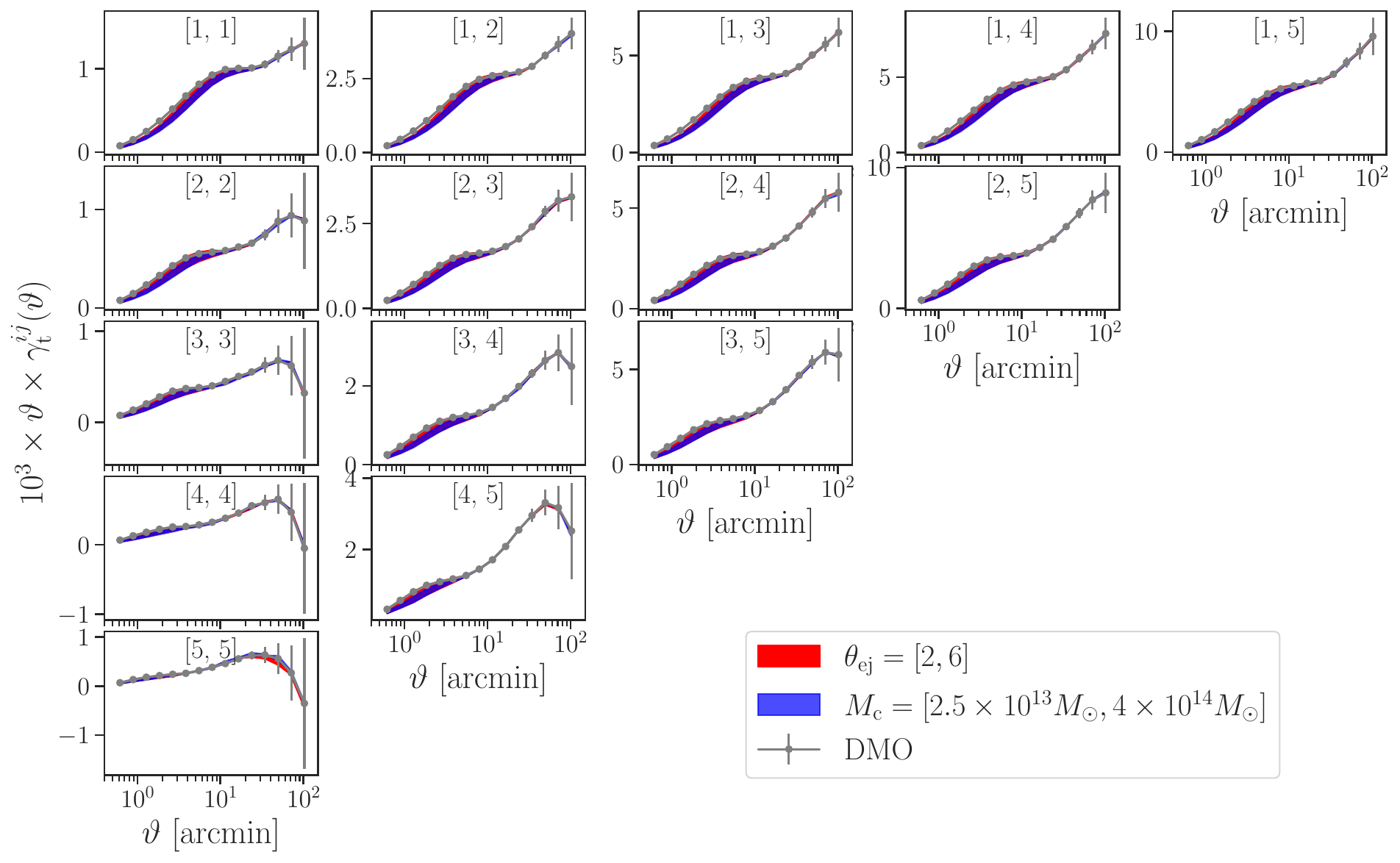}
        \caption{Galaxy-shear 2PCF ($\gamma_{\rm t}^{ij}(\vartheta))$.}
        \label{fig:gamma_t_dmb_dmo}
    \end{subfigure}
    \caption{The galaxy-galaxy 2PCF (a) and galaxy-shear 2PCF (b), shown for all tomographic bins combinations. The data points represent measurements obtained from catalogs constructed using the gravity-only (DMO) maps. The blue and the red bands illustrate the range spanned by the measurements corresponding to the minimum and maximum values of the parameters $M_{\rm c}$, and $\theta_{\rm ej}$, respectively. The bin combination is shown in parentheses in each plot.}
    \label{fig:w_gammat_combined}
\end{figure*}

\begin{figure*}
    \centering
    \begin{subfigure}[b]{\linewidth}
        \centering
        \includegraphics[width=\linewidth]{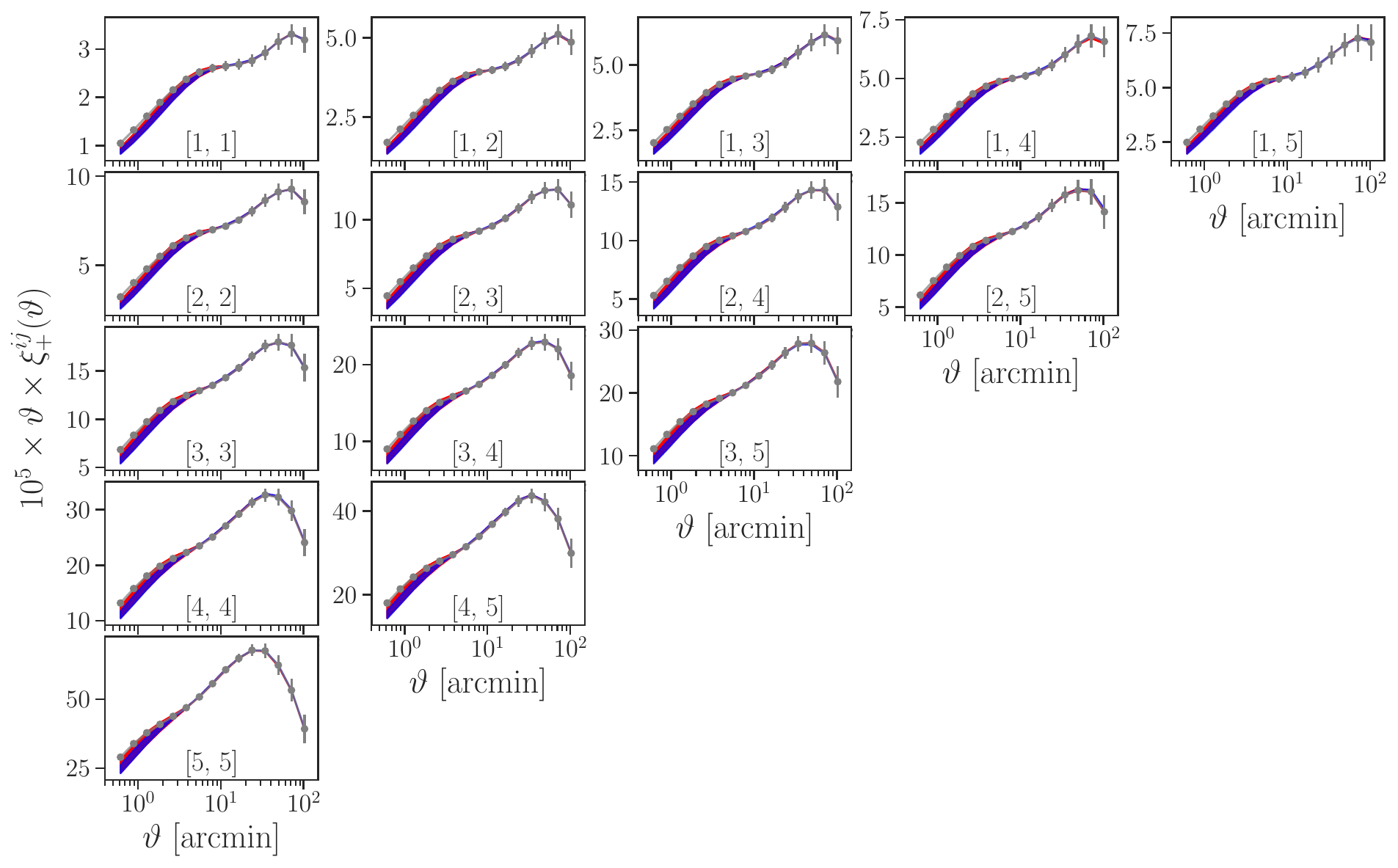}
        \caption{Shear-shear 2PCF $\xi_+$.}
        \label{fig:xi_plus_dmb_dmo}
    \end{subfigure}
    \hfill
    \begin{subfigure}[b]{\linewidth}
        \centering
        \includegraphics[width=\linewidth]{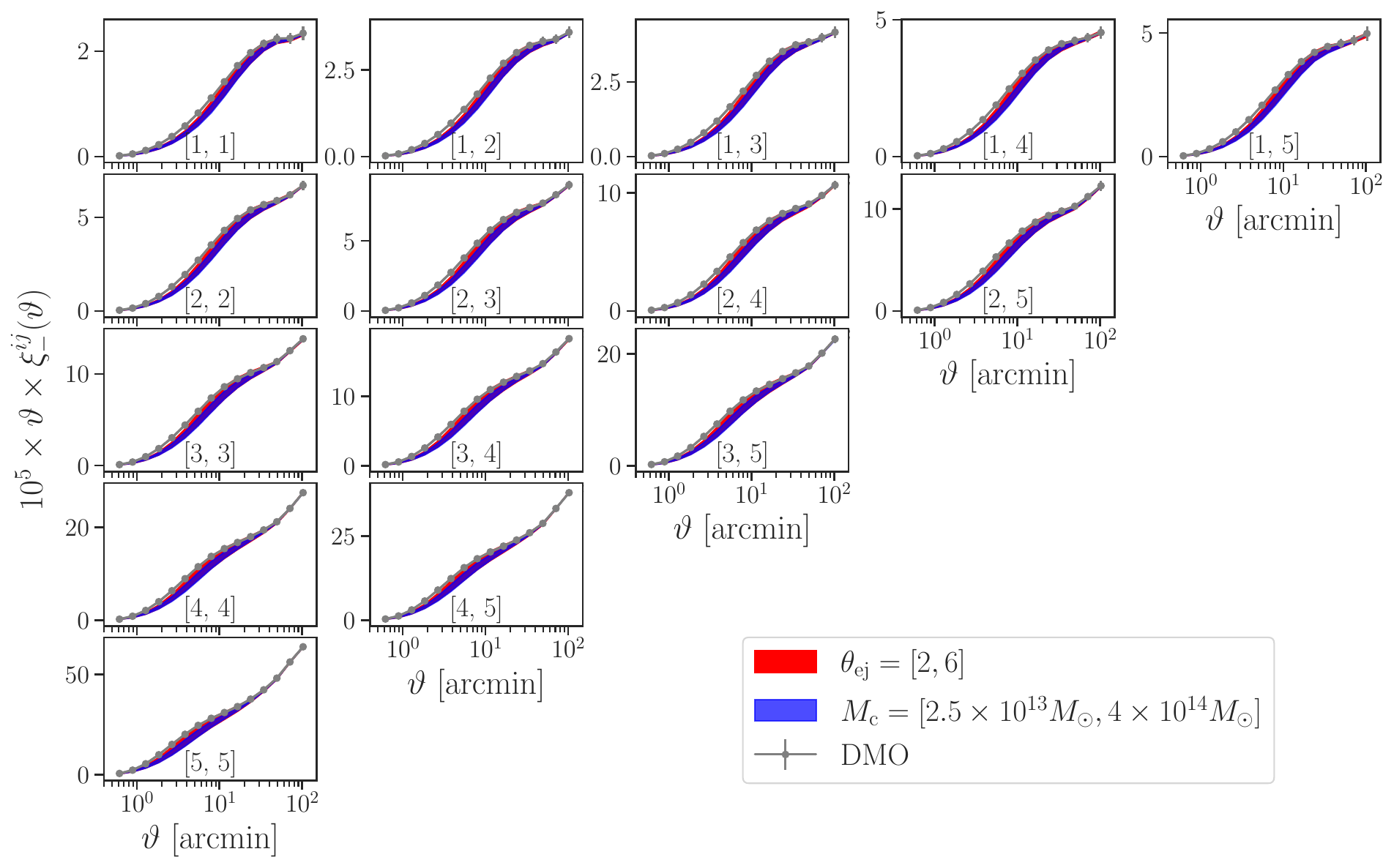}
        \caption{Shear-shear 2PCF $\xi_-$.}
        \label{fig:xi_minus_dmb_dmo}
    \end{subfigure}
    \caption{Same as \cref{fig:gamma_t_dmb_dmo}, but for $\xi_+$ (a), and $\xi_-$ (b).}
    \label{fig:xi_combined}
\end{figure*}

\begin{figure*}
    \centering
    \includegraphics[width=\linewidth]{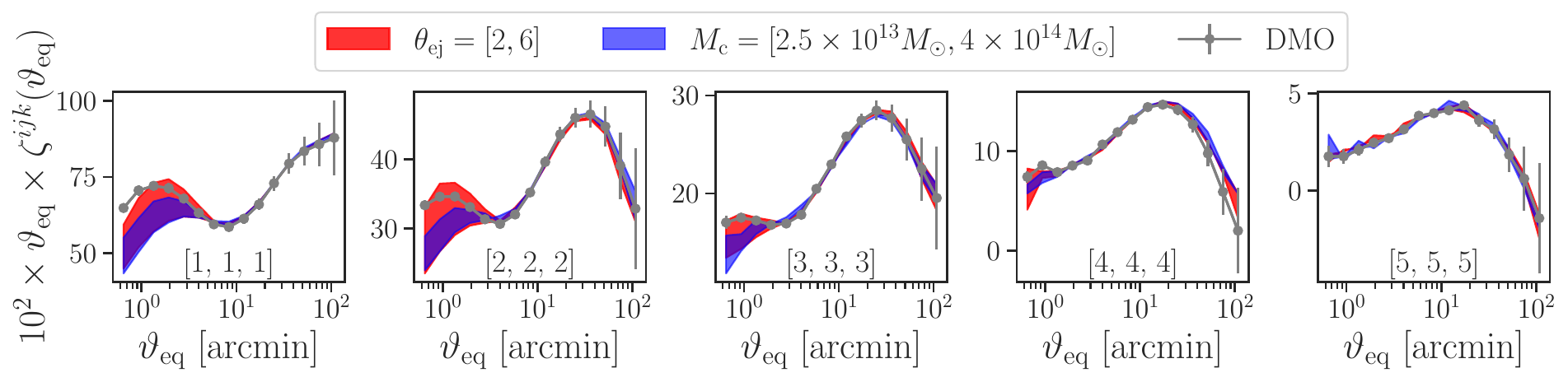}
    \caption{Same as \cref{fig:w_dmb_dmo}, but for galaxy--galaxy--galaxy 3PCF. }\label{fig:ggg_dmb_dmo_allbin}
\end{figure*}

\begin{figure*}
    \centering
    \includegraphics[width=0.75\linewidth]{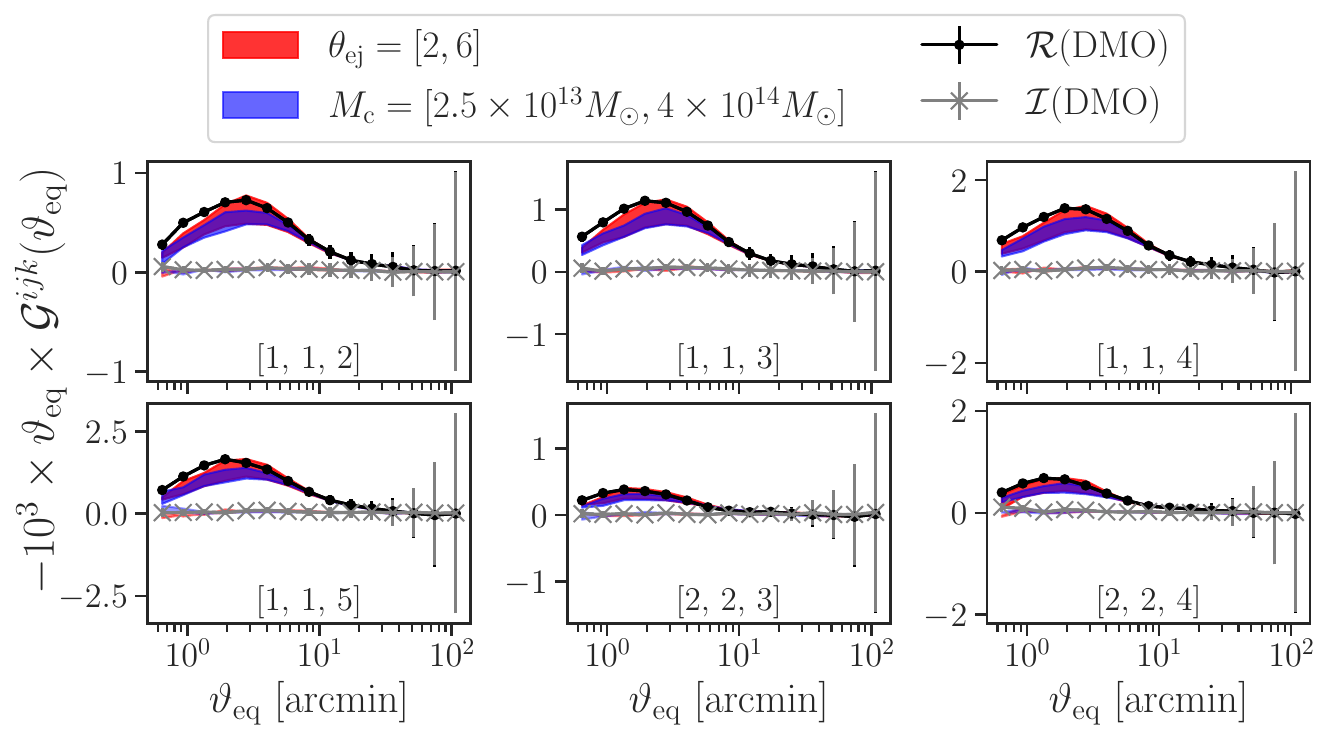}
    \caption{Same as \cref{fig:gamma_t_dmb_dmo}, but for galaxy--galaxy--shear 3PCF,  for a representative selection of tomographic bins. Data points shown in black and grey correspond to the real and imaginary components, respectively. }\label{fig:ggG_dmb_dmo_allbin}
\end{figure*}

\begin{figure*}
    \centering
    \includegraphics[width=0.6\linewidth]{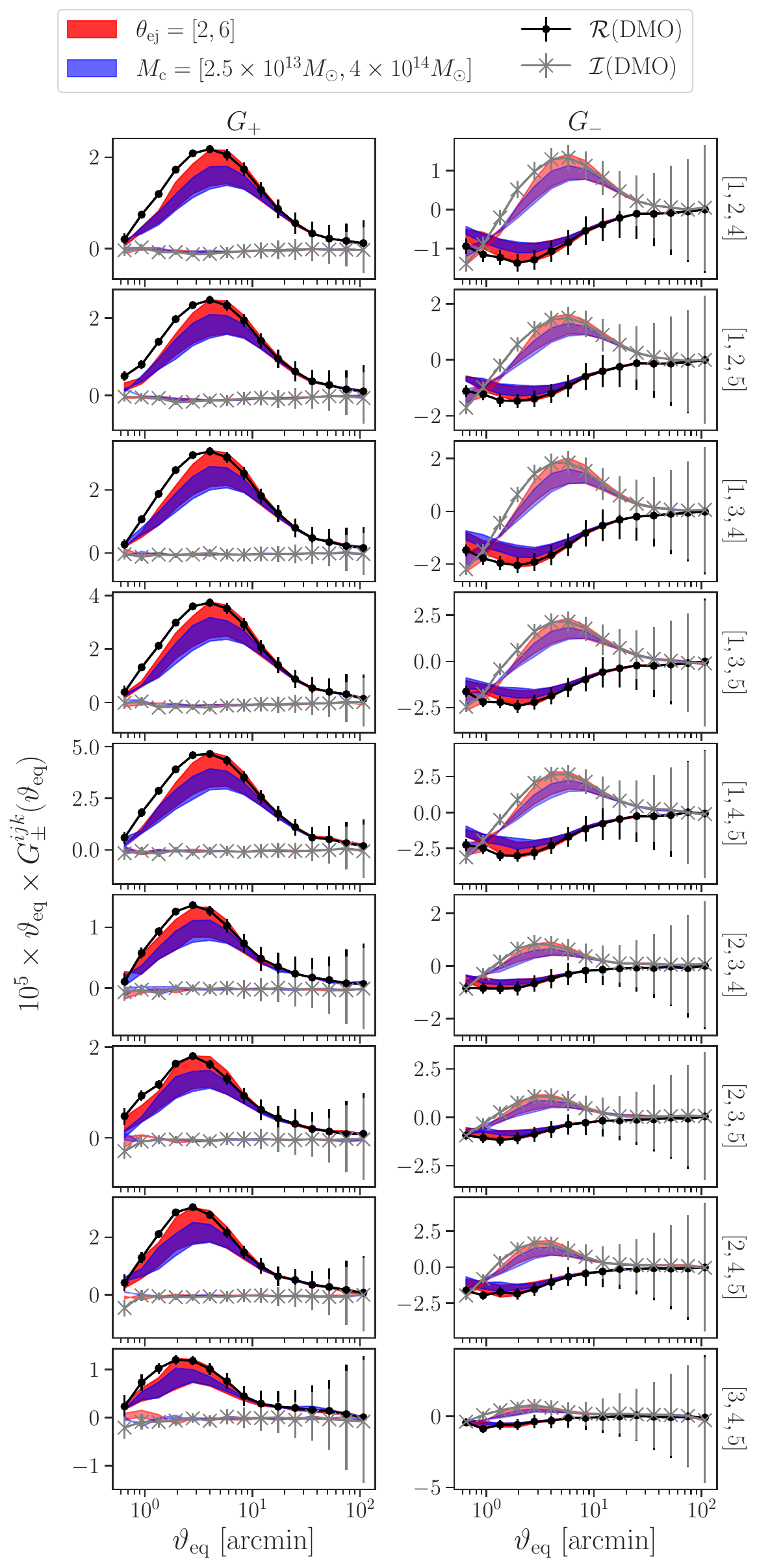}
    \caption{Same as \cref{fig:gamma_t_dmb_dmo}, but for galaxy--galaxy--shear 3PCF for a representative selection of  tomographic bins. The left and right columns correspond to $G_+$ and $G_-$, respectively. }
    \label{fig:gGG_dmb_dmo_allbin}
\end{figure*}

\begin{figure*}
    \centering
    \includegraphics[width=\linewidth]{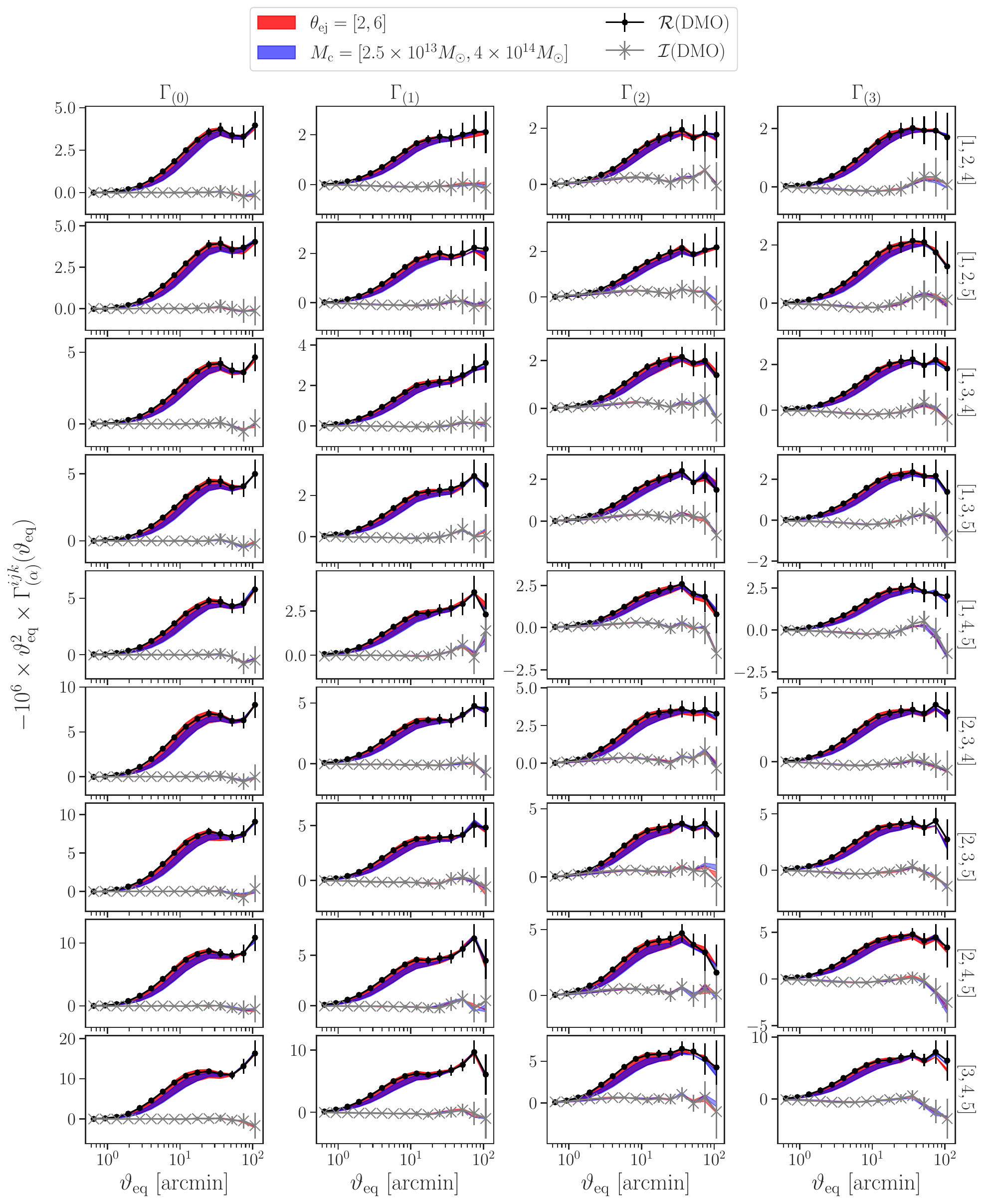}
    \caption{Same as \cref{fig:gamma_t_dmb_dmo}, but for shear--shear--shear 3PCF. Each column corresponds to a different natural component (0-3).}
    \label{fig:Gamma_dmb_dmo_allbin}
\end{figure*}

% %%%%%%%%%%%%%%%%%%%% REFERENCES %%%%%%%%%%%%%%%%%%

% % The best way to enter references is to use BibTeX:
% \bibliographystyle{unsrt}
\bibliographystyle{JHEP}
\bibliography{citation} % if your bibtex file is called example.bib

\end{document}